\documentclass[a4paper,11pt]{article}
\usepackage[utf8]{inputenc}
\usepackage{tikz}
\usepackage{amsmath,amsthm,amsfonts,amssymb,amscd}
\usepackage{amsmath}
\usepackage{physics}
\usepackage{caption}
\usepackage{multirow,booktabs}
\usepackage{xcolor}
\usepackage{fullpage}
\usepackage{slashed}
\usepackage{bbm}
\usepackage{enumitem}
\usepackage{fancyhdr}
\usepackage{mathrsfs}
\usepackage{wrapfig}
\usepackage{setspace}
\usepackage{calc}
\usepackage{multicol}
\usepackage{cancel}
\usepackage[retainorgcmds]{IEEEtrantools}
\usepackage{amsmath}
\usepackage{empheq}
\usepackage{jheppub}
\usepackage{comment}
\usepackage{framed}
\usepackage{tikz}
\usepackage{url}
\usetikzlibrary{shapes.misc}

\usepackage{xcolor}
\colorlet{shadecolor}{orange!15}
\theoremstyle{definition}

\newcommand{\af}[1]{\textcolor{orange}{[AF] #1}}

\title{How traversable is a traversable wormhole?}

\author[a]{Ben Freivogel,}
\author[a]{Alessandro Fumagalli,}
\author[b]{and Marija Toma\v{s}evi\'c}

\affiliation[a]{ITFA and GRAPPA, University of Amsterdam, Science Park 904, 1090 GL Amsterdam, The Netherlands}
\affiliation[b]{Department of Theoretical Physics, CERN, 1211 Meyrin, Switzerland}

\emailAdd{b.w.freivogel@uva.nl}
\emailAdd{a.fumagalli@uva.nl}
\emailAdd{marija.tomasevic@cern.ch}

\abstract{To answer the above question, we study low-frequency scattering in the four-dimensional traversable wormhole of Maldacena, Milekhin, and Popov. The resulting transmission probabilities reveal that wormhole traversability depends strongly on the nature of the probe.

For scalar probes, both neutral and charged, traversability depends on the time scale. On time scales of order the light-crossing time after sending in a signal, the transmission is parametrically suppressed, with most of the incoming signal reflected or temporarily trapped inside the wormhole throat. As time progresses, the trapped signal gradually leaks out, so that at late times the accumulated transmission cross-section approaches one half of the corresponding black hole absorption cross-section.  Despite this generic suppression at low frequencies, the transmission spectrum also exhibits resonant frequencies at which transmission becomes perfect. Charged massless fermions tell a very different story. Unlike scalars, they traverse the wormhole with essentially unit probability at low energies. The same mechanism underlies their efficient absorption by magnetic black holes and realizes a channel closely analogous to the Callan--Rubakov effect, revealing unexpected connections with monopole--fermion scattering.

Putting everything together, we conclude that scalar probes are best suited for uncovering distinct features of these magnetic wormholes, while charged massless fermions are the ideal carriers of information through them.}

\begin{document}\emergencystretch 3em
\hypersetup{pageanchor=false}
\makeatletter
\let\old@fpheader\@fpheader
\preprint{CERN-TH-2026-117}

\maketitle

\section{Introduction}
\label{sec:intro}

Observing a non-trivial topology of spacetime from asymptotic infinity was thought to be impossible for a very long time. A number of theorems were proven under physically reasonable assumptions, implying that non-trivial topology, including traversable wormholes, cannot be causally probed from asymptotic infinity \cite{10.1063/1.522498, Lee:1976topology, Galloway1983MinimalSS, PhysRevLett.71.1486, Graham:2007va, Minguzzi:2020ekb, Schinnerl:2021euf}. But every theorem has its assumptions, and over the years, we have understood how to get around them.

These loopholes have led to a variety of traversable wormhole constructions, as summarized below. Our focus will be on a particularly physical example, which was constructed by Maldacena, Milekhin, and Popov (MMP) \cite{Maldacena:2018milekhinpopov}. Their solution describes a 4d traversable wormhole with a single asymptotically flat region, that can be embedded in the Standard Model with Einstein gravity. The wormhole was found to be somewhat fragile, as energy larger than the inverse Schwarzschild radius destroys the wormhole. This, in turn, implied that only low-energy matter could go through the wormhole and not destroy it; we summarize the basic mechanism behind the wormhole construction and the corresponding gap in Sec.~\ref {sec:background}. What remained unclear, however, is whether such low-energy probes actually traverse the wormhole with appreciable probability. Equally important is a complementary, observational question. Namely, in the case of MMP, each wormhole mouth looks like a magnetically charged Reissner-Nordstrom black hole. So how could an observer near one of the mouths, but outside the wormhole throat, tell if she was detecting a wormhole mouth or just a black hole?

In this paper, we address both of these questions by physically probing the MMP wormhole with low-frequency waves. By studying a variety of probes, we determine how efficiently different forms of matter can traverse the wormhole. At the same time, the resulting scattering data reveal characteristic signatures of the wormhole geometry, allowing us to compare its response directly with that of a black hole. Thus, a single calculation provides a quantitative measure of traversability while simultaneously uncovering observable consequences of the wormhole geometry and its nontrivial topology. To this end, we will be using scalars, both neutral and charged, and charged fermions; we will comment on possible outcomes for the uncharged fermion case.

\vspace{3pt}

\textit{Wormhole traversability.}\; We show in Sec.~\ref{subsec:plane-wave} that neutral, massless scalars in the $s$-wave sector already exhibit interesting features that are not found for black holes. Whereas the black hole gives an absorption cross-section equal to the area of the horizon, the wormhole gives a frequency-dependent cross-section that, for generic frequencies, scales as
\begin{equation}
\label{sigma}
    \sigma \sim  \omega^2 r_e^2 \times \text{Area}, \qquad \omega r_e \ll 1.
\end{equation} 
Namely, plane wave scattering leads to a very small probability for generic low frequencies to pass. At the same time, the transmission spectrum contains a sequence of resonances at which reflection vanishes identically, and transmission becomes perfect. Thus, although generic low-frequency modes are unlikely to traverse the wormhole, special, finely tuned frequencies pass through with probability one\footnote{For the MMP wormhole, sending wave packets with higher frequencies than $r_e^{-1}$ would destabilize the wormhole, so in fact, there are no scalar wave packets that traverse the wormhole efficiently. Some of the other constructions, such as humanly traversable wormholes \cite{Maldacena:2020sxe}, are more robust, allowing for higher frequency probes. For these constructions, the inefficiency of low-frequency scattering does not rule out efficient traversability at higher frequencies.}.


The picture changes qualitatively once we consider wave packets, discussed in Sec.~\ref{subsec:wave-packet}. Since any physical wave packet necessarily samples a range of frequencies, the resonant structure must be averaged over. Surprisingly, if one waits sufficiently long times\footnote{Strictly speaking, observing such late-time behavior requires a modification of the original MMP setup, since our analysis neglects particles that miss the nearest mouth and instead travel through the ambient spacetime to the opposite mouth. For the purposes of this discussion, one may imagine surrounding the distant mouth with a reflective shield, thereby eliminating this source of backscattering.}, we can actually get \textit{half} of the black hole answer for the transmission cross-section. Traversability is therefore intrinsically time-dependent. We give a simple physical explanation for this phenomenon: most of the ingoing wave packet that is not initially reflected gets stuck in the wormhole throat and gradually seeps out with equal probability in both directions, resulting in the factor of $1/2$ in the transmission cross-section. 

The situation is similar in the higher angular momentum sector, and for charged scalars, although the transmission for generic low frequencies is even smaller than before; see Sec.~\ref{sec:charged-scalar}. However, the situation drastically changes once we consider charged fermions. Electrically charged fermions can be put in the lowest Landau level (unlike charged scalars---see Sec.~\ref{sec:fermions}), which makes them invisible to the angular momentum barrier induced by the gravitational potential. Furthermore, these fermions latch onto the magnetic field lines and are completely oblivious to the spacetime features of the throat---they effectively see only flat space. This fact makes charged fermion transmission (almost) completely unobstructed, and all fermions make it to the other end of the wormhole, with very little reflection. The physics of this process is very much like the Callan-Rubakov channel for magnetic monopoles; we comment on these connections in Sec.~\ref{sec:conclusion}.  

Taken together, our results reveal a sharp distinction between scalar and fermionic probes of traversable wormholes. Low-energy scalar transmission is strongly suppressed and highly time-dependent, while charged massless fermions traverse the wormhole with essentially unit probability. Traversability, therefore, depends sensitively on the nature of the probe: scalars experience the wormhole as a partially reflecting and resonant cavity, whereas charged fermions propagate through it almost unhindered.

\textit{Wormholes vs. black holes.}\; We can now take our results and see what they imply for distinguishing traversable wormholes from black holes. Using scalar probes, we see that the most efficient setup involves very long time scales, as it is only then that we obtain a parametrically significant reflection coefficient. However, with a very good detector, already at times of order $t \sim 2 \ell$, where $\ell$ is the length of the wormhole, one can see particles bouncing back from the throat, with a cross-section of the same order as \eqref{sigma}. Waiting for longer times only makes the experimenter more confident that she has a wormhole at hand. 
If one decides to use charged scalars, the physical picture is similar to that of the neutral ones, although there will be additional powers of $\omega$ in the reflection coefficient, making it significantly more difficult to detect. Finally, let us address the charged, massless fermions. Naively, one might think they would be poor distinguishers, as they encounter no low-energy barrier in either a black hole or a wormhole geometry. However, these fermions latch onto magnetic field lines, which thread the wormhole and close through the ambient spacetime. An observer near one of the mouths who injects a charged fermion into the throat would therefore see it return after traversing the magnetic loop. This behavior is impossible for a black hole, which would simply absorb the fermion. One may therefore distinguish the two geometries by injecting charged fermions and searching for delayed returns at intervals set by the length of the magnetic loop. In this sense, charged fermions provide a particularly direct probe of the global topology of the wormhole.

The scalar and fermionic probes, therefore, provide complementary diagnostics. Scalars reveal the presence of the wormhole through late-time leakage from the throat, while charged fermions probe the global topology more directly through their repeated returns along magnetic field lines.

\paragraph{No-go theorems for traversable wormholes.} Here we briefly review the classic references on topological censorship, and explain how the main ingredients needed to evade the theorems also play a crucial role in constructing traversable wormholes. We emphasize two types of such no-go theorems, which arrive at the conclusion that nontrivial topology cannot be probed from infinity through different, though ultimately related, mechanisms \cite{10.1063/1.522498, Graham:2007va}. 
One class of results showed that non-simply connected Cauchy surfaces lead to singular time evolution. Under cosmic censorship, such singularities would be hidden behind horizons, and so, causally disconnected from asymptotic infinity \cite{10.1063/1.522498}. A complementary approach used complete, achronal geodesics. Suppose a spacetime with non-trivial topology admits a causal curve that probes the topology and returns to asymptotic infinity. Then among all such causal curves, one could construct an earliest-arriving representative---a complete achronal null geodesic. However, assuming the average null energy condition (ANEC), such geodesics are necessarily homotopic to curves confined to the asymptotic region. Therefore, no causal curve can probe the non-trivial topology, establishing topological censorship \cite{Graham:2007va}. 

From the second theorem, naively, it seems easy to find a loophole: one could imagine probing the non-trivial topology using signals that are not the earliest-arriving ones\footnote{Technically, the authors of \cite{Graham:2007va} proved their theorem under the assumption of simply connected spacetimes, and they too acknowledge that the long-path loophole exists in non-simply connected spacetimes.}. This would imply that going through the non-trivial topology would take longer than going around it. In fact, this also touches upon another aspect of \cite{Graham:2007va}, relevant for us: traversable wormholes require the relevant null generators to experience net defocusing, which is achieved through ANEC violation. Therefore, if the path through the wormhole is not the earliest-arriving one, we can safely have ANEC-violating geodesics supporting the wormhole, opening the door to traversability.

We are still left with the first theorem, which uses different ingredients in its proof. The key to the proof lies in an uplift to the covering space: then the non-simply connected spacetime becomes simply connected, but with multiple asymptotic regions. In such simply connected spacetimes, it is easy to show that some null geodesics must be incomplete. 
Projecting back to the original spacetime, this null incompleteness is the singular evolution implied by the theorem.\footnote{Although see \cite{Galloway1983MinimalSS, Minguzzi:2020ekb} for caveats and refinements of the original theorems.} The loophole in this case is exactly the use of the covering space itself. Classically, this step is harmless, since Einstein's equations are local: if the original spacetime solves them, so does its cover. However, semiclassically, Einstein's equations are not completely local, as the matter source is given by the expectation value of the quantum stress tensor, which depends on the quantum state, boundary conditions, and global topology; Casimir energy is the canonical example. 
Therefore, we conclude from these two loopholes that, to be able to probe non-trivial topology of a spacetime, we must be in a situation where the probing takes longer than simply avoiding the topology, and where quantum effects play a crucial, topology-dependent role. 

\paragraph{Traversable wormhole constructions.} There have been a number of recent constructions of traversable wormholes, beginning with the seminal work in 3d by Gao, Jafferis, and Wall \cite{Gao:2016bin}, which used the non-local coupling of the CFTs, and two separate asymptotic (AdS) regions. Similar constructions were used by Maldacena and Qi to construct an eternal traversable wormhole in two dimensions \cite{Maldacena:2018lmt}, and an uplift to AdS$_4$ was found by Bintanja et al. in \cite{Bintanja:2021xfs}. Subsequent developments to the MMP wormhole include wormholes large enough for humans to traverse \cite{Maldacena:2020sxe}, multi-mouth wormholes \cite{Emparan:2020ldj}, pair-production mechanisms \cite{Horowitz:2019hgb}, and recent worldsheet realizations \cite{Zigdon:2026xal}; see also \cite{Fu:2018oaq, Maldacena:2018lmt, Fu_2019, Marolf:2019ojx, AlBalushi:2020kso, Cubrovic:2021puw, Harvey:2023oom, Bilotta:2023hwq, Kawamoto:2025oko, Begines:2026mlv}.

We should also clear up some claims regarding chronology protection and traversable wormholes. Namely, it was shown that traversable wormholes could lead to closed timelike curves \cite{PhysRevLett.61.1446}. 
Although the original paper dealt with short wormholes (which are now ruled out by \cite{Graham:2007va}), one can come up with a time travel paradox even with long wormholes! However, in that case, one can show that a violation of the achronal ANEC occurs before reaching the chronology-violating regime, thereby removing causality paradoxes outside the regime of validity of semiclassical gravity \cite{Tomasevic:2023ojy}.






\section{The background geometry}
\label{sec:background}

Let us first briefly review the salient features of the Maldacena-Milekhin-Popov (MMP) wormhole \cite{Maldacena:2018milekhinpopov}. The solution found by these authors is a wormhole in asymptotically flat space obtained by gluing two oppositely charged magnetic black holes, which, in the near-extremal limit, develop a long $\text{AdS}_2 \times \text{S}^2$ throat to a wormhole region described by a global cover of $\text{AdS}_2 \times \text{S}^2$. The black hole metric is 
\begin{align}\label{eq:BHmetric}
	ds^2= -f(r)dt^2+\frac{dr^2}{f(r)}+ r^2 d \Omega_2^2\,, \quad f(r)=\frac{(r-r_-)(r-r_+)}{r^2}\,,\quad A =\frac{Q_m}{2} \cos \theta d\varphi\,,
\end{align}
where $r_\pm= M G_N\pm \sqrt{M^2 G_N^2-r_e^2}$ with $r_e^2\equiv \frac{\pi^2 Q_m^2 G_N}{g^2}$ is a function of the integer charge of the black hole $Q_m$ and the $U(1)$ gauge coupling $g$. This metric is matched to the wormhole region, which is approximately described by the global AdS$_2$:
\begin{equation}\label{eq:WHmetric}
    ds^2 = r_e^2 \left(-(1 + \rho^2 + \gamma(\rho)) d\tau^2 + \frac{d\rho^2}{1 + \rho^2 + \gamma(\rho)} + (1 + \phi(\rho)) d\Omega^2\right),
\end{equation}
up to the functions $\gamma(\rho)$ and $\phi(\rho)$ that parameterize deviations from the exact global $\text{AdS}_2 \times \text{S}^2$ geometry. The relation between the wormhole coordinates and the black hole coordinates is
\begin{equation}\label{eq:coordinatematching}
    \tau = \frac{t}{\ell}, \hspace{10pt} \rho = \frac{\ell (r-r_e)}{r_e^2}\,,
\end{equation}
where $\ell$ is a free parameter which is fixed by the equations of motion. To stabilize the wormhole, the authors add a 4d fermion that, in the lowest Landau level, reduces to $Q_m$ 2d fermions which provide the necessary negative Casimir energy\footnote{Note that the zero-point energy of the massless charged fermions, which can be thought of as coming from virtual fermion loops, is crucial for supporting the wormhole. When we discuss scattering of fermions in Sec.~\ref{sec:fermions}, we are considering creating a real fermion to probe the geometry---this is an \textit{excitation} on top of the background.}. The functions $\gamma(\rho)$ and $\phi(\rho)$, as well as the parameter $\ell$, are determined by solving the backreaction problem taking into account this quantum contribution from the fermion. The solution for the functions is not relevant to our purpose, so we will not report it. The value of $\ell$ is fixed to be
\begin{align}\label{eq:lvalue}
	\ell= 16\frac{r_e^3}{Q_m G_N}\,.
\end{align}
This quantity is also what sets the energy gap of the throat. It is important to mention that these wormholes are quite delicate: the binding energy of the throat is of order $r_e^{-1}$, so when we send in waves, we need to keep the energy (i.e., the frequency) smaller than this scale $r_e \omega \ll 1$; otherwise, we would destroy it. The mouths of the two black holes are separated by a distance $d$ that must be smaller than the time that it takes for a signal to travel through the wormhole, which is approximately equal to $\pi \ell$; the causality bound is therefore $\pi \ell > d$. 

To make the two black holes stable, one must put the two black holes into a binary. This is done with an angular velocity that is related to $d$ in by the Kepler formula, including both gravitational and magnetic attraction of the two black holes: $\Omega=2 \sqrt{\frac{r_e}{d^3}}$. As in \cite{Maldacena:2018milekhinpopov}, we will work in a regime where this angular velocity is smaller than the energy gap scale of the throat $\Omega \ell \ll 1$, so that when probing the throat in our scattering experiment, we can consider the black holes as still. We refer the reader to \cite{Maldacena:2018milekhinpopov} for further details on the background. 

\section{Scattering neutral scalars}
\label{sec:scalar}

\begin{figure}
    \centering
    \includegraphics[width=0.6\linewidth]{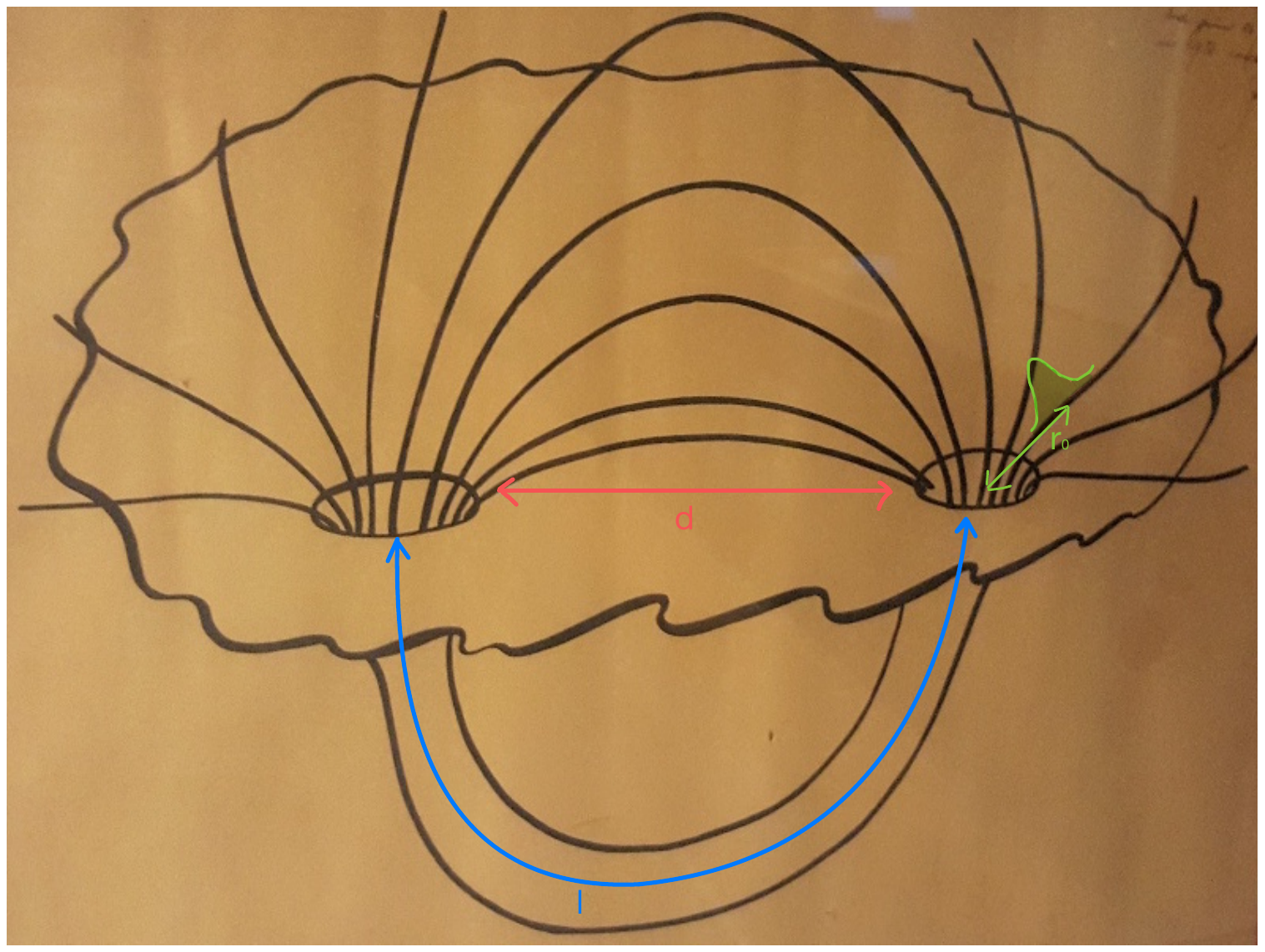}
    \caption{A cartoon of the scattering setup in the MMP wormhole. The figure is a drawing of a wormhole by John Wheeler in 1966, taken from \cite{Maldacena:2018milekhinpopov}. We will first study plane waves scattering, and then the propagation of a wave packet, which we draw in green in the figure. The wave packet will be prepared close to one mouth, traveling towards it. We will take its spread to be smaller than the distance from the mouth $r_0$ from which we send it in, as well as smaller than the distance between the two mouths.}
    \label{fig:wheeler}
\end{figure}

We will begin our analysis by considering the simplest possible probe, namely, neutral, massless scalars in the $s$-wave sector; we will generalize to non-zero angular sectors in Sec.~\ref{sec:charged-scalar}. To this purpose, we consider a minimally coupled massless scalar field and compute the transmission cross-section of plane waves and wave packets made out of this field. 
We imagine sending waves from one side of the wormhole, say the Far Away Right region (FAR), and measuring how much of the initial plane wave flux is transmitted through the wormhole to the other Far Away Left region (FAL). The setup is summarized in Fig.~\ref{fig:wheeler}.
The transmission cross-section that captures the physics of this process is defined in a similar way as the absorption cross-section for black holes \cite{Unruh:1976}
\begin{align}\label{eq:crosssectiondefinition}
     \sigma_{\text{tr}}= \frac{N_\text{tr}}{j_{z,\,in}}\,,\quad N_\text{tr}= r^2 \int d\Omega \, j_{r,{\text{tr}}}\,,
 \end{align}
where $j_{z,\,in}$ is the incoming Klein-Gordon current from the FAR region in the z direction, whereas $j_{r,{\text{tr}}}$ is the radial Klein-Gordon current in the FAL region
\begin{align}
    j_{r,tr}=\frac{1}{2i} \left( \Phi^* {\partial_{\tilde r}}\Phi- \Phi{\partial_{\tilde r}}\Phi^*\right)|_{FAL}\,,
    \end{align}
and $N_\text{tr}$ is the number of transmitted particles outside a two-sphere of radius $r$ in the FAL region.  We stick to a regime where the distance between the two mouths of the wormhole $d$, as well as the parameter $\ell$, are much larger than the horizon radius $r_+\simeq r_e$, so that we can send off waves from a distance $r_0$ from one mouth where the geometry is well approximated by flat space, but also close enough to one of the two mouths so that we can ignore the presence of the other:
\begin{align}
	r_e\ll r_0\ll d < \ell\,.
\end{align} 
Note that even in this regime, this cross-section measures conditional probabilities, since it answers the following question: given an incoming flux of particles close to the right mouth of the wormhole, how much of this flux is transmitted to the other side through the wormhole? In particular, it does not take into account the particles that either travel undisturbed outside the wormhole or the ones that get reflected and then travel to the other side. It would be interesting to include such propagation in the analysis but what we are discussing is sufficient to probe the traversability of the wormhole. One can think of neglecting this effect by surrounding one of the mouth with a reflecting shield.



\subsection{Plane wave scattering}
\label{subsec:plane-wave}

Let us start by considering the scattering of a plane wave, i.e., a solution of the equation of motion for the field $\Phi$ with fixed frequency $\omega$. The equation of motion is a Klein-Gordon equation $\Box \Phi=0$ in the wormhole background described in Sec.~\ref{sec:background}. We decompose the field using the spherical harmonics $\Phi_\omega(x^\mu)= e^{-i \omega t} R(r)Y_l^m(\theta,\varphi)$, and specialize to $l=0$. Outside the throat, where the metric is described by the RN solution \eqref{eq:BHmetric}, the equation for $R(r)$ is
\begin{equation}\label{eq:ReqRN}
f(r)\,\partial_{r}\!\left(f(r)\,r^{2}\,\partial_{r}R(r)\right)
+\omega^{2}r^2\,R(r)=0,
\end{equation}
This equation can be solved in the far away region $r \gg r_+$\footnote{We use $r_+$ here but recall that in the near extremal limit, we will have $r_+\simeq r_e$.}, where $f(r) \simeq 1$
\begin{align}\label{eq:genericFAsol}
    R_{FA}(r)= d_{in} \frac{e^{-i \omega r}}{r \omega}+d_{out} \frac{e^{i \omega r}}{r \omega}\,,
\end{align}
and in the in the low energy limit, $r\ll \omega^{-1}$, where we can drop the $\omega$ dependent term, and get
\begin{align}\label{eq:Romega0right}
R^{\omega \to 0}(r)= \alpha_1+\alpha_2 \frac{\log(\frac{r-r_+}{r-r_-})}{r_+-r_-}\simeq \alpha_1+\frac{\alpha_2}{r-r_e}\,,
\end{align}
where we used the near-extremal limit in the last step. These two regions of parameter space overlap if $\omega r_+ \ll 1$ since we can work in the regime where $r_+ \ll r\ll \omega^{-1}$, as can be seen by expanding for small $\omega r$ the equation \eqref{eq:genericFAsol} and for large $r$ the equation \eqref{eq:Romega0right}. Moreover, the near horizon region $r\simeq r_+$, where the metric is approximately $\text{AdS}_2 \times \text{S}^2$ and we match to the throat solution of the wormhole, is part of the low-energy region $r \ll \omega^{-1}$. This justifies the matching between the solution in the throat region \eqref{eq:WHmetric} at large radius, and the solution in far away region at small radius.  

\begin{figure}\label{fig:wormholescattering}
    \centering
    \includegraphics[width=0.7\textwidth]{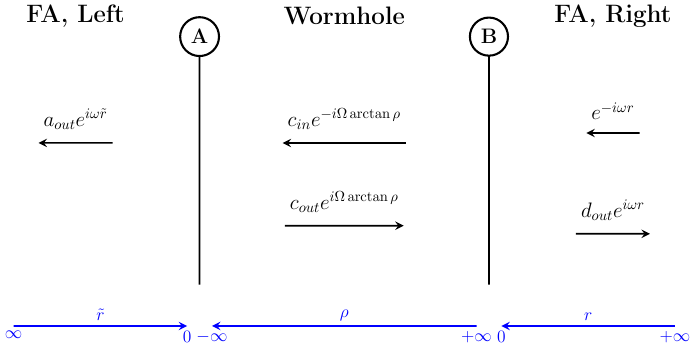}
    
    \caption{Scattering diagram for a neutral scalar field across a wormhole background. A and B denote the overlapping regions of these solutions, where we match the various coefficients.}
    \label{fig:wormhole_scattering}
\end{figure}

Let us set up the scattering problem; we will work in the near-extremal limit from now on: $r_+ = r_e$\footnote{Even if one cannot trust the exactly extremal solution classically \cite{Iliesiu:2020qvm}, for our purposes, nothing changes between using $r_+\simeq r_e$ or setting $r_+=r_e$ exactly.}. We imagine sending in a wave from the FAR region, where our solution is
\begin{align}\label{eq:RFAright}
	R_{FAR}(r)= d_{in} \frac{e^{-i \omega r}}{r \omega}+d_{out} \frac{e^{i \omega r}}{r \omega}\,.
\end{align}
The solution takes on the same form in the FAL, where the wave is transmitted, just with different coefficients,
\begin{align}\label{eq:RFAleft}
	R_{FAL}(r)= a_{in} \frac{e^{-i \omega \tilde r}}{\tilde r \omega}+a_{out} \frac{e^{i \omega \tilde r}}{\tilde r \omega}\,.
\end{align}
Since we will not have any reflected wave from the left region, we set $a_{in}=0$, and we can also set $d_{in}=1$ since this is just a choice in the normalization of the incoming wave. In the conditional problem we are considering, we can treat the left region as completely separated from the right region, so in principle we can consider the coordinate that appears in \eqref{eq:RFAleft} as independent from the one appearing in \eqref{eq:RFAright}; hence the use of the label $\tilde r$. In the wormhole throat, the Klein-Gordon equation upon considering $\Phi= e^{-i \Omega \tau}R(\rho) Y_l^m(\theta,\varphi)$ for $l=0$ takes on the form
\begin{align}\label{eq:KGwormhole}
	\rho ^4 \left(\left(\rho ^2+1\right) \left(\left(\rho ^2+1\right) R''(\rho )+2 \rho  R'(\rho )\right)+\Omega ^2 R(\rho )\right)=0,
\end{align}
where $\Omega$ is the frequency associated with the wormhole coordinate $\tau$. To get the relation with the frequency in the black hole region, we use eq.~\eqref{eq:coordinatematching}, which gives $\Omega=\omega \ell$. The solution of this equation is
\begin{align}\label{eq:Rwh}
	R_{WH}=c_{in} e^{-i\Omega  \arctan \rho }+c_{out} e^{i\Omega  \arctan\rho }.
\end{align}
In the positive large $\rho$ limit, we have
\begin{align}\label{eq:Rwlargerho}
	R_{WH}^{\rho \to \infty}\simeq e^{-\frac{i\pi \Omega}{2}} \left(c_{\text{in}} + c_{\text{out}} e^{i\pi \Omega}\right)
- \frac{i\, e^{-\frac{i\pi \Omega}{2}} \left(-c_{\text{in}} + c_{\text{out}} e^{i\pi \Omega}\right)\Omega}{\rho},
\end{align}
which is exactly the eq. \eqref{eq:Romega0right} upon substituting the relation between the RN coordinates and throat coordinates \eqref{eq:coordinatematching}:
\begin{align}
	R_{WH}^{\rho \to \infty}\simeq e^{-\frac{i\pi \Omega}{2}} \left(c_{\text{in}} + c_{\text{out}} e^{i\pi \Omega}\right)
- r_e^2\frac{i\, e^{-\frac{i\pi \Omega}{2}} \left(-c_{\text{in}} + c_{\text{out}} e^{i\pi \Omega}\right)\Omega}{r-r_e}\,,
\end{align}
showing that the near horizon overlaps with the low frequency regime, and in particular uniquely fixing the constants $\alpha_{1,2}$. As explained above, this solution has an overlapping regime with the FAR solution \eqref{eq:RFAright},  
so we can match the coefficients $c_{in,out}$ here and the coefficient $d_{out}$ in \eqref{eq:RFAright} by expanding $r-r_e \simeq r$ in the former. The full matching procedure is outlined in Fig.~\ref{fig:wormhole_scattering}.  
In the overlapping region (B), we match $R_{WH}^{\rho \to \infty}$ with $R_{FAR}^{r \to 0}$, and in the overlapping region (A) we match $R_{WH}^{\rho \to \infty}$ with $R_{FAR}^{r \to 0}$. We described above how the matching works in (B). In the region (A), we expand $R_{WH}$ for $\rho \to -\infty$, and keeping the absolute value, we get
\begin{align}
	R_{WH}^{\rho \to -\infty}&\simeq e^{-\frac{i\pi \Omega}{2}} \left(c_{\text{out}} + c_{\text{in}} e^{i\pi \Omega}\right)
- \frac{i\,\Omega}{|\rho|}\, e^{-\frac{i\pi \Omega}{2}}
\left(-c_{\text{out}} + c_{\text{in}} e^{i\pi \Omega}\right)=\\
&=e^{-\frac{i\pi \ell \omega}{2}}
\left(c_{\text{out}} + c_{\text{in}} e^{i\pi \ell \omega}\right)
- \frac{i\, r_e^2 \omega}{\tilde r}\,
e^{-\frac{i\pi \ell \omega}{2}}
\left(-c_{\text{out}} + c_{\text{in}} e^{i\pi \ell \omega}\right)
\end{align}
where we substituted the relation between the $\rho$ and $r$ coordinates and expanded $\tilde r-r_e \simeq r$. We then get two equations for the coefficients $c_{in,out}$ by equating this in the limit $\tilde r\gg r_e$ to \eqref{eq:RFAleft}. Solving these equations, together with the ones from the matching in (B), we get to the following result\footnote{One might be worried about the underlying instabilities diagnosed by the poles of the reflection coefficient in the upper half plane. However, one can see that there is no such poles within the low-frequency regime of our calculation.}
\begin{align}
\label{refl}
	\mathcal{R}\equiv d_{\text{out}} 
&= \frac{\left(-1 + r_e^4 \omega^4\right)\sin(\ell \pi \omega)}
{2 i\, r_e^2 \omega^2 \cos(\ell \pi \omega)
+ \left(1 + r_e^4 \omega^4\right)\sin(\ell \pi \omega)} \,, \\
\mathcal{T}\equiv a_{\text{out}} 
&= -\,\frac{2 i\, r_e^2 \omega^2}
{2 i\, r_e^2 \omega^2 \cos(\ell \pi \omega)
+ \left(1 + r_e^4 \omega^4\right)\sin(\ell \pi \omega)} \,.
\end{align}
These coefficients have the interpretation of reflection and transmission coefficients. From here, we can obtain the square of these coefficients,
\begin{equation}
\label{eq:coeffs}
|\mathcal{R}|^2 = \frac{\sin^2(\pi \omega \ell)(\omega^4 r_e^4 - 1)^2}{4 \omega^4 r_e^4 + \sin^2(\pi \omega \ell)(\omega^4 r_e^4 - 1)^2}, \qquad |\mathcal{T}|^2 = \frac{4 \omega^4 r_e^4}{4 \omega^4 r_e^4 + \sin^2(\pi \omega \ell)(\omega^4 r_e^4 - 1)^2}. 
\end{equation}
To calculate the transmission cross-section, one has to consider scattering a plane wave $\Phi\sim \mathcal{N} e^{i \omega(t-z)}$, where the normalization is determined by relating our choice $d_{in}=1$ in the radial problem to the amplitude of the plane wave using the standard plane wave decomposition. We can then compute the incoming current $j^z$, and the outgoing one $j^r$ using our solutions; using the eq.~\eqref{eq:crosssectiondefinition}, this gives the standard conversion formula
\begin{align}\label{eq:crosssectionplanewave}
	\sigma= \frac{\pi}{\omega^2} |\mathcal{T}|^2=\frac{\pi}{\omega^2}\frac{1}{1+\frac{\left(1-\omega ^4 r_e^4 \right){}^2 }{4 \omega ^4 r_e^4}\sin ^2(\pi  \ell \omega )}\,.
\end{align}
Now, recall that our regime of validity for this scattering problem involves a low-frequency regime. Therefore, we might be tempted to simply take a generic $\omega \to 0$ limit, for which we would obtain a fairly small transmission cross-section,
\begin{equation}
    \text{naive $\omega \to 0$ limit: } \qquad \sigma \approx  \left(\frac{r_e}{\pi \ell}\right)^2 A,
\end{equation}
where $A$ is the extremal area of the black hole, $A = 4\pi r_e^2$. This cross-section is small, as we require $\pi \ell \gg r_e$ for the wormhole to be well-defined. However, we must also worry about the combination $\omega \ell$: in order to solve the radial ODE, we required $\omega r_e \ll 1$, but $\ell \gg r_e$, so we cannot simply take $\omega \ell \ll 1$, which is what the naive small $\omega$ limit would imply. Moreover, even if we could satisfy this constraint, there is another point of worry: resonant peaks. Namely, when 
\begin{equation}
    \omega = \frac{n}{\ell}, \qquad n \in \mathbb{Z},
\end{equation}
we get a complete transmission of the signal
\begin{equation}
    |\mathcal{R}|^2 = 0, \qquad |\mathcal{T}|^2 = 1.
\end{equation}
On the other hand, when we are away from the resonances, i.e. the function $\sin^2 (\pi \omega \ell)$ is order one, we get
\begin{equation}
    |\mathcal{R}|^2 \approx 1 - \frac{4 \omega ^4 r_e^4}{\sin ^2(\pi  \omega  \ell )}, \qquad |\mathcal{T}|^2 \approx  \frac{4 \omega ^4 r_e^4}{\sin ^2(\pi  \omega  \ell )},
\end{equation}
which implies almost total reflection, unless we hit the resonant frequencies.
In fact, in this case, the transmission cross-section is made small due to the low frequency,
\begin{equation}
    \sigma \approx \left(\omega r_e\right)^2 A.
\end{equation}
Therefore, although the wormhole was initially deemed traversable, it, in fact, allows only for a very small amount of energy to pass through. However, there is one more thing we can do: given that fine-tuning of frequencies is involved, which might not always be available to an observer trying to set up her scattering experiment, let us consider averaging over frequencies instead. We can average over a window of the size of the sine period, for instance. To that end, we use the delta function representation,
\begin{equation}
\label{eq:delta}
    \delta(x) = \frac{1}{\pi} \lim_{\epsilon\to 0} \frac{\epsilon}{\epsilon^2 + x^2}, \hspace{10pt} \forall x,
\end{equation}
to rewrite our equation for the transmission cross-section \eqref{eq:crosssectionplanewave} as
\begin{equation}
    \sigma = 2\pi r_e^2 \frac{\epsilon}{\epsilon^2 + \sin^2{\pi z}} = 2\pi^2 r_e^2 \delta(\sin{\pi z}),
\end{equation}
where we set $\epsilon = 2 r_e^2 \omega^2$, and $z = \ell \omega$. Even though $\epsilon$ is small, $z$ need not be. The averaged cross-section over a period is then
\begin{equation}
    \bar{\sigma} = \frac{\ell}{m} \int_{\omega_0+m/\ell}^{\omega_0+(m+1)/\ell} d\omega\; \sigma(\omega) = \int_{z_0+m}^{z_0+m+1} dz\; \sigma(z), \hspace{10pt} m \in \mathbb{Z},
\end{equation}
where we considered a central frequency $\omega_0$ away from the delta function peaks. This yields
\begin{align}
    \bar{\sigma}=2\pi^2 r_e^2= \frac{A}{2}\,.
\end{align}
Therefore, performing this sort of an average, we obtain \textit{half} of the standard black hole result \cite{Das:1996we}. Although it might seem somewhat arbitrary to perform such an average here, we will see in the next section that such averages arise when considering wave packet scattering.

\subsubsection*{General argument for total transmission resonances}

The resonant peaks found above are not special to traversable wormholes; indeed, one can argue for their existence from a simple quantum mechanics problem of scattering with two barriers.\footnote{We thank David Berenstein for discussions on this point.} We find it convenient to give the explicit argument because we have not found the exact result of interest in the literature. 

Note that for each angular momentum mode, the problem becomes one-dimensional. By making field redefinitions and changing coordinates, the problem of finding the radial wavefunctions can be exactly mapped to a one-dimensional time-independent Schrodinger problem, with an effective potential determined by the geometry. In this case, the geometry will consist of two angular momentum barriers separated by the long wormhole throat. In the asymptotic regions, as well as in the throat, the potential becomes very small.

Then one can show the following. Given a parity-symmetric potential with a classically allowed region between two barriers, the reflection coefficient has exact zeros at real frequencies. When the classically allowed region has long-lived quasi-bound states, there is one total transmission resonance near each quasi-bound state.

We show this by first analyzing scattering in the even- and odd-parity sector. We assume the potential goes to zero sufficiently fast at large $x$.  In the even parity sector, the eigenstates are, at large $|x|$
\begin{equation}
    \psi_+(x) = \cos (k |x| - \delta_+(k))  \ \ \  \ \ \ |x| \to \infty.
\end{equation}
Since scattering does not mix parity-even and parity-odd sectors, scattering in each sector must give a phase.
In the odd sector, the asymptotic solution is similarly
\begin{equation}
     \psi_-(x) = \text{sgn}(x) \sin (k |x| - \delta_-(k)) \ \ \  \ \ \ |x| \to \infty.
\end{equation}
We can combine these solutions in order to calculate the conventional transmission and reflection coefficients from the phases $\delta_+$ and $\delta_-$. 

To do this, we construct a scattering solution as a superposition of even and odd sectors:
\begin{equation}
    \psi(x) = a \psi_+(x) + b \psi_-(x)
\end{equation}
where all quantities depend on the momentum $k$. Demanding that there is  no incident wave from $x=\infty$, corresponding to scattering from the left, we obtain
\begin{equation}
    a e^{i \delta_+} + i b e^{i \delta_-} = 0.
\end{equation}
We can then calculate the reflection coefficient. We find
\begin{equation}
   r = {\frac{1}{2}}(e^{-2 i \delta_+} - e^{- 2 i \delta_-}).
\end{equation}
Therefore, total transmission occurs at the resonant frequencies
\begin{equation}
    \delta_+(k) - \delta_-(k)= n \pi, \qquad n \in \mathbb Z.
\end{equation}
So far, this is general and does not make use of the classically allowed region between the barriers. The key conclusion of the above analysis is that total transmission requires solving only a single real equation, so it is generic that they occur in parity-symmetric potentials.

We now need to use this additional physics to determine the $k$ dependence of the phases, in order to determine the spacing between resonances. In our regime of interest, the frequency is low compared to the height of the barriers, since the barrier height scales as  $1/r_e$, and we assume $\omega r_e \ll 1$. Therefore, the throat is a potential well with nearly stable quasi-bound states. In this regime, the phase $\delta_+$ changes rapidly by $\pi$ every time the frequency crosses the frequency of an even quasi-bound state, while $\delta_-$ does the same near odd quasi-bound states. Away from these resonant frequencies the phases change slowly. Therefore, $\delta_+ - \delta_- = n \pi$ is satisfied near each resonance of the potential, since one of the $\delta_i$ varies rapidly by $\pi$ and so certainly hits resonance condition. 

The spacing of resonances $\Delta \omega$ is therefore equal to the spacing between quasi-bound states of the well. In general this depends on the details of the potential, which depend on the details of the probe. For frequencies large compared to the scale of the potential in the throat, but still in our small frequency regime, the throat regime is well approximated by free propagation and the spacing is simply $
\ell \Delta \omega = \pi   
$
so the spacing of resonances is set by the length $\ell$ of the wormhole. 
We will find these resonances explicitly when we scatter charged scalars in Sec.~\ref{sec:charged-scalar}.

\subsection{Wave packet scattering}
\label{subsec:wave-packet}

In the previous section, we obtained some curious features for the transmission cross-section. In particular, the existence of finely tuned resonant frequencies implies that we should perhaps consider a more physical setup in which we scatter proper wave packets off the wormhole. Such a scattering procedure necessarily involves averaging over frequencies, so we might hope to see qualitatively different behavior. 
In particular, we will compute the total cross-section by sending a wave packet from the FAR region and measuring the full transmitted signal received in the FAL region. By total, we mean that we wait for an infinite amount of time to collect all the possible outgoing radiation from the `exit' mouth. Interestingly, this will give a cross-section equal to half the black hole cross-section, $\sigma_{tr}=\text{Area}/2$, akin to the calculation by the end of the previous section. Moreover, we will calculate the flux of particles that exit the wormhole at \textit{finite} times, for instance, of the order of several wormhole traversals, $t \sim n\pi \ell$. This calculation is useful since it gives an idea of how information propagation occurs inside the wormhole. We comment on a plausible picture of the full traversal process by the end of this section.


\subsubsection{Preparing the wave packets}

Let us start by defining our wave packets. We prepare a Gaussian position space envelope centered around a point $\vec x_0$ (which we take to be in the FAR region) at time $t_0=0$, traveling in the direction given by the momentum $\vec k_0$, 
\begin{align}\label{eq:wavepacket-x}
    \psi_{in}(\vec{x}) = \mathcal{N}_{in}\, e^{i \vec{k}_0 \cdot (\vec{x} - \vec{x}_0)} e^{-\frac{(\vec{x} - \vec{x}_0)^2}{2 \sigma^2}}\,,
\end{align}
where $\vec x$ is a position vector with spherical coordinates $(r,\theta,\varphi)$. Since we consider massless probes, the dispersion relation is linear, $\omega=|k|$,
so all frequency components propagate with the same group velocity in the asymptotic region. Consequently, there is no wave packet spreading for free propagation. We can align our coordinate system such that $\vec k_0$ is aligned with the $z$ axis,  and choose the initial position $\vec x_0$ to be aligned with this axis as well. Thus, $\vec{x}_0 = (0, 0, r_0)$ and $\vec{k}_0 = (0, 0, -\omega_0)$, as we take our wave packet to be incoming. We show in App.~~\ref{app:radialvsthreed} that projecting onto the incoming $s$-wave sector of this wave packet results in a radial Gaussian envelope of the form 
\begin{align}
\label{eq:wavepacketradial}
	\psi_{r_0,\omega_0}(t=0, r)= \mathcal{N} e^{-\frac{(r-r_0)^2}{2 \sigma^2 }} \frac{1}{r}e^{-i \omega_0 r} \,,
\end{align}
where $\omega_0= |\vec{k}_0|$, $r_0= |\vec{r}_0|$, and where the constant $\mathcal N$ absorbs the overall normalization and the constant phase obtained in the angular projection. We consider a regime in which this wave packet is localized sufficiently far from the right mouth to ensure the geometry is approximately flat, but also far away from the other mouth. The initial radial position must satisfy
\begin{align}\label{eq:wavepacketregimes}
	r_e\ll r_0\ll d < \ell\, ,
\end{align} 
where $d$ is the ambient space distance between the two mouths, and $\ell$ is the wormhole length. A cartoon of the wave packet is shown in Fig.~\ref{fig:wheeler}. Moreover, the wave packet cannot be too spread out, nor too small, leading to a set of constraints
\begin{align}
     r_e <\sigma \ll r_0-r_e \ll d-r_0\,.
\end{align}
With these constraints in mind, the wave packet is localized in the FAR region, and we can just describe it using the $r$ coordinate. We will furthermore restrict to a regime where the initial wave packet has a frequency close to $\omega_0$, which means $\sigma \omega_0  \gg 1$. To probe the resonances we have seen in the previous section, we need $\omega_0 = n/\ell$ with $n$ being a positive integer. To satisfy the above constraints, we also need $1\ll n \ll r_e^{-1}$.

To compute the quantities of interest, we want to see how the wave packet evolves at later times in the wormhole spacetime. This can be easily done by decomposing it in a complete basis of solutions of the KG equation in the full wormhole spacetime, which are nothing but the solutions $\Phi_\omega$ we derived in the previous section. We do this in detail in App.~\ref{app:packetevol}, and here we report on relevant results. The transmitted wave packet in the FAL region at some radius $\tilde r$ and time t is
\begin{align}\label{eq:psiLgaussian}
	\psi^L_{r_0,\omega_0}(t,\tilde r)=\frac{1}{\tilde r }\int_0^\infty d\omega \frac{\mathcal N   \sigma}{\sqrt{2 \pi} } \,e^{-\frac{1}{2} \sigma ^2 \left(\omega -\omega _0\right){}^2}\,e^{-i \omega_0 r_0}\,\mathcal T_\omega\,e^{i  \omega\left( r_0+\tilde r-t\right)}\,.
\end{align}
Observe that we started with a radial Gaussian wave packet \eqref{eq:wavepacketradial}, and we end up with frequency averaged over a normalized (up to the initial amplitude $\mathcal N$) distribution. We could have chosen some other envelope whose Fourier transform is known, for instance some $g_\omega$; in our calculations below, we will focus on Gaussian envelopes, but we write more general expressions where possible,
\begin{align}
\label{eq:left-ing}
	\psi^L_{r_0,\omega_0}(t,\tilde r)= \frac{1}{\tilde r}\int_0^\infty d\omega \, g_{\omega} \,\mathcal T_\omega\, e^{i  \omega\left( r_0+\tilde r-t\right)}\,, \qquad g_\omega = \frac{\mathcal N \sigma}{\sqrt{2\pi}} e^{-\frac{\sigma^2}{2}(\omega-\omega_0)^2} e^{-i\omega_0 r_0},
\end{align}
where we note that $g_\omega = g_\omega(r_0, \omega_0, \sigma)$. We can also look at how the wave packet looks like in the FAR region at generic $t$. Its ingoing component is then given by
\begin{align}
	\psi^{R, in}_{r_0,\omega_0}(t,r)&=\frac{e^{-i \omega_0 r_0}}{r}\int_0^\infty d \omega \frac{\mathcal N \sigma\sqrt{2 \pi}}{2 \pi }\,e^{-\frac{1}{2} \sigma ^2 \left(\omega -\omega _0\right){}^2}  e^{-i \omega(r+t-r_0)}=\\
	&=\mathcal N \frac{e^{-i \omega_0 (r+t)}}{r} e^{-\frac{\left(r-r_0+t\right){}^2}{2 \sigma ^2}}\,.
\end{align}
Compared with the initial wave packet, we see that the peak of the distribution is now at a smaller radius $r=r_0-t$ instead of at $r=r_0$, since the particles have traveled towards the center of the mouth. Translating back to the plane wave envelope (in the $s$-wave sector), this gives
\begin{align} \label{eq:wavepacketingenerict}
	\psi_{in}(t,\vec{x}) 
	=\mathcal{N}_{in} e^{-i \omega_0 t}e^{-i\omega_0 (z - r_0)} e^{-\frac{(|\vec{x} - \vec{x}_0|-t)^2}{2 \sigma^2}}\,.
\end{align}
We will now use these formulae to derive the transmission cross-section for wave packet scattering.

\subsubsection{Transmission cross-section integrated over all times}

We can now compute the currents coming from these wave packets, as we have done in the plane wave section. The transmitted radial current in the FAL region at time $t$ is 
\begin{align}
    j_{r,tr}(t,\tilde r) = 
     \frac{1}{\tilde r^2}\int d\omega d\omega' \frac{\omega+\omega'}{2} g^*_{\omega} g_{\omega'}\, \mathcal T_\omega^* \mathcal T_{\omega'}\,e^{-i (\tilde r+r_0-t) \left(\omega -\omega '\right)}\,,
    \end{align}
where we used \eqref{eq:left-ing} to evaluate the current. The transmitted flux at time $t$, $N_{tr}(t)$, is then given by the same formula as in \eqref{eq:crosssectiondefinition}, just with $r\to \tilde{r}$. To compute the incoming current at time $t$, we use the wave packet \eqref{eq:wavepacketingenerict}, which gives
\begin{align}
 	|j^z_{in}(t,\vec x)| = 4 \omega_0^3 \abs{\mathcal N}^2 e^{-\frac{(|\vec{x} - \vec{x}_0|-t)^2}{ \sigma^2}}\,.
 \end{align}
When projecting onto the $s$-wave sector, we also replace  $|\vec{x} - \vec{x}_0|$ with $r-r_0$. Unlike in the plane wave scattering,  these currents are time-dependent, so to define a cross-section, we integrate both the number of transmitted particles and the incoming current for all times, which gives us
\begin{align}
	\sigma_{tr}= \frac{\int dt N_{tr}(t)}{\int dt\, |j^z_{in}(\vec x,t)|} = \frac{\pi}{\omega_0^2} \int d\omega |\mathcal T_\omega|^2\,  \frac{\sigma}{\sqrt{\pi} } \,e^{-\sigma ^2 \left(\omega -\omega _0\right){}^2}\,,
\end{align}
where we used the fact that our wave packet is localized around $\omega \simeq \omega_0$. For the transmission coefficient \eqref{eq:coeffs}, we can use the delta function definition \eqref{eq:delta} to write
  \begin{align}
 	|T_\omega|^2=\frac{\epsilon^2}{\epsilon^2+\sin ^2(\pi  \ell \omega )}\simeq \epsilon \pi \delta(\sin \pi \ell \omega)\,,
 \end{align}
where we defined $\epsilon=2 \omega^2 r_e^2$, and took a small $\omega r_e$ limit. Note that $\epsilon$ is slowly varying, and we can evaluate it at $\omega \simeq \omega_0$. Using the same methods as by the end of Sec.~\ref{subsec:plane-wave}, we obtain
 \begin{align}
 	\sigma_{tr} =\frac{\pi}{\omega_0^2}  \epsilon= 2 \pi r_e^2=\frac{A}{2},
 \end{align}
where $A$ is the area of the corresponding black hole horizon. This result can be interpreted as follows. The radiation coming from the FAR region is absorbed by the mouth with a cross-section given by the horizon area, as it happens in the black hole case, since the mouth looks like a black hole from outside. Then, if we wait long enough, everything leaks out of the wormhole either towards the right or to the left mouth of the wormhole. Only half of what entered the wormhole is transmitted to the left mouth, giving the additional factor of one half.

\subsubsection{Looking at the transmitted signal after a finite time}

The discussion above shows that a significant fraction of the probe radiation is eventually transmitted through the wormhole. However, this conclusion only applies after waiting for sufficiently long times. To determine whether the probe traverses the wormhole within a finite time interval, it is useful to study the time dependence of the transmitted signal. 

We therefore consider the quantity $N_{tr}(t)$ at finite values of $t$, normalized by the initial incoming flux $j^{z}_{in}(t=0,\vec x=\vec x_0)$, so that it still has the dimensionality of a cross section and it is independent on the amplitude of the packet we send in. This defines the quantity\footnote{Strictly speaking, this is not a cross-section in the usual asymptotic sense. Nevertheless, it captures the same physical information while additionally resolving the time dependence of the transmitted signal. In particular, it measures how much of the initial wave packet emerges from the opposite mouth at a given time.} 
\begin{align}
	\Sigma_{tr}(t)&=\frac{N_{tr}(t)}{j^z_{in}(0, r_0)}= \frac{\pi}{\omega_0^2} \int d\omega d\omega' g^*_{\omega} g_{\omega'} \mathcal T_\omega^* \mathcal T_{\omega'}\,e^{-i (r+r_0-t) \left(\omega -\omega '\right)},
\end{align}
where we used the fact that, for a sufficiently sharply peaked envelope, the dominant contribution comes from frequencies near $\omega=\omega'=\omega_0$.
The resulting expression is therefore the standard $\pi/\omega_0^2$ conversion factor between spherical and plane-wave cross-sections, multiplied by a normalized average of $\mathcal T_\omega^*\mathcal T_{\omega'}$ over the frequency-space wave packet profile. We can now look at this quantity at different timescales. 

\paragraph{Short timescales.} Let us start by considering times of order $t\simeq r+r_0$. Given that we generically have $r,r_0 \ll \ell$, we expect to see no transmission at all. Since the imaginary exponent is small, we can compute the cross-section as
\begin{align}
 	\Sigma_{tr}(t\simeq \tilde r+r_0)&=\frac{\pi}{\omega_0^2} \int d\omega d\omega' g^*_{\omega} g_{\omega'}  \mathcal T_\omega^* \mathcal T_{\omega'}=\frac{\pi}{\omega_0^2} |\expval{T_\omega}_{g}|^2\,.
 \end{align}
where the notation indicates that we should first take the expectation value in the $g$ distribution, and then square it. Because we are studying a time-dependent process at fixed values of $t$ and $r$, spacetime translation invariance is explicitly broken. As a result, these time-dependent, averaged coefficients do not satisfy the usual unitarity relation, $\abs{\expval{\mathcal T}}^2+\abs{\expval{\mathcal R}}^2\neq 1$. 

From eq.~\eqref{refl}, we see that $\mathcal T_\omega$ is a slowly varying function of $\omega r_e$ in the low-frequency regime $\omega r_e\ll 1$, while its dependence on $\omega\ell$ is highly oscillatory through the trigonometric functions. We may therefore approximate $x=r_e^2\omega^2$ as constant over the support of the wave packet. In this approximation, $\mathcal T_\omega$ becomes periodic in $\omega$ with period $2/\ell$.  We can take the envelope $g_{\omega}$ to be a rectangle function equal to $g_{\omega}=\frac{2}{\ell}$ when $\omega \in (\omega_0,\omega_0+\frac{2}{\ell})$, and zero elsewhere. This corresponds to averaging the function $T_\omega$ over its period, i.e., computing
 \begin{align}
 	\expval{\mathcal T_\omega}_{g}=\frac{\ell}{2}\int_{\omega_0}^{\omega_0+\frac{2}{\ell}} d\omega \left(-\,\frac{2 i\, x}
{2 i\, x \cos(\ell \pi \omega)
+ \left(1 + x^2\right)\sin(\ell \pi \omega)} \right),
 \end{align}
where we take $x=r_e^2 \omega^2$ to be a constant. Changing the variables to $z=e^{i \pi \omega \ell}$ and applying the residue theorem, one finds that the integral vanishes. In other words, at this timescale, nothing gets through the whole wormhole, as expected.
 
\paragraph{Long timescales.} Another timescale of interest is $t =t_\ell=\tilde r+ r_0+ \pi \ell$, which corresponds to the earliest time at which we might expect the transmitted signal to begin emerging from the opposite mouth of the wormhole. In this regime, we can write that the same quantity $\Sigma_{tr}$ evaluates to
  \begin{align}
 	\Sigma_{tr}(t_\ell,\tilde r)&=\frac{\pi}{\omega_0^2} \int d\omega d\omega' g^*_{\omega} g_{\omega'} (\mathcal T_\omega e^{-i \pi  \ell \omega})^* T_{\omega'}e^{-i \pi  \ell \omega'}\,e^{-i (r+r_0-t+\pi \ell) \left(\omega -\omega '\right)}=\\
 	&\simeq \frac{\pi}{\omega_0^2} \left|\expval{\mathcal T_\omega e^{-i \pi  \ell \omega}}_{g}\right|^2,
 \end{align}
where in the last step we used the fact that the phase in the exponential vanishes when $t=\tilde r+ r_0+ \pi \ell$. We can compute this average similarly as before to obtain
 \begin{align}
 	\Sigma_{tr}(t_\ell,\tilde r)\simeq 4 \pi  r_e^2 \left(4 \omega_0 ^2 r_e^2\right)= A \times \left(4 \omega_0 ^2 r_e^2\right). 
 \end{align}
Therefore, we find that the transmitted signal at this timescale is suppressed relative to the geometric area scale by an additional factor of $4\omega_0^2 r_e^2 \ll 1$. Physically, this means that the observer begins to detect the first transmitted radiation emerging from the opposite mouth of the wormhole. However, the amount of transmitted flux is still parametrically smaller than the total incoming flux, whose scale is set by the horizon area.

More generally, we can consider any time of the form $t_{n\ell}=\tilde r+r_0+n \pi \ell$, we get zero for even $n$  and the same answer as before for odd $n$:
\begin{align}
 	&\Sigma_{tr}(t_{n\ell},\tilde r)=\begin{cases} 
       A\times \left(4 \omega_0^2 r_e^2\right) & \text{for odd } n \\[1ex]
        0 & \text{for even } n
    \end{cases}\,.
 \end{align}
Since the signal starts on the right at $t=0$, corresponding roughly to $n=0$, we can interpret this result as saying that the wormhole leaks out a little bit of the signal every time the wave packet bounces off the left side, i.e., at roughly $t=\tilde r+r_0+n\pi \ell$ for $n$ odd. Since $r_e \omega_0 \ll 1$, the amount of radiation that comes out at any time-scale we observed is very small. We can also calculate a similar quantity for the reflected radiation, for which we can use
 \begin{align}
 	N_{ref}(t)= r^2 \int d\Omega^2  j^r_{R,out} \,,\quad j^r_{R,out}=\frac{1}{2i} \left( \psi^* {\partial_ {\tilde r}}\psi- \psi{\partial_{\tilde r}}\psi^*\right)|_{FAR, out}\,.
 \end{align}
 At times of order $t_{n\ell}=r+r_0+n \pi \ell$, we get
 \begin{align}
\Sigma_{ref}&=\frac{\pi}{\omega_0^2}\left|\expval{\mathcal R_\omega e^{-i n \ell \pi \omega }}_g\right|^2=\begin{cases} 
0 & \text{for odd } n\\[1ex]
        \text{Area} \times \left(4 \omega_0^2 r_e^2\right) & \text{for even } n             \end{cases}\,.
 \end{align}
This indeed indicates that when $n$ is odd, i.e., when the wave packet is localized mostly on the left side, there is no reflected wave on the right side, and when $n$ is even, we get a little bit of radiation exiting the right mouth.

The analysis in this section shows that uncharged scalar probes traverse the wormhole rather inefficiently. Only a small fraction of the signal is transmitted within a finite time, and obtaining a sizeable transmission cross-section requires collecting the outgoing radiation over parametrically long timescales. We thus conclude that neutral scalar fields are poor probes for traversing the wormhole. On the other hand, precisely because their transmission differs significantly from the black hole case, they provide a sensitive diagnostic for distinguishing a traversable wormhole from a black hole.

\section{Scattering charged particles}
\label{sec:charged-particles}

Let us now turn to a seemingly innocent change in the scattering experiment, where we scatter \textit{electrically} charged objects off magnetically charged ones. We will first cover the case of a charged scalar; in Sec.~\ref{sec:fermions}, we will scatter electrically charged fermions. A crucial difference with respect to the neutral scattering lies in the fact that both black holes and wormhole mouths look like magnetic monopoles to charged particles, at least from far away. This means that we can use the symmetries and the physics of magnetic monopoles to understand some of the features that we obtain. To that end, we will first briefly review the relevant aspects of magnetic monopoles, before moving on to our scattering experiments.

\subsection{Short review on magnetic monopoles}
\label{sec:monopole}

The idea that magnetic monopoles, stable particles carrying magnetic charges $q_m$, must exist is supported by two, solid arguments. First of all, as put forth by Dirac, the existence of magnetic monopoles directly implies quantization of electric charge $q_e$. A second, related argument, was given by 't Hooft and Polyakov, who discovered that magnetic monopoles naturally occur in non-Abelian gauge
theories, making them a robust prediction of grand unified theories. Therefore, while Dirac has demonstrated the consistency of magnetic monopoles with quantum electrodynamics, 't Hooft and Polyakov have proven the necessity of magnetic monopoles in grand unified theories \cite{Preskill:1984gd}. 

The quantization of charge can most easily be seen through the constraint that the Dirac string should not be observable\footnote{Or, more amusingly, through the gedanken hoax of Coleman \cite{Coleman:1982cx}.}, or, in the Wu-Yang gauge, that the overlapping region of north and south vector potentials has a unique answer for the wavefunction. However, the quantization can be shown purely from the angular momentum constraint as well. In the charge–monopole system, the conserved total angular momentum includes the angular momentum stored in the electromagnetic field, so we do not have only the mechanical orbital term. Therefore, the generalized angular momentum is the sum of orbital, spin, and electromagnetic field contributions, and can be written as
\begin{equation}
    \vec{J} = \vec{L} + \vec{S} -  \mu \hat{r}, \qquad \mu = \frac{q_e q_m}{2},
\end{equation}
where $\vec{L} = m \vec{r} \times \vec{v}$ is the standard angular momentum, $\vec{S} = \frac{1}{2} \vec{\sigma}$ is the spin, $\hat{r} = \frac{\vec{r}}{r}$ is the unit normal connecting the two charges, and $q_e$ and $q_m$ are the electric and magnetic charge, respectively. We have normalized the magnetic charge such that the vector potential has the form $A_\phi = \frac{q_m}{2}\cos\theta$. Once we go to a quantum description, we must impose commutation relations from which we obtain that the product of charges must be an integer\footnote{One might worry that, since both charges flow, this relation holds only at the tree level. However, there have been multiple arguments in favor of anti-screening for the magnetic monopole. For instance, one can see explicitly that this relation holds for a heavy monopole at the one-loop level in \cite{Coleman:1982cx, Craigie:1985ti}.},
\begin{equation}
    q_e q_m = n, \qquad n \in \mathbb{Z}.
\end{equation}
Note that this situation has nothing in common with the standard approach to quantization, in which the quantized parameter comes from the discrete part of the spectrum of eigenvalues of a Hermitian operator.
In our case, the quantized parameter---the product of electric and magnetic charges-–-is not an eigenvalue of any quantum-mechanical operator; instead, the charge quantization condition has a \textit{topological} origin. This is the source of many fascinating features of magnetic monopoles; interested readers can see classic textbooks for more details \cite{Weinberg:2012pjx, Shnir:2005vvi}.

In our calculations below, we will be interested in the scattering of electrically charged particles off magnetically charged (gravitational) objects. Interestingly, this is exactly where some puzzling features of magnetic monopoles come into play. We can get some intuition for why this might be so from the generalized angular momentum. Let us imagine an electrically charged particle heading straight towards the monopole. If this particle were to somehow pass through the monopole, the orientation of $\hat{r}$ would flip naturally. However, for the angular momentum to remain conserved, we see that either the charge or the helicity of the charged particle would have to flip as well. One might think that we would always have some effective barrier so that we do not get such puzzling situations. Indeed, for charged scalars, this is essentially what happens. The allowed monopole harmonics (see Sec.~\ref{sec:charged-scalar}) satisfy $j=|\mu|,|\mu|+1,\ldots$, where $j$ is the total angular momentum quantum number, so there is no genuine $s$-wave channel. Therefore, even the lowest angular momentum sector carries a non-trivial centrifugal barrier. Consequently, charged scalars cannot reach the monopole core without suppression. However, this is exactly where the difference with respect to charged, massless \textit{fermions} comes in. In the presence of magnetic flux, the Dirac operator on the sphere admits zero modes. These occur in the sector $j= |\mu| - 1/2$ and give rise to a special radial channel with no centrifugal barrier (see Sec.~\ref{sec:fermions}). Consequently, charged massless fermions can reach the monopole core without obstruction.


How does one deal then with the impossible choice of forfeiting either helicity or the charge of the fermion?  This question has been at the center of research for many decades, including in recent times \cite{Kitano:2021pwt, Hamada:2022eiv, vanBeest:2023dbu,  vanBeest:2023mbs, Brennan:2023tae, Khoze:2023kiu, Loladze:2024ayk, Csaki:2024ajo, Tachikawa:2026cxd}. The main issue lies in the fact that we are dealing with a featureless, singular monopole---an Abelian one, with no core structure.  
In non-Abelian settings, a smooth 't Hooft-Polyakov monopole can carry charge (and become a dyon), so the incoming fermion can, in principle, simply deposit the charge onto it. But even in the non-Abelian case, one has a problem, since, for a heavy monopole, the process of monopole-to-dyon transition is energetically disfavored, so at low energies, one must take more care \cite{Craigie:1985ti}. In fact, one can view the Abelian monopole as the low-energy description of the 't Hooft-Polyakov monopole, as nicely explained in \cite{Bolognesi:2024kkb}.

A crucial ingredient for understanding these puzzles is the axial anomaly. In the presence of a monopole background, fermion number and chirality are tied to topological gauge field configurations, and the low-energy dynamics can violate naive chiral charge conservation through the ABJ anomaly. This phenomenon underlies the classic Callan–Rubakov effect, in which monopoles catalyze processes that would otherwise appear forbidden \cite{10.1063/1.2947547, VARubakov_1988, PhysRevD.25.2141, RUBAKOV1982311}. More recently, the problem has been revisited from the perspective of generalized global symmetries and non-invertible defects, leading to a refined understanding of the so-called charge–flavor puzzle in monopole scattering, in which one can preserve the charge or the flavor in the scattering process, but not both. In particular, recent work by \cite{vanBeest:2023dbu, vanBeest:2023mbs} has emphasized that part of the apparent paradox is resolved by carefully accounting for \textit{topological} operators in the infrared theory. Essentially, they have shown that there must exist topological lines (or more precisely, codimension-one surfaces) that attach to the outgoing fermion state, effectively setting it in the twisted sector of the theory; in 4d, these are known as ABJ defects \cite{Cordova:2022ieu, Choi:2022jqy}. These topological structures are such that they exactly compensate for the missing charge (as flavor is set to be preserved); see App.~\ref{komargodski} for a brief review of these recent developments and the statement of the charge-flavor puzzle. Although the above papers dealt with elastic scattering (in which the monopole state cannot be changed), there exist toy models that account for monopole dynamics \cite{POLCHINSKI1984345}, and that also show evidence of such topologically twisted states \cite{Loladze:2025jsq}. We will come back to this point in Sec.~\ref{sec:conclusion} and comment on a possible relation with magnetic black holes.

\subsection{Scalars with mass, charge, and angular momentum}
\label{sec:charged-scalar}

Having established the basics of monopole-charge interaction, let us now turn to our very similar problem. Consider a minimally coupled scalar field with electric charge $q_e$, which obeys
\begin{equation}\label{chwaveeq}
D_\mu\, D^\mu \varphi = 0 \,, \qquad D_\mu = \nabla_\mu - i q_e A_\mu \,, \qquad A = \frac{Q_m}{2} \cos\theta d\phi\;, \qquad F = -\frac{Q_m}{2} \sin\theta \, d\theta \wedge d\phi\;.
\end{equation}
The equations of motion are then given explicitly by\footnote{Adding a scalar mass $\mathfrak{m}$ would simply shift the separation constant to $\tilde{\lambda} = \lambda + \mathfrak{m}^2$. }, 
\begin{equation}
    D^\mu D_\mu \varphi = \Box \varphi -2i q_e A^\mu \nabla_\mu \varphi - q_e^2 A^\mu A_\mu \varphi = 0,
\end{equation}
where we set $\nabla^\mu A_\mu = 0$ since there are no non-zero components of this expression for our metric. We can make an ansatz for our field
\begin{equation}
    \varphi = e^{-i\omega t} R(r) \Theta(\theta) e^{im\phi},
\end{equation}
for which it is easy to see that the separated equation becomes
\begin{equation}
    \partial_r \left(r^2 f(r) \partial_r\right)R(r)  +\left(\frac{\omega^2 r^2}{f(r)} - \lambda\right)R(r) =0 ,
\end{equation}
\begin{equation}
\label{monopole-eqn}
    \Theta''(\theta) + \cot{\theta} \Theta'(\theta) +\left(\lambda -\frac{(m - \mu \cos\theta)^2}{\sin ^2{\theta }}\right)\Theta(\theta) = 0,
\end{equation}
where $\lambda$ is the separation constant, and we defined $\mu = \frac{q_e Q_m}{2}$. We see that the radial equation remains the same as for the neutral scalar scattering, which one could have deduced simply from the fact that $A = A_\phi d\phi$. The solution to the $\Theta$ equation is given by a set of Jacobi polynomials,
\begin{equation}
\begin{split}
\Theta(x = \cos\theta) &= 2^m \mathcal{N}
\left(\sin\frac\theta2\right)^{m-\mu}
\left(\cos\frac\theta2\right)^{m+\mu}
P^{(m-\mu,\;m+\mu)}_{\,j-m}(\cos\theta) \\
&= \mathcal{N} (1-x)^{\frac{m-\mu}{2}} (1+x)^{\frac{m+\mu}{2}} P^{(m-\mu, m+\mu)}_n(x),
\end{split}
\end{equation}
where the normalization constant $\mathcal{N}$ has not yet been determined. The range of $j$, which can be integer or half-integer, is lower-bounded by the charge, $j = |\mu|, |\mu| + 1, \dots$, and for each value of $j$, $m = -j, -j+1, \dots, j$. These solutions have been dubbed \textit{monopole harmonics} by Wu and Yang in \cite{Wu:1976ge}. Let us determine the normalization constant now. We can use the normalization of Jacobi polynomials,
\begin{equation}
    \int_{-1}^1 dx\; (1-x)^a (1-x)^b P^{(a,b)}_n(x) P^{(a,b)}_{n'}(x) = \frac{2^{a+b+1}}{2n + a + b + 1} \frac{\Gamma(n+a+1) \Gamma(n+b+1)}{\Gamma(n+a+b+1) \;n!} \delta_{nn'}.
\end{equation}
Our full solution is $ Y(x, \phi) = e^{im\phi} \Theta(x)$, and we demand that 
\begin{equation}
    \int_0^{2\pi} d\phi \int_{-1}^1 dx \; Y^*(x, \phi)\, Y(x, \phi) = 1,
\end{equation}
so plugging the above expression for $Y$ and imposing the orthonormality of Jacobi polynomials, we obtain 
\begin{equation}
    \mathcal{N} = \sqrt{\frac{2j+1}{4\pi 2^{2m}}} \sqrt{\frac{(j+m)! (j-m)!}{(j+\mu)! (j-\mu)!}},
\end{equation}
up to an overall phase. One can write the final result in terms of Wigner $d$ matrices, given that they are related to these spin-weighted spherical harmonics via a constant prefactor. Our final result is then
\begin{align}
\label{mag-mon}
   & Y^{(\mu)}_{jm} (\theta, \phi) = \\
&=\sqrt{\frac{2j+1}{4\pi}} \sqrt{\frac{(j+m)! (j-m)!}{(j+\mu)! (j-\mu)!}} \left(\sin\frac\theta2\right)^{|m-\mu|}
\left(\cos\frac\theta2\right)^{|m+\mu|}
P^{(|m-\mu|,\;|m+\mu|)}_{\,n}(\cos\theta) e^{im\phi}.
\end{align}
Now, this is the angular part of the solution for \textit{all} values of $r$, so at asymptotic infinity, we will not have spherical harmonics but instead the monopole harmonics. 

The matching procedure will be the same as for the neutral scalar, with $\lambda$ a slightly different constant now,
\begin{equation}
    \lambda=j(j+1)-\mu^2.
\end{equation}
However, there is an important difference: from the above conditions, we see that $j \ge \mu$, which means that the smallest value of $\lambda$ is no longer zero, but equal to $\mu$. Therefore, for the charged scalar, there will always be an angular momentum barrier.

\subsubsection{The far region} 

Let us then solve the radial part for the black hole, and for the wormhole, with $\lambda$ turned on. The full radial equation in the far region is given by
\begin{equation}
    \partial_r (f r^2 \partial_r)R(r) + \left(\frac{\omega^2 r^2}{f} - \lambda\right) R(r) = 0, \qquad f(r)  \to 1,
\end{equation}
for which the solution is
\begin{equation}
    R_{AF}(r) = c_1 j_{\kappa}(r w)+c_2 y_{\kappa}(r w), \qquad \kappa = \frac{1}{2} \left(\sqrt{4 \lambda +1}-1\right), \qquad  i_n(z) = \sqrt{\frac{\pi}{2 z}} I_{n+1/2}(z), 
\end{equation}
where the above are spherical Bessel functions of the first and second kind, respectively. For our purposes, we will need the expressions in the very far limit $r \to \infty$ and in the limit close to the black hole/wormhole mouth, which here is denoted as $r\to 0$. The relevant solutions are written below, and the details can be found in App.~\ref{app:charged-scalar}. The small $r$ expansion for general, non-integer $\kappa$ is given by 
\begin{equation}
    R_{AF}^{r\to 0} = \tilde{N}_1 (c_1 - \tan(\kappa \pi) c_2) r^\kappa + \tilde{N}_2 c_2 r^{-\kappa-1},
\end{equation}
with
\begin{equation}
    \tilde{N}_1 = \frac{\sqrt{\pi}}{2\Gamma(\kappa + 3/2)} \left(\frac{\omega}{2}\right)^\kappa, \qquad \tilde{N}_2 = -\frac{\Gamma(\kappa +1/2)}{2\sqrt{\pi}} \left(\frac{\omega}{2}\right)^{-\kappa-1}.
\end{equation}
The large $r$ limit gives
\begin{equation}\label{eq:chargedFAsol}
    R_{AF}^{r\to \infty} = \frac{{\alpha}_{in}}{2r\omega} e^{-i\omega r} + \frac{{\alpha}_{out}}{2r\omega} e^{i\omega r},
\end{equation}
with 
\begin{equation}
    {\alpha}_{in} = e^{i\kappa\pi/2} (ic_1 - c_2), \qquad {\alpha}_{out} = -e^{-i\kappa \pi/2} (i c_1 + c_2).
\end{equation}
Both of these will be the same for the black hole and for the wormhole, as they cannot be distinguished asymptotically. The reflection coefficient is then calculated as
\begin{equation}
    \mathcal{R} = \frac{\alpha_{out}}{\alpha_{in}} = e^{-i\pi \kappa} \frac{c_2 + ic_1}{c_2 - ic_1}.
\end{equation}
This expression is valid for both integer and non-integer values of $\kappa$.

\subsubsection{The near region} 

The near-region equation can be put in the form
\begin{equation}
    (\rho^2 \pm 1) R''(\rho) + 2\rho R'(\rho) + \left(\frac{\Omega^2}{\rho^2 \pm 1} - \lambda\right)R(\rho) = 0,
\end{equation}
where the + sign is for the wormhole metric, and the - sign is for the black hole metric. The frequency has been appropriately rescaled $\Omega = \omega \ell$. 

\subsection*{The black hole case}
We are mostly interested in the scattering off the wormhole; however, we find it useful to report the result for the black hole first and compare the wormhole calculation with it (see \cite{Maldacena:2020skw} for more details on magnetic black holes). In this case, the solution is simply given by 
\begin{equation}
    R_{NH}^{bh}(\rho) = c_3 P_{\kappa}^{(i \Omega)}(\rho )+c_4 Q_{\kappa}^{(i \Omega)}(\rho ),
\end{equation}
where $P$ and $Q$ are the associated Legendre polynomials of the first and second kind, respectively. The details of the calculation are given in App.~\ref{app:charged-scalar}; here we only note the final result for the reflection coefficient.

One can explicitly check that $|\mathcal{R}|^2$ goes to $1- 4 r_e^2 \omega^2$ for $\kappa\to0$ and small $\omega$, as it should in order to match with \eqref{refl} and give the area dependence for the absorption cross-section. For general $\kappa$, the result is a bit messy, but it simplifies for $\kappa = 1$ (which for standard scattering corresponds to $l = 1$), in which case we get
\begin{equation}
    |\mathcal{R}|^2 = 1-\frac{4 r_e^6 \omega^4}{9 \ell^2}, \qquad \ell = \frac{1}{2\pi T},
\end{equation}
which gives
\begin{equation}
    \sigma = \frac{\pi}{\omega^2}|\mathcal{T}|^2  =A\times \frac{r_e^4 \omega^4}{9 \omega^2 \ell^2} \propto A^3 \omega^2,
\end{equation}
where $A$ is the area of the black hole, and we used the relationship between the length of the throat and the black hole temperature \cite{Maldacena:2018milekhinpopov}. For non-integer values of $\kappa$, the result is a bit complicated, but one can nevertheless obtain the full form of the solution; see App.~\ref{app:charged-scalar}. We will just note that there are no special values of $\omega$ for which the reflection coefficient goes to zero---this is very much unlike the wormhole case, to which we turn now.

\subsection*{The wormhole case}

The solution in the wormhole background is given by 
\begin{equation}
    R_{WH}(\rho) = \hat{b}_1 P_{\kappa}^{(\Omega)}(i \rho )+ \hat{b}_2 Q_{\kappa}^{(\Omega)}(i\rho ),
\end{equation}
with the same meaning of functions as in the black hole case. Going through the same procedure as before (the details of which can be found in App.~\ref{app:charged-scalar}), we can obtain a pretty hefty expression for the reflection coefficient for generic values of $\kappa$ (see \eqref{eq:hefty}). To simplify, we can set $\kappa = 1$, for instance, to obtain
\begin{equation}
    \abs{\mathcal R}^2=\frac{\sin ^2(\pi  \omega  \ell ) \left(\omega ^8 r_e^{12}+\ell ^4 \left(\omega ^{12} r_e^{12}-81\right)-2 \omega ^{10} \ell ^2 r_e^{12}\right){}^2}{324 \omega ^8 \ell ^4 r_e^{12} \left(\omega ^2 \ell ^2-1\right)^2+\sin ^2(\pi  \omega  \ell ) \left(\omega ^8 r_e^{12}+\ell ^4 \left(\omega ^{12} r_e^{12}-81\right)-2 \omega ^{10} \ell ^2 r_e^{12}\right){}^2}\,.
\end{equation}
Looking at this form of the reflection coefficient squared, we can notice some properties similar to the scalar case. In particular, we can see that for the frequencies
\begin{align}
    \omega \ell = n+1\,,\quad n \in \mathbb N\,,\quad n>0\,,
\end{align}
we obtain perfect transmission. In App.~\ref{app:charged-scalar} we show that this generalizes for any integer $\kappa$: we get resonances at
\begin{align}\label{eq:integerkresonances}
    \omega \ell = n+\kappa\,,\quad n,\kappa \in \mathbb N\,,\quad n>0 \,.
\end{align}
Note that there are additional special frequencies at which this quantity vanishes, which depend on $r_e$, and not on $\ell$ only; however, they are outside the regime of validity of the solution\footnote{These frequencies satisfy $\omega ^4 r_e^4=\frac{\ell ^2}{r_e^2}$, which would imply $\ell\ll r_e$, which violates the causality condition.}. The non-integer $\kappa$ case is more involved, but it can be shown that there still exist resonant frequencies at
\begin{align}
    \omega \ell=n+\kappa+\epsilon\,,\quad n\in \mathbb{N}\,,\;n>0, \quad \epsilon \sim (\omega r_e)^{4\kappa+2}\,.
\end{align}
For generic frequencies, which do not make the trigonometric functions vanish, we can expand the reflection coefficient in small $r_e \omega$. Let us first present the $\kappa=1$ case:
\begin{align}
    \abs{\mathcal R}^2\simeq 1-(\omega r_e)^{12}\frac{4 \left(\omega ^2 \ell ^2-1\right)^2 }{81 \omega^4 \ell ^4}\frac{1}{\sin^2(\pi \omega \ell)}\,.
\end{align}
We see that for these generic frequencies the transmission cross-section is much smaller than the scalar case, where it was of the order $(\omega r_e)^4$. For generic values of $\kappa$, both integer and non-integer one similarly gets
\begin{align}\label{eq:integerkawayfromzeroes}
    \abs{\mathcal R}^2\simeq 1-(\omega r_e)^{8\kappa+4} g(\omega \ell, \kappa)\,,
\end{align}
where $g(\omega \ell, \kappa)$ is a function of $\omega \ell $ and $\kappa$ that we derive in App.~\ref{app:charged-scalar}; this function is also finite away from the resonant frequencies. 

We therefore conclude that, for the charged case, there are also resonant frequencies for which we have unobstructed transmission. Since the calculation would go in a similar way as in the neutral scalar, we can infer that if we scatter wave packets, we would obtain similar results. In particular, if we collect radiation for a very long time, we expect the answer to be half of the black hole cross-section:
\begin{align}
    \sigma_{tr}= \frac{A}{2} \times (\omega r_e)^{4\kappa} f(\omega \ell)\,,
\end{align}
where this time we expect that there could be a function of $f(\omega \ell)$. Instead, for early times, one would get a much smaller cross-section
\begin{align}
    \Sigma_{tr}\propto A \times (\omega r_e)^{8\kappa+ 2}  \tilde f(\omega \ell)\,,
\end{align}
where again $\tilde f(\omega \ell)$ is another function of $\omega \ell$ only. So both for long times and early times we find answers that are suppressed with respect to the neutral case by a positive power of the small quantity $\omega r_e$.
\subsubsection{The transmission cross-section}

In this section, we derive the analog of the connection formula of the cross-section with the transmission coefficient for the charged scalar case. But first, we have to derive the partial wave decomposition.

\paragraph{Partial wave decomposition.} Let us see how to do partial wave decomposition now. We first check that the above reduces to the standard spherical harmonics in the $\mu \to 0$ limit. Indeed, in that case $j \to \ell$, and using the Rodrigues' formula for the associated Legendre polynomials, it is easy to show that one obtains
\begin{equation}
    Y_{\ell m}(\theta, \phi) = (-1)^m \sqrt{\frac{2\ell + 1}{4\pi}} \sqrt{\frac{(\ell-m)!}{(\ell+m)!}} P_\ell^m(\cos\theta) e^{im\phi},
\end{equation}
which is the standard quantum-mechanical notation for spherical harmonics. 

To derive the partial wave decomposition (PWD) for monopole harmonics, let us first recall how one does it in the spherical harmonics case. The general formula for PWD is written as 
\begin{equation}
e^{i\vec k\cdot \vec r}
= \sum_{\ell=0}^{\infty}
(2\ell+1)i^\ell j_\ell(kr)
P_\ell(\hat k\cdot \hat r) =  4\pi
\sum_{\ell=0}^{\infty}
\sum_{m=-\ell}^{\ell}
i^\ell j_\ell(kr)
Y_{\ell m}(\theta, \phi)Y_{\ell m}^{*}(\theta_k, \phi_k),
\end{equation}
where $j_\ell(kr)$ are spherical Bessel functions, and where each unit vector has its own angles associated with it, $\vec{k} = (\theta_k, \phi_k)$ and $\vec{r} = (\theta, \phi)$. To make calculations simpler, we can choose the incoming wave direction to be along the $z$-axis, $\hat k = \hat z$. As 
\begin{equation}
    \hat k =
(\sin\theta_k\cos\phi_k,\; \sin\theta_k\sin\phi_k,\; \cos\theta_k),
\end{equation}
this means we set $\theta_k = 0$. However, $Y^*_{\ell m} (\theta_k = 0, \phi_k) = 0$, unless we set $m = 0$, so this is why the PWD collapses to a simpler expression,
\begin{equation}
    e^{i \vec{k} \cdot \vec{r}} = e^{ikr\cos\theta} =  \sum_{\ell=0}^\infty\sqrt{4\pi (2\ell + 1)} i^\ell j_\ell (kr) Y_{\ell 0}(\theta, \phi),
\end{equation}
since $P_\ell(1) = 1$. 

Now we have to perform a similar analysis for the monopole harmonics. The main difference now lies in the singular nature of the monopole. In particular, in the presence of a monopole, a wavefunction of a charged particle cannot be described globally by one ordinary complex function on the sphere; instead, one usually resorts to writing two separate wavefunctions on the north and the south pole, and then determining the relevant transition functions in the overlap region---this is exactly what Wu and Yang did in their seminal paper \cite{Wu:1976ge}. The fact that the full wavefunction is glued with a twist in the overlap region, $\Psi_N = e^{i\mu \phi} \Psi_S$, means that we have a section of a complex line bundle, not a global function. Physically, this means that the phase of a charged particle is gauge-dependent, so when one moves around the monopole, the gauge field has non-trivial holonomy. The wavefunction is therefore only locally a function; globally, it remembers the magnetic flux through the sphere. The monopole harmonics are precisely the angular eigenfunctions adapted to this bundle structure: they transform with the same transition function as the charged wavefunction, and therefore form the correct angular basis for charged particles moving in a monopole background. 

In our gauge $A = \frac{Q_m}{2}\cos\theta d\phi$, the monopole potential is instead written as a single expression, but this comes at the price of making the gauge field singular at the polar axis, since $d\phi$ is ill-defined there. This does not change the physics, as we have the same monopole line bundle. The patching information that was explicit in the Wu-Yang description is now hidden in the behavior of the wavefunction near the Dirac string. Thus, the charged wavefunction may be written locally as an ordinary function on the punctured sphere, but globally it should still be understood as a section of the monopole line bundle. This is why the appropriate angular eigenfunctions are still monopole harmonics, only expressed in this singular gauge. Consequently, the ordinary plane wave $e^{i\vec{k}\cdot \vec{r}}$, which is a globally defined scalar function, is not strictly the correct incoming state in the monopole background. Instead, one should think of the incoming state as a monopole-covariant analogue of a plane wave, whose partial wave decomposition is naturally written in terms of monopole harmonics.

Therefore, we write a more general form on the LHS that plays the role of a globally well-defined ``plane wave'',
\begin{equation}
    \Psi^{(\mu)}_{\vec k}(\vec r)  = 4\pi \sum_{j = |\mu|}^\infty \sum_{m = -j}^j i^j j_j(kr) Y_{j m}^{(\mu)}(\theta, \phi) Y_{j m}^{(\mu)*} (\theta_k, \phi_k),
\end{equation}
which, as shown above, reduces to the standard spherical harmonics case when $\mu = 0$. Setting $\hat k = \hat z$ would similarly lead to $Y^{(\mu)}_{j m}(0, \phi_k) = 0$ for generic $m$, but now the one non-zero mode is given by $m = \mu$, as one can see from the expression \eqref{mag-mon}. Therefore, we have
\begin{equation}
     \Psi^{(\mu)}_{\hat{k} = \hat{z}}(\vec r) = \sum_{j = |\mu|}^\infty \sqrt{4\pi (2j+1)} i^j j_j(k r) Y^{(\mu)}_{j\mu}(\theta, \phi),
\end{equation}
where we technically have also a phase factor $e^{-i\mu \phi_k}$, but we can set $\phi_k = 0$ without loss of generality. A similar (up to conventions) expression can be found in \cite{Shnir:2005vvi}. In the expressions below, we will slightly abuse the notation and write the standard plane wave for the LHS, although one should always understand it in the way explained above.

Let us then start with a `plane' wave $\mathcal N\,e^{i\omega z}$. Using the monopole harmonics expansion, we can write
\begin{equation}\label{eq:psiz}
\psi_z=\mathcal N\,e^{i\omega z}=\mathcal N\,e^{i\omega r\cos\theta}= \mathcal N\sum_{j=\mu}^{\infty}\sqrt{4\pi(2j+1)}\,i^j\,j_j(\omega r)\,Y^{(\mu)}_{j\mu}(\theta,\phi)\,,
\end{equation}
with
\begin{equation}
\mu=\frac{q_eQ_m}{2},\qquad j\ge \mu,\qquad \lambda=j(j+1)-\mu^2.
\end{equation}
Previously, we considered the following asymptotic incoming wave \eqref{eq:chargedFAsol}:
\begin{equation}\label{eq:asymptoticinwave}
\phi^{(\mu)}_{jm,\,\rm in}\simeq e^{-i\omega t}\left(\frac{\alpha_{\rm in}}{2r\omega}e^{-i\omega r}+\frac{\alpha_{\rm out}}{2r\omega}e^{i\omega r}\right)Y^{(\mu)}_{jm}(\theta,\phi),
\end{equation}
which means that we are considering a particular $j$ channel, and in that channel, we should consider $m=\mu$. The fact that each of these terms solves the asymptotic KG equation means that the plane wave $\psi_z$ also solves it. Even if we did not do so in the calculation above, we could have used a different normalization where we set $\alpha_{\rm in}=1$. Using this convention, the normalization will come from matching the incoming $s$-wave part of the large-$r$ Bessel function in \eqref{eq:psiz} with the asymptotic solution
\eqref{eq:asymptoticinwave}, giving
\begin{equation}
\mathcal N_j^{-1}=\sqrt{4\pi(2j+1)}\,i^j e^{-i\left(\frac{j\pi}{2}+\frac{\pi}{2}\right)},
\qquad
|\mathcal N_j|^2=\frac{1}{4\pi(2j+1)}.
\end{equation}
The asymptotic incoming flux in the $z$ direction is then
\begin{equation}
j_z=\frac{1}{2i}\left(\psi_z^*\partial_z\psi_z-\psi_z\partial_z\psi_z^*\right)=|\mathcal N|^2\,\omega = \sum_{j = \mu}^\infty |\mathcal{N}_j|^2 \omega, \qquad \psi_z = \mathcal{N} e^{i\omega z}.
\end{equation}
We can focus on a single $j$ mode below, so we work with $\mathcal{N}_j$. For the transmitted outgoing mode,
\begin{equation}
\phi^{(\mu)}_{jm,\,\rm out}\simeq e^{-i\omega t}\,\frac{T_{j}}{2r\omega}e^{i\omega r}Y^{(\mu)}_{jm}(\theta,\phi),
\end{equation}
we get
\begin{equation}
j_r=\frac{1}{2i}\left(\phi^*\partial_r\phi-\phi\partial_r\phi^*\right)=\frac{|T_{j}|^2}{4r^2\omega}\,\big|Y^{(\mu)}_{jm}(\theta,\phi)\big|^2,
\end{equation}
where we dropped the subleading terms in $r$. Therefore,
\begin{equation}\label{eq:pwd}
\sigma_j=\frac{r^2\int d\Omega\,j_r}{j_z}
=\frac{1}{|\mathcal N_j|^2}\frac{|T_{j}|^2}{4\omega^2}
=\frac{\pi}{\omega^2}(2j+1)|T_{j}|^2,
\end{equation}
where we used $\int d\Omega\,|Y^{(\mu)}_{j}|^2=1$. In the neutral limit, $\mu=0$ and $j=0$, so this reduces to the standard spherical expression.

\subsection{Charged, massless fermions}
\label{sec:fermions}

In this case, we are going to be a little bit more general and assume a generic spherically symmetric setup, which will then encompass the wormhole, the black hole, and any such spherically symmetric magnetic object. To that end, we will parametrize our background as 
\begin{equation}
    ds^2 = e^{2\sigma(t,x)}(-dt^2 + dx^2) + R^2(x)d\Omega^2, \hspace{15pt} A = \frac{Q_m}{2} \cos\theta d\phi.
\end{equation}
As for the Dirac spinor, we choose a tensor product representation that factorizes into a 2d spinor that lives on the $(t,x)$ part and a 2d spinor that lives on the sphere; this gives us 
\begin{equation}
    \gamma^1 = i\sigma_x \otimes \mathbbm{1}, \hspace{10pt} \gamma^2 = \sigma_y \otimes \mathbbm{1}, \hspace{10pt} \gamma^3 = \sigma_z \otimes \sigma_x, \hspace{10pt} \gamma^4 = \sigma_z \otimes \sigma_y,
\end{equation}
where $\sigma_i$ are Pauli matrices. We can then similarly decompose our ansatz for the field, expanding the 4d spinor in a basis of eigenspinors of the angular Dirac operator on the sphere $S^2$ (which we denote by $\mathcal{D}_2$) in the ``monopole'' background,
\begin{equation}
   \chi(t,x,\theta,\phi)=\frac{e^{-\sigma/2}}{R}
\sum_n \sum_{m=-j_n}^{j_n}
\psi_{n,m}(t,x)\otimes \eta_{n,m}(\theta,\phi).,
\end{equation}
where
\begin{equation}
\mathcal{D}_2\eta_{n,m}=\lambda_n\eta_{n,m},\qquad
\int d\Omega_2\,\eta_{n,m}^\dagger\eta_{n',m'}=\delta_{nn'}\delta_{mm'}.
\end{equation}
In this notation, $n$ labels the Landau level, $m = -j_n, -j_n+1, \dots, j_n$ labels the degeneracy inside that level, and $d_n = 2 j_n + 1$ is the degeneracy of the $n$-th level. The Dirac operator on the sphere can be written as (see details in App.~\ref{app:charged-fermion}),
\begin{equation}
    \mathcal{D}_2 \eta_{n,m} = \left(\frac{\sigma_y}{\sin\theta}\left(\partial_\phi  - i A_\phi\right) + \sigma_x \left(\partial_\theta + \frac{\cot\theta}{2}\right)\right)\eta_{n,m} = \lambda_n \eta_{n,m}.
\end{equation}
From here, we can see that the effective action for the fermions will be given by
\begin{equation}
\sum_n\sum_{m=-j_n}^{j_n}
\int d^2x\,\sqrt{-g}\;
\bar\psi_{n,m}\big(e^{-2\sigma} \left(i\sigma_x (\partial_t - i A_t) + \sigma_y(\partial_x - i A_x)\right)+\lambda_n \frac{e^{-\sigma}}{R} \sigma_z\big)\psi_{n,m}.
\end{equation}
Here we see that the first term provides the kinetic term for the 2d fermions, whereas the second term gives the mass to the fermions, $m \sim \frac{\lambda_n}{R}$. Therefore, generically, there will be an effective barrier for fermions, given by their Landau level mass. The only way we could have an unobstructed transmission is if we have zero modes, $\lambda_0 = 0$; let us see if that is possible and what it would imply. In App.~\ref{app:charged-fermion}, we give a slightly different derivation, akin to the original MMP calculation.

\subsubsection*{The spherical equation}

To obtain the spectrum of the Dirac spinor on S$^2$,  we will have to solve two equations,
\begin{equation}
    P_- f_- = \lambda f_+, \qquad P_+ f_+ = \lambda f_-,
\end{equation}
with
\begin{equation}
    P_\pm = \partial_\theta \mp B(\theta) + C(\theta), \qquad B(\theta) = \frac{m - \mu \cos\theta}{\sin\theta}, \qquad C(\theta) = \frac{\cot\theta}{2},
\end{equation}
where we imposed the ansatz that $\eta_\pm(\theta, \phi) = f_\pm(\theta) e^{im\phi}$. Now we can obtain
\begin{equation}
    P_-(P_+f_+) = \lambda^2 f_+, \qquad P_+(P_- f_-) = \lambda^2 f_-,
\end{equation}
and we can solve this ODE. Calculating the above gives us 
\begin{equation}
    f_+'' + \cot\theta f'_+ - \frac{(m - \left(\mu+\frac{1}{2}\right) \cos\theta)^2}{\sin^2\theta} f_+ - \left(\lambda^2+ \mu + \frac{1}{2}\right)f_+ = 0,
\end{equation}
\begin{equation}
    f_-'' + \cot\theta f'_- - \frac{(m - \left(\mu-\frac{1}{2}\right) \cos\theta)^2}{\sin^2\theta} f_- - \left(\lambda^2 - \mu + \frac{1}{2}\right)f_- = 0.
\end{equation}
We see that we obtain exactly the same equation as for the charged scalar \eqref{monopole-eqn}, just with shifted parameters; this is, of course, as expected, since we are solving for monopole harmonics after all. The shift in the charge $\mu$ comes from the spin connection $C(\theta)$. From the positivity of the angular momentum operator, we obtain that from $f_+$, we have $\lambda^2 \ge |\mu + \frac{1}{2}|$ and from $f_-$ that $\lambda^2 \ge |\mu - \frac{1}{2}|$. Connecting with the monopole analysis of the charged scalar, we can obtain the expression for the Landau level,
\begin{equation}
    \lambda^2 = \nu(\nu + 1) - j(j+1), \qquad \nu \equiv \mu - \frac{1}{2},
\end{equation}
which is the expression coming from both equations. However, if we now try to solve for the lowest Landau level ($\lambda = 0$), we see that we run into an interesting consequence: the lowest Landau level (LLL) implies $j = \nu = \mu - \frac{1}{2}$, but this condition can only be satisfied by one of the modes, depending on the sign of the charge. Namely, if $\mu > 0$, then the LLL condition is compatible with the $f_-$ condition, but if $\mu < 0$, then it is compatible only with the $f_+$ condition. Physically, this is the usual monopole zero-mode chirality statement: the lowest angular momentum mode exists in only one spinor component, with the surviving component determined by the sign of the monopole charge $\mu$. Therefore, without loss of generality, we will decide to keep $f_-$ and set $f_+ = 0$. Crucially, this means we only get to keep a single chirality.

Note also that to have the LLL condition satisfied, we did not need to be in the $s$-wave sector: had we been (and therefore set $j = 0$), we would obtain a minimally charged monopole, $\mu = \frac{1}{2}$. In our case, we want our charge of the black hole to be large, as we want macroscopic black holes (which are close to extremality), and because this way we obtain a large number of 2d fermions, as the Landau level degeneracy comes as a number of flavors of 2d fermions. The number of zero modes is given by $d_n = 2 j_n + 1 = |q|$ in this case, consistent with the index theorem.

We can also see broadly what would happen if we had kept the mass of the 4d fermion. The difference will lie in an extra term determining the mass, which will be present at the bosonized level as $\sim m \cos\phi$; see \cite{Tong}, for instance. One can now also add a $\theta$ term, which is redundant in the massless case. These new parameters lead to an interesting phase diagram, as depending on their interplay, one can have a phase transition akin to the physics of confinement. In particular, in the limit of very small mass, for the standard massive fermion bosonization (meaning a single Dirac fermion that does not come from dimensional reduction), we obtain a gapless Schwinger model described by 2d Ising CFT. This happens for a specific value of $\theta = \pi$, as the gauge field generates a confining quadratic potential for the boson, while the fermion mass generates a cosine potential, and the two can compete. 
It would be interesting to further study this model in our setup, and account for the small mass effects.

Note also that, if $\mu = 0$, and we have neutral fermions instead, then we cannot have $\lambda = 0$, as it would imply $j = -\frac{1}{2}$, which is, of course, outside the regime of validity of $j$. This implies that it is only the charged fermions that allow us to use LLL's, and which ultimately lead to a free 2d Dirac equation. For the neutral case, we will obtain a massive fermion equation, with the mass determined by $\lambda_n/R$, implying a non-zero reflection coefficient. This action can be exactly solved via bosonization, just like we will do for the massless Dirac equation in Sec.~\ref{subsec:boso}. Crucially, however, since the fermions do not couple to the gauge field, we will not have a rich phase diagram as above for the charged case. Instead, we will find a gapped model, known as a sine-Gordon model, with a potential given by a similar cosine term. This barrier, however, can disappear if we add additional interactions that tune the reduced 2d theory to a gapless point or produce lighter collective degrees of freedom. It would be interesting to study this further.

Before discussing the radial equation, we want to estimate how many such zero modes we have. For this, we use the index theorem, which says
\begin{equation}
    \mathrm{ind}\,\mathcal{D}_2 = n_+ - n_- = \frac{1}{2\pi}\int_{S^2} F = |Q_ m|,
\end{equation}
where $n_\pm$ are the numbers of zero-modes of positive and negative chirality, and $F$ is the magnetic flux associated with the vector potential of the black hole.
We have only one chirality, so $n_\pm = |Q_m|$ and $n_\mp = 0$, so the total number of zero modes is equal to $|Q_m|$. Therefore, our effective fermion action becomes
\begin{equation}
   \sum_{m=1}^N \int d^2x\;\bar{\psi}_{m} \left(i\sigma_x (\partial_t - i A_t) + \sigma_y(\partial_x - i A_x)\right) \psi_m.
\end{equation}
where we denoted $|Q_m| = N$, as we have a large number of zero modes due to the large charge of the black hole. 

\subsubsection*{The radial equation}

We can now focus on the $(t, r)$ part of the equation. Setting the background gauge field to zero, $A_\mu(t,x) = 0$, and with $\psi = \begin{pmatrix}
    \psi_+ \\ \psi_-
\end{pmatrix}$,
\begin{equation}
    (\partial_t - \partial_x) \psi_- = 0, \qquad (\partial_t + \partial_x) \psi_+ = 0.
\end{equation}
We can make an ansatz, 
\begin{equation}
    \psi_+ = \sum_k \alpha_k e^{-i \omega_k (t-x)}, \hspace{15pt}  \psi_- = \sum_k \beta_k e^{-i \omega_k (t+x)},
\end{equation}
which solves the equations of motion. Note that in this $s$-wave sector, the solutions are chiral, and each chirality has an associated direction: $\psi_+$ will be outgoing, and $\psi_-$ ingoing. This simple fact is behind many of the puzzles of the monopole-fermion scattering problem.

Let us take a moment to appreciate what happened here. We obtained a \textit{free} Dirac equation in 2d, regardless of the background metric; the only ingredient we needed was a magnetic object for which we can be in the lowest Landau sector. This means that the fermions will propagate with no obstruction to the core of the ``monopole'', regardless of the monopole being a black hole, a wormhole, or just an actual monopole. This was, of course, foreshadowed by our angular momentum reasoning in Sec.~\ref{sec:monopole}; here we see the calculation explicitly.  As one effectively gets $|\mathcal T|^2 =1$, this implies an infinite transmission cross-section via \eqref{eq:pwd}. 

This conclusion rests on the assumption that we can neglect the gauge field fluctuations. In fact, such fluctuations can sometimes greatly alter the above conclusion: a particularly relevant example was given by Alford and Strominger in \cite{Alford:1992ef}. Alford and Strominger studied the s-wave scattering of an electrically charged fermion off a magnetically charged extremal, \textit{dilatonic} black hole. Their main claim was that classically, meaning in a fixed background and ignoring backreaction, the $s$-wave mode is special because it is absorbed perfectly---just like in our case. However, quantum mechanically, that is, once the effective low-energy dynamics is treated properly, that same channel develops an effective barrier, and all of the fermions get reflected back. 

Their setup leads to a single-flavor Schwinger model for the 2d fermions, with a dilaton-dependent coupling. Crucially, because they have a linear dilaton background in the throat of their black hole, the dilaton blows up as one goes further down the throat. This divergence
translates into the effective Schwinger mass blowing up, and consequently, the fermions get reflected from this exponential barrier. Our case is different in two important aspects: we have a multi-flavor Schwinger model (due to the large Landau level degeneracy), and the dilaton does not blow up as the size of the spheres stabilizes to a constant value down the throat. This means that we should expect a barrier, although not as impenetrable as in the case of \cite{Alford:1992ef}. As we show below, in fact, only one ``center-of-mass'' mode feels the effective barrier, while the N-1 modes continue with no barrier. 

\subsubsection{Bosonization of the 2d model} 
\label{subsec:boso}

Let us recall first what the relevant dictionary is for the standard Schwinger model in flat space. We have the Dirac fermion coupled to the gauge field,
\begin{equation}
    \mathcal{L}_\psi = -\frac{1}{4} F_{\mu\nu} F^{\mu\nu} + i \bar{\psi} \gamma^\mu \left(\partial_\mu + i e A_\mu\right) \psi , \qquad j^\mu = \bar{\psi} \gamma^\mu \psi.
\end{equation}
Using the standard bosonization dictionary (see e.g. \cite{Tong}), we have
\begin{equation}
    j^\mu_V \longleftrightarrow -\frac{1}{\sqrt{\pi}} \epsilon^{\mu\nu} \partial_\nu \phi, \qquad j^\mu_A \longleftrightarrow  \frac{1}{\sqrt{\pi}} \partial^\mu \phi, \qquad i\bar{\psi} \gamma^\mu \partial_\mu \psi \longleftrightarrow \frac{1}{2} (\partial_\mu \phi) (\partial^\mu \phi),
\end{equation}
where we set $\epsilon^{01} = +1$ and the field $\phi$ is understood to be a scalar. The subscripts V and A stand for vector and axial, as both symmetries are preserved at the classical level. In 2d, there is only one independent component of the field strength, which we can choose to be $F_{01} \equiv E$. Therefore, $-1/4 F^2 = 1/2 E^2$. The action will now take the form
\begin{equation}
    \mathcal{L}_{\phi} = 
    \frac{1}{2} E^2 + \frac{1}{2} (\partial \phi)^2 + \frac{e}{\sqrt{\pi}} \epsilon^{\mu\nu} A_\mu \partial_\nu \phi =   \frac{1}{2} E^2 + \frac{1}{2} (\partial \phi)^2 + \frac{e}{\sqrt{\pi}} E \phi.
\end{equation}
The equation of motion for the boson gives us something interesting, as
\begin{equation}
     \partial_\mu j^\mu_A = \frac{e}{\pi} E = \frac{e}{2\pi} \epsilon^{\mu\nu} F_{\mu\nu}.
\end{equation}
This is exactly the axial anomaly of the QED$_2$ Schwinger model, which breaks the classical chiral symmetry. Although this is a quantum anomaly, here we obtained it as a result of a simple equation of motion. This illustrates that the bosonized model incorporates quantum effects. Note that the vector current is conserved, as it should be, $\partial_\mu j^\mu_V = 0$. 
The equation of motion for the field strength will give us
\begin{equation}
    \partial_\mu F^{\mu\nu} = e j^\nu = -\frac{e}{\sqrt{\pi}} \epsilon^{\nu\alpha} \partial_\alpha \phi,
\end{equation}
from which we determine that 
\begin{equation}
    E = -\frac{e}{\sqrt{\pi}}\phi.
\end{equation}
Plugging this back into the action, we obtain the bosonized version of the Schwinger model
\begin{equation}
    \mathcal{L}_\phi = \frac{1}{2} \left(\partial \phi\right)^2 - \frac{e^2}{2 \pi} \phi^2,
\end{equation}
from which we see that we obtained an effective mass for the scalar,
\begin{equation}
    m = \frac{e}{\sqrt{\pi}}.
\end{equation}
In our case, we will have an action that is a result of dimensional reduction, and therefore, the $F^2$ term will come with a dilaton (see App.~\ref{app:charged-fermion}), although the fermions will still see flat space. Another change will come from the $N$ flavors of fermions. 
When we have $N$ flavors, something interesting happens. Namely, only the ``center-of-mass'' mode will couple to the gauge field, while the rest will remain free (and therefore, massless once we integrate out the gauge field). To see this, we first note that the dictionary gives
\begin{equation}
    j_i^\mu = \bar{\psi}_i \gamma^\mu \psi_i  \longleftrightarrow - \frac{1}{\sqrt{\pi}} \epsilon^{\mu\nu} \partial_\nu \varphi_i, \qquad i\sum_{i=1}^N \bar{\psi}_i \gamma^\mu \partial_\mu \psi_i \longleftrightarrow \frac{1}{2} \sum_{i=1}^N (\partial_\mu \varphi_i) (\partial^\mu \varphi_i).
\end{equation}
We define the center-of-mass mode as
\begin{equation}
    \phi \equiv \frac{1}{\sqrt{N}}\sum_{i=1}^N \varphi_i, 
\end{equation}
with which we can rewrite the sum over the $N$ bosonic fields as a sum over this collective mode and a $N-1$ number of massless fields, which we denote as $\chi_a$, $a = 1, \dots, N-1$. The massless fields are composed as linear combinations of the original fields $\varphi_i$. For instance, for $N=3$ we will have
\begin{equation}
    \phi = \frac{\varphi_1 + \varphi_2 + \varphi_3}{\sqrt{3}}, \qquad \chi_1 = \frac{\varphi_1 - \varphi_2}{\sqrt{2}}, \qquad \chi_2 = \frac{\varphi_1 + \varphi_2 - 2 \varphi_3}{\sqrt{6}},
\end{equation}
where straightforward algebra shows that
\begin{equation}
    \sum_{i=1}^3 (\partial \varphi_i)^2 = (\partial \phi)^2 + \sum_{a=1}^2 (\partial \chi_a)^2.
\end{equation}
Therefore, we can write
\begin{equation}
     \sum_{i=1}^N (\partial \varphi_i)^2 = (\partial \phi)^2 + \sum_{a=1}^{N-1} (\partial \chi_a)^2,
\end{equation}
where for any $N$, we will have a suitable linear combination for $\chi$'s. Writing the action in terms of the collective mode and the rest of the modes, while including the dilaton dependence, we get
\begin{equation}
    \mathcal{L}_\phi = -e^{2\sigma}\frac{f(\Phi)}{4} F_{\mu\nu}F^{\mu\nu} + \frac{1}{2} (\partial \phi)^2 + \frac{1}{2} \sum_{a=1}^{N-1} (\partial \chi_a)^2 + e \sqrt{\frac{N}{\pi}} \epsilon^{\mu\nu} A_\mu \partial_\nu \phi,
\end{equation}
where we set $f(\Phi) = \Omega_2 \Phi^{3/2}$, and $e^{2\sigma}$ accounts for the $\sqrt{-g}$ factor. We see now explicitly that the gauge field couples only to the collective mode. At this point, we need the equations of motion, which will now be different due to the dilaton term, 
\begin{equation}
    \partial_\mu \left(e^{2\sigma} f(\Phi) F^{\mu\nu} \right)= e j^\nu,
\end{equation}
where we note $j^\nu = \sum_{i=1}^N j_i^\nu$. From here, we get
\begin{equation}
    e^{-2\sigma}f(\Phi) E = -e \sqrt{\frac{N}{\pi}} \phi.
\end{equation}
Our bosonized action will then have the form
\begin{equation}
    \mathcal{L}_\phi = \frac{1}{2} (\partial \phi)^2 + \frac{1}{2} \sum_{a=1}^{N-1} (\partial \chi_a)^2 - e^{2\sigma}\frac{e^2 N}{2 \pi f(\Phi)} \phi^2,
\end{equation}
so our (covariant, so we write $\sqrt{-g} = e^{2\sigma}$) mass term now depends on the dilaton and the number of flavors, 
\begin{equation}
    m = \frac{e}{\sqrt{\pi}} \sqrt{\frac{N}{f(\Phi)}}.
\end{equation}
The dilaton dependence will be simple in our case, as it does not vary too much (and it becomes a constant deep in the throat). As mentioned before, this is in contrast with Alford-Strominger, where the dilaton blows up as one goes further into the throat, creating an exponential barrier for the bosonized mode. The fact that even gauge field fluctuations do not deter the fermions from going straight to the core is something that \cite{Maldacena:2018milekhinpopov} already addressed in their appendix A. This was important for them as they used such fermions to support the wormhole with negative energy. For us, it simply means these fermions will go through the wormhole with an $\order{1}$ probability\footnote{One could wonder if such fermions could be used for other purposes as well. For instance, one can imagine probing the microstructure of supersymmetric black holes with these fermions.}.

Let us make some final remarks. We used the bosonized version of the Schwinger model, which captures a lot of the quantum theory of 2d fermions in QED. However, it is not clear if this model captures everything. First of all, it rewrites all the relevant quantities in terms of fermion bilinears. If we wanted to see a correlator that needs to distinguish between the fermions, we would not be able to get a useful result from the bosonization technique. One can still calculate such fine-grained observables; for instance, in \cite{Anninos:2024fty}, one uses a different method that deals directly with the fermion action\footnote{For them, it was important to look at such fine-grained correlators, as they contained crucial information about the S$^2$ instantons; we will not have such instanton contributions here. We thank Tarek Anous for discussions on this point.}. It would be interesting to see if such observables can tell us something useful in our setup. 

Another caveat is that the 2d action, which we obtained via dimensional reduction, is not the most general action consistent with the symmetries of the system. As mentioned in \cite{vanBeest:2023dbu}, one can add a plethora of terms (e.g., quartic couplings, $\theta$-terms, \dots) that would be consistent with the theory, and that could, in principle, change the $s$-wave story that we relied on here. It would be very interesting to account for some of those terms and to see if our conclusions still hold\footnote{We thank Luigi Tizzano for discussions on this point.}. Finally, we used Abelian bosonization; one can also perform a more adequate non-Abelian bosonization, as explained in \cite{Witten:1983ar}. The additional information contained in the non-Abelian formulation is the full flavor current algebra and the classification of allowed monopole boundary conditions. For our purposes of showing that $N-1$ modes remain massless, the Abelian procedure was enough. However, to elucidate the charge-flavor puzzle, discussed in Sec.~\ref{sec:monopole}, one should really perform the non-Abelian procedure; see \cite{AFFLECK1994374} in which they used exactly that to sharpen the BCFT picture of monopole-fermion scattering (see also Sec.~\ref{sec:conclusion}).

\section{Discussion}
\label{sec:conclusion}

In this paper, we studied the scattering of low-frequency waves off magnetically charged black holes and traversable wormholes. We analyzed the transmission and reflection probabilities for both neutral and charged scalar fields propagating through the wormhole geometry. Our setup is intrinsically one-sided: an observer sends radiation into one mouth of the wormhole and subsequently measures the reflected signal. We assume that the second mouth does not participate in the experiment—for instance, it could be shielded by a perfectly absorbing screen.

Within this setup, we find that the reflected signal received after short times (though still long enough for the radiation to traverse the wormhole and return) is highly suppressed, scaling as $\sigma_r \sim A(\omega r_e)^2$, with $A$ the area of the wormhole mouth. At late times, however, the situation changes qualitatively. Radiation trapped in the long wormhole throat gradually leaks out, and the accumulated reflected flux approaches $\sigma_r \sim A/2$ for neutral scalars. Physically, the incoming wave repeatedly bounces inside the throat and eventually emerges through either mouth with approximately equal probability. As a result, an observer who waits sufficiently long recovers a reflected cross-section of order one-half of the black hole area. For charged scalars, the corresponding late-time cross-section is reduced due to the restriction on the allowed partial waves.

We then carried out an analogous analysis for massless fermions. While neutral fermions behave similarly to scalars and encounter an effective potential barrier, charged fermions exhibit a qualitatively different behavior: they traverse the wormhole with $\mathcal O(1)$ probability. This is consistent with the mechanism that renders the wormhole traversable in the first place, namely the presence of charged fermions occupying the lowest Landau level. Indeed, this feature is not specific to traversable wormholes. Any magnetically charged object—including black holes and ordinary magnetic monopoles—supports the same lowest Landau level sector, providing an unobstructed channel for charged massless fermions.

These results highlight a sort of a hierarchy of traversability. Neutral scalar probes are strongly affected by the wormhole geometry, exhibiting suppressed and highly time-dependent transmission together with characteristic resonances. Charged massless fermions, by contrast, traverse the wormhole with essentially unit probability at low energies. At the same time, both classes of probes encode observable signatures of the wormhole geometry, albeit in qualitatively different ways.

\subsection*{Future directions}

There are plenty of distinct future directions one can take from here. 

\paragraph{Distinguishing a wormhole from a black hole.} The most obvious future direction is to use the reflection and transmission coefficients to properly study the different ways in which wormholes are distinct from black holes. Even though some progress in this direction has been made (e.g. \cite{Konoplya:2016hmd, Biswas:2022wah, Bao:2022iaz, Yang:2024prm, Chen:2024tss}), the particularly physical wormhole of MMP has not yet been studied in detail. The exception lies in the scalar wave case, for which \cite{Mondal:2025tht} performed a numerical analysis in a slightly different setup: they construct wave packets deep in the throat and study their propagation. It would be very interesting to flesh out these calculations of, say, Love numbers, quasi-normal modes, echoes, and other possible signatures, for all types of low-energy waves.

More broadly, our results naturally connect to the modern effective field theory description of compact objects and gravitational-wave observables. The reflection and transmission coefficients computed here are related to the quantities that appear in the scattering amplitudes, and worldline-EFT approaches
\cite{Goldberger:2004jt,Kosower:2018adc,Kalin:2020mvi,Cheung:2018wkq,Jones:2023ugm,Bautista:2023sdf,Ivanov:2024sds,Correia:2024jgr, Aoude:2024sve} (see \cite{Buonanno:2022pgc,Correia:2024yfx} for reviews). From this perspective, traversable wormholes differ qualitatively from black holes by possessing a finite transmission channel rather than a purely absorptive horizon. It would be particularly interesting to understand how the distinct features of a traversable wormhole are encoded in the analytic structure of the response functions \cite{Correia:2025enx}, for example, through distinctive poles and resonances. Moreover, it would be interesting to see whether such methods could allow to treat the full scattering problem (instead of sending off waves close to one mouth): our setup would be similar to the scattering of a wave off two black holes, with additional features given by the existence of the throat.

\paragraph{Large quantum gravity fluctuations.}

The wormhole we consider is constructed from near-extremal, magnetic black holes, and it crucially features a very long throat, which plays the role of the extremality parameter. One might then be worried that Schwarzian effects could become large in such a setup, as they naturally occur for near-extremal black holes. The natural low-energy theory for such a system has already been identified in the 2d Maldacena–Qi (MQ) wormhole \cite{Maldacena:2018lmt}, whose effective action consists of two coupled Schwarzian modes. Since an appropriate dimensional reduction of the 4d setup should lead to a similar description, this theory provides a useful guide for assessing the importance of quantum gravity fluctuations.

Namely, in the decoupled near-extremal case, the Schwarzian action is small at low temperature, so the boundary reparametrization becomes strongly fluctuating. In the MQ wormhole, however, the two sides are coupled by a relevant interaction that remains finite in the extremal limit. Rather than leaving two independent soft modes, the coupling locks the relative reparametrization and generates an effective potential for the corresponding degree of freedom, driving the system into a gapped wormhole phase. Consequently, the large fluctuations associated with two independent near-extremal throats are expected to be suppressed. A complementary perspective comes directly from the MMP construction. Namely, we saw that the charged, massless fermions, which render the wormhole traversable, effectively see only flat space, so they would not care about a large $\ell$ parameter, and would, therefore, be unaffected by the extremal limit.\footnote{We thank Juan Maldacena for bringing this point to us. See also \cite{ALM} for a different calculation indicating the same outcome---namely that the Schwarzian effects do not always become large for near-extremal magnetic black holes coupled to a large-$c$ CFT. } 

Nevertheless, it would be very interesting to make this story more precise and to see if there exists a regime of parameters that could change the above expectation. Moreover, it is not clear what will happen once we add additional matter (like our scalar probes) into the system. A natural follow-up to our work is to perform the analogous analysis as \cite{Emparan:2025sao, Biggs:2025nzs, Emparan:2025qqf, Betzios:2025sct} and see how Schwarzian effects alter the transmission cross-section. Although the expectation may be that such corrections will be small, large corrections could still appear near resonances, thresholds, or in observables that couple directly to the remaining collective mode. This will be discussed in detail in a separate work.

\paragraph{BCFT description and monopoles.} Finally, we mentioned throughout the paper the relationship with magnetic monopole analyses. As noted in Sec.~\ref{sec:monopole}, monopole-fermion scattering is littered with intriguing puzzles regarding the outgoing scattering states, as one naively cannot write down a state conserving all the relevant quantum numbers. As it stands, the problem has not been fully resolved (see App.~\ref{komargodski}), although there is strong evidence from several directions that new, topological sectors must be involved in the discussion. Let us assume, for the purposes of this section, that the full resolution will indeed involve topological operators that latch onto the outgoing states. Can we expect a similar story to occur for magnetically charged \textit{black holes}? Naively, the answer is no: the black hole will simply eat the incoming fermion, with all of its charges, and become a dyonic object. Here is where we already see a notable difference with respect to the standard monopole: for it, the dyon transition is energetically constrained, and only those particles that have high enough energy can excite the dyonic degrees of freedom; at low energies (which is our setup), such a process is forbidden. This is in stark contrast with the black hole setup, as the black hole has no such gap and can always become a dyon. The outgoing state will then be mixed in the subsequent Hawking radiation, which allows for all sorts of particles.\footnote{This is, of course, assuming that the black hole does not evaporate its magnetic charge first. If the black hole retains magnetic charge, then there will be a magnetic monopole remnant at the end of the evaporation process, at which point, we are back in the standard monopole regime \cite{Lee:1991qs, Maldacena:2020skw, Coviello:2025slv}. In fact, by the end of the evaporation process, the monopole can even \textit{stick out} from the black hole horizon \cite{Lee:1991vy}.}  Another way of saying this is that we do not expect global symmetries to be preserved in quantum gravity, so the charge-flavor puzzle no longer applies.

However, the relevant 2d physics of black holes can be represented via the BCFT description (see, for instance, \cite{Almheiri:2019hni}). Say that we have a near-extremal magnetic black hole, whose near-horizon region is described by AdS$_2$ physics. Down the throat, we can put charged, massless fermions, and as they see flat space, this is effectively described by a CFT$_2$. But these kinds of setups are exactly found in Karch-Randall descriptions, where we have an AdS brane, coupled to a large $N$ CFT (which, for us, is a large number of Landau level degeneracy). Such branes are then interpreted as dualizing the BCFT on the boundary. The `bath CFT' then plays the role of the asymptotic region that connects to the magnetic throat. Once we have this language, we can more directly connect to the recent monopole discussion, as there too one reduces the system to a BCFT, be it for the purposes of the Callan-Rubakov effect \cite{AFFLECK1994374}, or the Kondo problem \cite{Maldacena:1995pq}. The question of preserving all of the relevant quantum numbers then reduces to finding the correct monopole boundary condition (that is, correct boundary CFT state). Now we can state our question precisely: for the BCFT state that incorporates the black hole physics, is there such a boundary condition that indeed preserves all the quantum numbers? Some work in this direction is already being pursued in \cite{ALM}. Since global symmetries can be preserved in QFT, we cannot rely on their violation to get our answer. We think it would be extremely interesting to answer this question, as it would directly connect to a possibly new ingredient in the black hole evaporation process (can replica wormholes affect possible topologically twisted states?), or at least, to novel boundary states in BCFTs.

We thank Roberto Emparan and Luigi Tizzano for comments on the draft and useful discussions. We also thank Ahmed Almheiri, Tarek Anous, Alex Belin, David Berenstein, Mariana Carrillo Gonzalez, Craig Clark, Gabriel Cuomo, Joe Davighi, Markus Dierigl, Daniel Harlow, Shota Komatsu, Juan Maldacena, Alexey Milekhin, Mehrdad Mirbabayi, Miguel Montero,  Kyriakos Papadodimas, Marko Simonovi\'{c}, Joaquin Turiaci, Ben Withers, Anna Wolz, and Zhenbin Yang for useful discussions. AF and MT thank the Galileo Galilei Institute for Theoretical Physics for hospitality during the workshop ``Pathways to Quantum Black Holes: from Effective Theories to Exact Methods,'' where part of this work was completed. MT also thanks the Perimeter Institute for Theoretical Physics for hospitality during her Emmy Noether Fellowship.

\appendix

\section{Details for the neutral scalar scattering}
\label{app:uncharged-scalar}

In this appendix, we will first show that the $s$-wave reduction of a standard wave packet leads to a radial envelope that we will use throughout the main text. We will then show how to decompose such wave packets in frequency amplitudes, for which the time evolution is simpler to write down.

\subsection{Obtaining the radial wave packet envelope} 
\label{app:radialvsthreed}

We would like to show that the $s$-wave reduction of a standard wave packet results in a radial envelope of the type
\begin{align}
    \psi(\vec{x}) = \mathcal{N}_{in} e^{i \vec{k} \cdot (\vec{x} - \vec{x}_0)} e^{-\frac{(\vec{x} - \vec{x}_0)^2}{2 \sigma^2}} \qquad \xrightarrow[]{s-\text{wave}} \qquad \psi_s(r) = -\frac{\mathcal{N}_{in}}{2  i k} \frac{1}{r}e^{-i k (r - r_0)} e^{-\frac{(r-r_0)^2}{2 \sigma^2}}\,.
\end{align}
To that end, we align our coordinate system such that the momentum is aligned with the $z$ axis, and we choose $r_0$ to be aligned with this axis as well. Thus, $\vec{x}_0 = (0, 0, r_0)$ and $\vec{k} = (0, 0, -k)$. For an arbitrary point $\vec{x} = (r, \theta, \phi)$ in spherical coordinates, the $z$-component is $z = r \cos\theta$, and the phase term becomes $i\vec{k}\cdot (\vec{x} - \vec{x}_0) = -i k (r \cos\theta - r_0)$. We can write the distance as $|\vec{x} - \vec{x}_0|^2 = (r - r_0)^2 + 2 r r_0 (1 - \cos\theta)$.
To find the spherically symmetric $s$-wave sector, we project the wavefunction by averaging over the solid angle\footnote{This is equivalent to expanding in spherical Bessel functions and keeping only the $l=0$ term.},
\begin{align}
    \frac{1}{4\pi}\int_0^{2\pi} d\phi \int_0^\pi e^{\gamma \cos\theta} \sin\theta \,d\theta = \frac{e^\gamma - e^{-\gamma}}{2\gamma},
\end{align}
where we defined $\gamma = r\left(\frac{r_0}{\sigma^2} - ik\right)$. For $r_0/\sigma^2\ll k$, we can choose the incoming mode $e^{-ikr}$, and finally obtain our expression of interest,
\begin{align}
    \psi_s(r) = \mathcal{N}e^{-\frac{(r-r_0)^2}{2 \sigma^2}}  \frac{1}{r} e^{-i k r} \,,
\end{align}
which is precisely the radial envelope we considered in the text \eqref{eq:wavepacketradial}, with $k = \omega_0$.

\subsection{Deriving the evolution of the wave packet} \label{app:packetevol}

We now explain how to evolve an initially localized wave packet through the wormhole geometry and derive the wave packets observed in the FAL and FAR regions at later times. We begin with an ingoing Gaussian wave packet localized in the FAR region at time $t = 0$,
\begin{align}
	\psi_{r_0,\omega_0}(r_0,t=0)=  \mathcal{N} e^{-\frac{(r-r_0)^2}{2 \sigma^2 }} \frac{1}{r}e^{-i \omega_0 r} \,. 
\end{align}
The factor $e^{-i\omega_0 r}$ implies that the packet is approximately ingoing, propagating toward smaller $r$, while $\sigma$ determines its spatial width and $\omega_0$ its central frequency. Rather than evolving this packet directly, it is convenient to decompose it into a complete basis of exact scattering solutions of the Klein–Gordon equation in our background. The advantage of this basis is that each mode has the radial profile discussed in Sec.~\ref{subsec:plane-wave}, and the time evolution of the initial wave packet becomes straightforward: each frequency mode will evolve with a simple phase $e^{i\omega t}$.

We denote by $u_\omega(\mathrm r)$ the solutions corresponding to waves incident from the FAR region, and by $\hat{u}_\omega(\mathrm r)$ the spatially reversed solutions, corresponding to waves incident from the FAL region. Here, $\mathrm r$ collectively denotes the coordinates covering the entire spacetime, including both asymptotic regions and the wormhole throat. The wave packet can then be expanded as a superposition of exact waves with all energies $\omega$,
\begin{align}
	\psi_{r_0,\omega_0}(\mathrm r,t=0) =\int_0^\infty d\omega\, u_\omega (\mathrm r) f_{r_0,\omega_0}(\omega)+\int_0^\infty d\omega\, \hat u_\omega (\mathrm r) \hat f_{r_0,\omega_0}(\omega)\,.
\end{align}
The functions $f(\omega)$ and $\hat{f}(\omega)$ are the spectral amplitudes of the wave packet in this scattering basis. Physically, $f(\omega)$ measures how much of the initial packet is composed of modes incoming from the FAR side, while $\hat{f}(\omega)$ measures the overlap with modes incoming from the opposite asymptotic region. The functions $u_\omega(\mathrm r,t)$ are nothing but the solutions $\Phi_\omega$ obtained in the Sec.~\ref{subsec:plane-wave} (we denote them as $u_\omega$ here just because we are restricting to the $s$-wave sector):
\begin{align} \label{eq:scalarfourierdecomposition}
u_\omega(\mathrm r,t) = 
\begin{cases}
   \frac{1}{r \omega}(e^{-i \omega r} + R_\omega e^{i \omega r})\, e^{-i \omega t}\,, & \text{(FAR)}\,, \\[4pt]
   \text{Wormhole region: }\rho \text{ coordinates,} & \\[4pt]
   \frac{T_\omega}{\tilde r \omega}\, e^{i \omega \tilde r} e^{-i \omega t}\,, &\text{(FAL)}\,.
\end{cases}
\end{align}
We do not report on the solution in the wormhole region since it will not be important in the discussion here. Each mode describes a unit incoming wave from the FAR region, which is partially reflected into the FAR and partially transmitted into the FAL region. The functions $\hat u_\omega(\mathrm r)$ are the spatially reversed ones with a similar interpretation,
\begin{align} \label{eq:scalarfourierdecomposition2}
\hat u_\omega(\mathrm r,t) = 
\begin{cases}
    \frac{\hat T_\omega}{r \omega}\, e^{i \omega r} e^{-i \omega t}\,, &\text{(FAR)}\,,\\[4pt]
   \text{Wormhole region: }\rho \text{ coordinates,} & \\[4pt]
   \frac{1}{r \omega}(e^{-i \omega \tilde r} + \hat R_\omega e^{i \omega \tilde r})\, e^{-i \omega t}\,, & \text{(FAL)}\,.
\end{cases}
\end{align}
These hatted coefficients satisfy the following relations, which follow from conservation of the Klein-Gordon current:
\begin{align}
	\hat T_\omega=T_\omega\,,\quad \hat R_\omega=R_\omega^*\frac{T_\omega^*}{T_\omega}\,.
\end{align}
In our particular case, one can also check that $R=\hat R$. These modes form an orthogonal basis with respect to the Klein–Gordon inner product,
\begin{equation}
    \int d \mathrm r\,\mathrm r^2 u^*_\omega(\mathrm  r) u_{\omega'}(\mathrm r) = C_\omega \delta_{\omega \omega'}, \qquad  C_\omega = \frac{2\pi}{\omega^2},
\end{equation}
and similarly for the hatted modes.
The amplitudes are then obtained by projection onto this basis,
\begin{align}
	f_{r_0,\omega_0}(\omega)= \frac{1}{C_\omega}\int d \mathrm r\, \mathrm r^2  u^*_\omega (\mathrm r)\psi_{r_0,\omega_0}(\mathrm r,0) ,
\end{align}
with similar expressions for $\hat{f}$, and where the integration is over the full region comprised of the wormhole and asymptotic regions. Since the initial packet is localized far in the FAR region and propagates inward, its overlap with the opposite-incident modes $\hat{u}_\omega$ is exponentially suppressed. We may therefore focus only on the coefficients $f(\omega)$.
 
Approximating the packet as sharply localized around $r=r_0$, which is guaranteed in the regime discussed in the text (see eq.~\eqref{eq:wavepacketregimes}), we can simply integrate over the right region coordinate $r$, and extend the integral to the whole real $r$ axis. Therefore, we just have to compute a Gaussian integral, ending up with
\begin{align}\label{eq:ftot}
	f_{r_0,\omega_0}(\omega)&= \frac{\mathcal N \sigma\sqrt{2 \pi}}{\omega C_\omega }\left( e^{-\frac{1}{2} \sigma ^2 \left(\omega -\omega _0\right){}^2+i r_0 \left(\omega -\omega _0\right)}+R_\omega e^{-\frac{1}{2} \sigma ^2 \left(\omega +\omega _0\right){}^2-i r_0 \left(\omega +\omega _0\right)}\right)\,.
\end{align}
We now assume $\sigma \omega_0 \simeq \sigma \omega \gg 1$ so that the packet is sharply peaked in frequency around $\omega_0$. To be compatible with the previous inequalities, we need $\omega^{-1}\ll \sigma \ll r_0-r_e$ which, since we have $r_e \omega \ll 1$, implies 
\begin{align}
r_0 \omega \gg 1\,,\quad d \, \omega \gg 1\,.
\end{align}
In this limit, the second term is exponentially suppressed because it is centered around $\omega \simeq -\omega_0$, outside the support of the positive-frequency integral; physically, this means that the packet is almost purely ingoing. The coefficients $\hat f_{r_0,\omega_0}(\omega)$ are suppressed in the same exponential fashion.  For the other term, we can take $\omega -\omega_0$ to scale as $\sigma^{-1}$. Dropping the exponentially suppressed term, we get
\begin{align}\label{eq:ftot2}
	f_{r_0,\omega_0}(\omega)&= \frac{\mathcal N \sigma \sqrt{2 \pi}}{\omega C_\omega }\,e^{-\frac{1}{2} \sigma ^2 \left(\omega -\omega _0\right){}^2+i r_0 \left(\omega -\omega _0\right)}\,.
\end{align}
Having decomposed the initial packet into stationary scattering states, its time evolution becomes straightforward: each frequency mode simply evolves with a phase $e^{i\omega t}$. In the FAL region, we have
\begin{align}
\label{eq:FALwp}
	\psi^L_{r_0,\omega_0}(\tilde r,t)&= \int_0^\infty d\omega\, u_{\omega}(\tilde r)f_{r_0,\omega_0}(\omega) =\\
	&=\frac{1}{\tilde r }\int_0^\infty d\omega \frac{\mathcal N   \sigma}{\sqrt{2 \pi} } \,e^{-\frac{1}{2} \sigma ^2 \left(\omega -\omega _0\right){}^2}\,e^{-i \omega_0 r_0}\,T_\omega\,e^{+i  \omega\left( r_0+\tilde r-t\right)}\,,
\end{align}
where we used $C_\omega \omega^2=2 \pi $. This expression makes the physical picture transparent: the initial packet is decomposed into frequency eigenmodes, each mode scatters through the wormhole with transmission amplitude $T_\omega$, and the final transmitted packet is obtained by recombining all transmitted frequencies. Knowing the spectral amplitude, it is also straightforward to obtain the wave packet in the FAR region at generic time $t$, whose ingoing component is
\begin{align}
	\psi^{R, in}_{r_0,\omega_0}(t,r)&=\frac{e^{-i \omega_0 r_0}}{r}\int_0^\infty d \omega \frac{\mathcal N \sigma\sqrt{2 \pi}}{2 \pi }\,e^{-\frac{1}{2} \sigma ^2 \left(\omega -\omega _0\right){}^2}  e^{-i \omega(r+t-r_0)}=\\
	&=\mathcal N \frac{e^{-i \omega_0 (r+t)}}{r} e^{-\frac{\left(r-r_0+t\right){}^2}{2 \sigma ^2}}\,.
\end{align}

\section{Details for the charged scalar scattering}
\label{app:charged-scalar}

\subsection{The far region details}

We start with the solution in the far region, given by
\begin{equation}
    R_{AF}(r) = c_1 j_{\kappa}(r \omega)+c_2 y_{\kappa}(r \omega), \qquad \kappa = \frac{1}{2} \left(\sqrt{4 \lambda +1}-1\right), \qquad  i_n(z) = \sqrt{\frac{\pi}{2 z}} I_{n+1/2}(z), 
\end{equation}
where the above are spherical Bessel functions of the first and second kind, respectively, which are related to the standard Bessels with $i = y,j$ and $I = J, Y$. All constants are denoted by $c_n$, $n \in \mathbb{Z}$. 

\vspace{5pt}

\paragraph{Integer $\kappa$.} To obtain proper limits of this solution, we use 
\begin{equation}
    y_n(x) = (-1)^{n+1} j_{-n-1} (x),
\end{equation}
and write
\begin{equation}
\label{aff}
    R_{AF} = \sqrt{\frac{\pi}{2r\omega}} \left(c_1 J_{\kappa+1/2} (r\omega) - c_2 (-1)^\kappa J_{-\kappa-1/2}(r\omega)\right).
\end{equation}
To perform the small $r$ expansion, we use
\begin{equation}
    J_\alpha (z) = \frac{1}{\Gamma(1+\alpha)} \left(\frac{z}{2}\right)^\alpha,
\end{equation}
which is valid for all $\alpha$ except when $\alpha$ is a negative integer. Then we get
\begin{equation}
\label{afbh}
\begin{split}
    R_{AF}^{r\to0} &= \sqrt{\frac{\pi}{2r\omega}} \left(\frac{c_1}{\Gamma(\kappa+3/2)} \left(\frac{r\omega}{2}\right)^{\kappa+1/2} - \frac{(-1)^\kappa c_2 }{\Gamma(-\kappa+1/2)}\left(\frac{r\omega}{2}\right)^{-\kappa-1/2} \right) \\
    &= N_1 c_1 r^\kappa + N_2 c_2 r^{-\kappa-1},
\end{split}
\end{equation}
with 
\begin{equation}
    N_1 = \frac{\sqrt{\pi}}{2 \Gamma(\kappa + 3/2)} \left(\frac{\omega}{2}\right)^{\kappa}, \qquad N_2 = - \frac{(-1)^\kappa \sqrt{\pi}}{2\Gamma(-\kappa + 1/2)} \left(\frac{\omega}{2}\right)^{-\kappa-1}.
\end{equation}

\vspace{5pt}

For the asymptotic infinity expansion, we take again \eqref{aff}, and expand for large $r$, for which we use
\begin{equation}
    J_\alpha(z) = \sqrt{\frac{2}{\pi z}} \cos\left(z - \frac{\alpha \pi}{2} - \frac{\pi}{4}\right), \qquad \text{arg}\, z < \pi,
\end{equation}
that gives
\begin{equation}
    R_{AF}^{r\to\infty} = \frac{1}{r \omega} \left(c_1 \sin\left(r \omega - \frac{\kappa \pi}{2}\right) - c_2 (-1)^\kappa \cos\left(r \omega + \frac{\kappa \pi}{2} \right)\right)
\end{equation}
where $\kappa$ is not necessarily an integer: for the charged scalar, $\lambda = j(j+1) - \mu^2$, while for the neutral scalar $\mu = 0$ and $\kappa \in \mathbb{Z}$. Rewriting in terms of exponentials, we obtain
\begin{equation}
    R_{AF}^{r\to\infty} = \frac{\alpha_{in}}{2r\omega} e^{-ir\omega} + \frac{\alpha_{out}}{2r\omega} e^{ir\omega}, 
\end{equation}
with 
\begin{equation}
\label{alphas}
    \alpha_{in} = e^{i\pi \kappa/2} (i c_1 - c_2), \qquad \alpha_{out} = -e^{-i\pi\kappa/2} (ic_1 + c_2).
\end{equation}

\paragraph{Non-integer $\kappa$. } For non-integer values of $\kappa$, as will be the case for the charged scalar, we use the relation (from Wikipedia)
\begin{equation}
    Y_\alpha(z) = \frac{J_\alpha(z) \cos(\alpha \pi) - J_{-\alpha}(z)}{\sin(\alpha \pi)},
\end{equation}
so for the small $r$ expansion, we use the same relation as before (which is valid whenever $\alpha = \kappa + 1/2$ is not a negative integer),
\begin{equation}
    J_\alpha(z) = \frac{1}{\Gamma(1+\alpha)} \left(\frac{z}{2}\right)^\alpha, \qquad J_{-\alpha}(z) = \frac{1}{\Gamma(1-\alpha)} \left(\frac{z}{2}\right)^{-\alpha},
\end{equation}
which gives
\begin{equation}
    Y_\alpha(z\to 0) = \frac{\cot(\alpha \pi)}{\Gamma(1+\alpha)} \left(\frac{z}{2}\right)^\alpha - \frac{\Gamma(\alpha)}{\pi} \left(\frac{z}{2}\right)^{-\alpha}, \qquad \Gamma(\alpha) \Gamma(1-\alpha) = \frac{\pi}{\sin(\alpha \pi)}.
\end{equation}
Therefore, our full solution in the small $r$ expansion will be
\begin{equation}
\begin{split}
    R_{AF}^{r\to 0} &= \sqrt{\frac{\pi}{2 r \omega}} \left(\frac{c_1 - \tan(\kappa\pi) c_2}{\Gamma(\kappa + 3/2)} \left(\frac{r \omega}{2}\right)^{\kappa +1/2} - \frac{c_2 \Gamma(\kappa + 1/2)}{\pi}\left(\frac{r \omega}{2}\right)^{-\kappa -1/2} \right) \\
    &= \tilde{N}_1 (c_1 - \tan{(\kappa \pi)} c_2) r^\kappa + \tilde{N}_2 c_2 r^{-\kappa -1},
\end{split}
\end{equation}
where
\begin{equation}
    \tilde{N}_1 = \frac{\sqrt{\pi}}{2\Gamma(\kappa + 3/2)} \left(\frac{\omega}{2}\right)^\kappa, \qquad \tilde{N}_2 = -\frac{\Gamma(\kappa +1/2)}{2\sqrt{\pi}} \left(\frac{\omega}{2}\right)^{-\kappa-1}.
\end{equation}
Note that when $\kappa \in \mathbb{Z}$, $\tan{(\kappa \pi)} = 0$, and we recover the expressions from before. Now we can look at the asymptotic limit $r\to\infty$, in which case we use 
\begin{equation}
    J_\alpha(z) = \sqrt{\frac{2}{\pi z}} \cos\left(z - \frac{\alpha \pi}{2} - \frac{\pi}{4}\right), \qquad Y_\alpha(z) = \sqrt{\frac{2}{\pi z}} \sin\left(z - \frac{\alpha \pi}{2} - \frac{\pi}{4}\right),
\end{equation}
which gives
\begin{equation}
    R_{AF}^{r\to \infty} = \frac{1}{r\omega}\left(c_1 \sin\left(r \omega - \frac{\kappa \pi}{2}\right) - c_2 \cos\left(r \omega - \frac{\kappa \pi}{2}\right)\right),
\end{equation}
that is, when we rewrite in terms of exponentials,
\begin{equation}\label{eq:chargedFAsol2}
    R_{AF}^{r\to \infty} = \frac{{\alpha}_{in}}{2r\omega} e^{-i\omega r} + \frac{{\alpha}_{out}}{2r\omega} e^{i\omega r},
\end{equation}
with 
\begin{equation}
    {\alpha}_{in} = e^{i\kappa\pi/2} (ic_1 - c_2), \qquad {\alpha}_{out} = -e^{-i\kappa \pi/2} (i c_1 + c_2),
\end{equation}
where we see that they agree with those of integer $\kappa$ \eqref{alphas}.

\subsection{The black hole case}

The solution to the near region equation is given by 
\begin{equation}
    R_{NH}^{bh}(\rho) = c_3 P_{\kappa}^{(i \Omega)}(\rho )+c_4 Q_{\kappa}^{(i \Omega)}(\rho ),
\end{equation}
where $P$ and $Q$ are the associated Legendre polynomials of the first and second kind, respectively. To determine which solution is ingoing and which outgoing, we note that in the tortoise coordinate,
\begin{equation}
    \frac{dr_*}{d\rho}=\frac{1}{\rho^2-1}
\quad\Rightarrow\quad
r_*=\frac12\ln\!\left(\frac{\rho-1}{\rho+1}\right),
\end{equation}
the near-horizon limit gives 
\begin{equation}
    r_* \sim \frac12\ln(\rho-1)+\text{const},\qquad r_*\to -\infty.
\end{equation}
Now, evolving with $e^{-i\Omega t}$, we see that for EF coordinates, $u = t- r_*$ and $v = t + r_*$, it's $e^{-i\omega v}$ that's the ingoing solution. This is because the outgoing solution should be the one that goes to infinity as time goes to infinity, and the ingoing solution is the one for which when $t\to\infty$, we go towards the horizon, so $r_*\to-\infty$. From here, we see that the ingoing solution is the one for which $t = -r_*$, as we are evolving null rays, $dt^2 = dr_*^2$. Therefore,
\begin{equation}
    R_{in} \sim e^{-i\Omega r_*} \sim (\rho-1)^{-i\Omega/2}.
\end{equation}
We should now expand our associated Legendre polynomials to see which terms to keep. From DLMF \href{https://dlmf.nist.gov/14.8\#E12}{relations}, we see that for $\rho\to1^+$
\begin{equation}
    P_\kappa^\mu(\rho) = \frac{1}{\Gamma(1-\mu)} \left(\frac{2}{\rho-1}\right)^{\mu/2},
\end{equation}
while for the other one, we have 
\begin{equation}
    Q_\kappa^\mu(\rho) \sim a_1 \left({\rho-1}\right)^{\mu/2} + a_2 \left({\rho-1}\right)^{-\mu/2},
\end{equation}
where $a_i$ are a specific combination of gamma functions, but since we see that the outgoing branch exists here, we will set the overall coefficient $c_4 = 0$\footnote{Technically, in DLMF, they only list the term with $a_1$, but from their relation 14.9.14, we see that there is a symmetry $Q^\mu_\nu(x) = Q^{-\mu}_\nu(x)$, which implies we must take the other branch into account as well.}. To match with the asymptotic region, we must take the $\rho\to\infty$ limit, for which we use the DLMF again\footnote{DLMF tells us about two regimes which seem to be exclusive, namely $\text{Re}(\kappa)$ bigger or smaller than $-1/2$. However, it is only a statement about which behaviour is dominant: using the connection formulae, e.g. 14.9.1, we can see that both branches contribute, as they should if we want the matching to work.},
\begin{equation}
    P^{i\omega}_{\kappa}(\rho)\sim
\frac{\Gamma(\kappa+\tfrac12)}{\sqrt{\pi}\,\Gamma(\kappa-i\Omega +1)}(2\rho)^{\kappa}
\;+\;
\frac{\Gamma(-\kappa-\tfrac12)}{\sqrt{\pi}\,\Gamma(-\kappa-i\Omega)}(2\rho)^{-\kappa-1}.
\end{equation}
To connect with the asymptotically far region, we will also need to rescale
\begin{equation}
    \Omega = \omega \ell, \qquad \rho = \frac{r-r_e}{r_e^2} \ell \sim \frac{r \ell}{r_e^2}.
\end{equation}
Putting everything together, the $\rho\to\infty$ solution is then
\begin{equation}
\begin{split}
    R_{in}^{bh}(\rho) &= c_3 \left(\frac{\Gamma(\kappa + 1/2)}{\sqrt{\pi} \,\Gamma(\kappa - i \omega \ell + 1)} \left(\frac{2 r \ell}{r_e^2}\right)^\kappa + \frac{\Gamma(-\kappa-1/2)}{\sqrt{\pi} \Gamma(-\kappa - i \omega \ell)} \left(\frac{2r\ell}{r_e^2}\right)^{-\kappa-1}\right) \\
    & = N_3 c_3 r^\kappa + N_4 c_3 r^{-\kappa -1},
\end{split}
\end{equation}
with
\begin{equation}
    N_3 = \frac{\Gamma(\kappa + 1/2)}{\sqrt{\pi} \,\Gamma(\kappa - i \omega \ell + 1)} \left(\frac{2 \ell}{r_e^2}\right)^\kappa, \qquad N_4 = \frac{\Gamma(-\kappa-1/2)}{\sqrt{\pi} \Gamma(-\kappa - i \omega \ell)} \left(\frac{2\ell}{r_e^2}\right)^{-\kappa-1}.
\end{equation}
\paragraph{Matching with integer $\kappa$.}Now we can finally perform the matching with \eqref{afbh}. Matching requires us to have 
\begin{equation}
    N_1 c_1 = N_3 c_3, \qquad N_2 c_2 = N_4 c_3,
\end{equation}
so we obtain
\begin{equation}
    c_1 = \frac{2 \Gamma(\kappa + 1/2) \Gamma(\kappa + 3/2)}{\pi \Gamma(\kappa - i\omega \ell + 1)} \left(\frac{4\ell}{\omega r_e^2}\right)^\kappa c_3, 
    \end{equation}
\begin{equation}
     c_2 = \frac{2\Gamma(-\kappa + 1/2) \Gamma(-\kappa - 1/2)}{\pi \Gamma(-\kappa - i\omega \ell) } \left(\frac{-4\ell}{\omega r_e^2}\right)^{-\kappa-1} c_3.
\end{equation}
The reflection coefficient is determined by the ratio of the asymptotic ingoing and outgoing coefficients, so
\begin{equation}
\mathcal{R} = \frac{\alpha_{out}}{\alpha_{in}} = e^{-i\pi \kappa} \frac{c_2 + ic_1}{c_2 - ic_1}.
\end{equation}
One can explicitly check that $|\mathcal{R}|^2$ goes to $1- 4 r_e^2 \omega^2$ for $\kappa\to0$ and small $\omega$, as it should in order to match with \eqref{refl} and give the area dependence for the absorption cross-section. For general $\kappa$, the result is a bit messy, but it simplifies for $\kappa = 1$ (which for standard scattering corresponds to $l = 1$), in which case we get
\begin{equation}
    |\mathcal{R}|^2 = 1-\frac{4 r_e^6 \omega^4}{9 \ell^2},
\end{equation}
which gives
\begin{equation}
    \sigma = \frac{\pi}{\omega^2}|\mathcal{T}|^2\propto A^3 \omega^2,
\end{equation}
where $A$ is the area of the black hole.

\paragraph{Matching with non-integer $\kappa$. } For non-integer $\kappa$, we have 
\begin{equation}
    \tilde{N}_1  (c_1 - \tan(\kappa \pi) c_2) = N_3 c_3, \qquad \tilde{N}_2 c_2 = N_4 c_3,
\end{equation}
from which we get
\begin{equation}
    c_1 = \left(\frac{2\Gamma(\kappa + 1/2) \Gamma(\kappa + 3/2)}{\pi \Gamma(\kappa - i\omega \ell + 1)} \left(\frac{4\ell}{\omega r_e^2}\right)^\kappa - \frac{2 \tan(\kappa \pi) \Gamma(-\kappa -1/2)}{\Gamma(\kappa + 1/2) \Gamma(-\kappa - i\omega \ell)}\left(\frac{4\ell}{\omega r_e^2}\right)^{-\kappa-1}\right) c_3,
\end{equation}
\begin{equation}
    c_2 = -\frac{2 \Gamma(-\kappa -1/2)}{\Gamma(\kappa + 1/2) \Gamma(-\kappa - i\omega \ell)}\left(\frac{4\ell}{r_e^2}\right)^{-\kappa-1} c_3.
\end{equation}
The reflection coefficient is again given by 
\begin{equation}
\mathcal{R} = \frac{\alpha_{out}}{\alpha_{in}} = e^{-i\pi \kappa} \frac{c_2 + ic_1}{c_2 - ic_1}.
\end{equation}
For exactly half-integer values, the expression blows up (due to the various gamma functions), but we do recover the results for integer values of $\kappa$. It is hard to say more about what happens for generic non-integer values of $\kappa$, although we see that all orders in $\omega$ appear, including the linear order (and further fractional orders since $\omega$ scales with $\kappa$). We will also note that there are no special values of $\omega$ for which the reflection coefficient goes to zero: essentially because the $\omega$ dependence in the gamma factors comes with an $i$, so it cannot be compensated by $\kappa$. 

\subsection{The wormhole case}

Let us now apply the same techniques to the traversable wormhole. We will first cover the case of neutral scalars in the $s$-wave sector (with $\kappa = 0$), and then move on to the more generic case. The notation is such that tilded quantities are to be understood in the FAL region (that is, matching is done on the exit mouth, $\rho \to -\infty$), while those without a tilde will be matched at the entry mouth, $\rho \to \infty$.

\paragraph{Case with $\kappa = 0$.} The equation that we will solve is
\begin{equation}
    (\rho^2 + 1) R''(\rho) + 2\rho R'(\rho) + 
    \frac{\Omega^2}{\rho^2 + 1} R(\rho) = 0,
\end{equation}
for which the solution is given by 
\begin{equation}
    R_W(\rho) = b_1 e^{i \Omega \arctan{\rho}} + b_2 e^{-i\Omega \arctan{\rho}}.
\end{equation}
Now we need two limits of this solution, $\rho \to \pm \infty$,
\begin{equation}
    R_W(\rho\to\infty) = \gamma^b_1 + \gamma^b_2 r^{-1}, \qquad R_W(\rho\to-\infty) = \tilde{\gamma}^b_1 + \tilde{\gamma}^b_2 r^{-1},
\end{equation}
with 
\begin{equation}
    \gamma^b_1 = b_1 e^{i\pi \omega \ell/2} + b_2 e^{-i\pi \omega\ell/2}, \qquad \gamma^b_2 = \left(-i b_1 \omega\ell e^{i\pi \omega\ell/2} + i b_2 \omega\ell e^{-i\pi \omega\ell/2}\right) \frac{r_e^2}{\ell},
\end{equation}
\begin{equation}
    \tilde{\gamma}^b_1 = b_1 e^{-i\pi \omega\ell/2} + b_2 e^{i\pi \omega\ell/2}, \qquad \tilde{\gamma}^b_2 = \left(-i b_1 \omega\ell e^{-i\pi \omega\ell/2} + i b_2 \omega\ell e^{i\pi \omega\ell/2}\right)\frac{r_e^2}{\ell},
\end{equation}
where we used the matching conditions
\begin{equation}
    \Omega = \omega\ell, \qquad |\rho| = \frac{r\ell}{r_e^2}.
\end{equation}
We will match the solution for $\rho \to -\infty$ with the L asymptotic region, and the solution with $\rho \to \infty$ with the R asymptotic region. The equations for the asymptotic regions have been obtained above. The L asymptotic region will have its ingoing sector set to zero,
\begin{equation}
\begin{split}
    R_{AF}(r) &= \sqrt{\frac{\pi}{2 r \omega}} \left(a_1 J_{\kappa + 1/2}(r \omega) + a_2 Y_{\kappa+1/2}(r\omega)\right) \\
    &= \sqrt{\frac{\pi}{2 r \omega}} \left(a_1 J_{\kappa + 1/2}(r \omega) + a_2 \left(\frac{J_{\kappa+1/2}(r\omega) \cos(\pi(\kappa+1/2)) - J_{-\kappa-1/2}(r\omega)}{\sin(\pi(\kappa+1/2))}\right)\right) \\
    &= \sqrt{\frac{\pi}{2 r \omega}} \left((a_1 - \tan(\pi\kappa) a_2)J_{\kappa+1/2}(r\omega) - \frac{a_2}{\cos(\pi\kappa)} J_{-\kappa-1/2}(r\omega)\right),
\end{split}
\end{equation}
which in the small $r$ limit gives
\begin{equation}
    R_{AF}(r\to 0) = \tilde{\gamma}_1^a r^\kappa + \tilde{\gamma}_2^a r^{-\kappa-1},
\end{equation}
with
\begin{equation}
    \tilde{\gamma}^a_1 = \frac{\sqrt{\pi}(a_1 - \tan(\pi\kappa)a_2)}{2 \Gamma(3/2+\kappa)} \left(\frac{\omega}{2}\right)^\kappa, \qquad \tilde{\gamma}_2^a = -\frac{\sqrt{\pi} a_2}{2 \cos(\pi\kappa) \Gamma(1/2-\kappa)}\left(\frac{\omega}{2}\right)^{-\kappa-1}.
\end{equation}
The far limit $r\to\infty$ gives
\begin{equation}
    R_{AF}(r\to\infty) = \frac{\alpha_{in}}{2r\omega} e^{-ir\omega} + \frac{\alpha_{out}}{2r\omega} e^{ir\omega},
\end{equation}
with
\begin{equation}
    \alpha_{in} = e^{i\pi\kappa/2}(i a_1 - a_2), \qquad  \alpha_{out} =  -e^{-i\pi\kappa/2} (ia_1 + a_2).
\end{equation}
Now, we want to impose that nothing goes back in the wormhole, so we set $\alpha_{in} = 0$, which imposes
\begin{equation}
    a_2 = ia_1.
\end{equation}
Note that this is regardless of the value of $\kappa$. Imposing the above condition also affects the small $r$ coefficients, and the matching condition is now
\begin{equation}
    \tilde{\gamma}_1^a =  \tilde{\gamma}_1^b, \qquad  \tilde{\gamma}_2^a =  \tilde{\gamma}_2^b,
\end{equation}
and for $\kappa=0$ we get
\begin{equation}
    b_1 = \frac{a_1}{2} e^{i\pi w\ell/2} \left(1 - \frac{1}{w^2 r_e^2}\right), \qquad b_2= \frac{a_1}{2} e^{-i\pi w\ell/2} \left(1 + \frac{1}{w^2 r_e^2}\right).
\end{equation}
Note that we matched with $-r$ so as to account for the sign in $\rho$. From here we see that $\gamma_1^b $ and $\gamma_2^b$ must also be changed
\begin{equation}
    \gamma_1^b = a_1 \cos(\pi \omega\ell)  - \frac{i a_1}{\omega^2 r_e^2} \sin(\pi \omega\ell), \qquad \gamma_2^b = a_1 \omega r_e^2 \sin(\pi \omega\ell) + \frac{ia_1 }{\omega} \cos(\pi \omega\ell).
    \end{equation}
Matching with the other mouth requires us to have 
\begin{equation}
    \gamma^b_1 = \gamma^c_1, \qquad \gamma_2^b = \gamma_2^c,
\end{equation}
where 
\begin{equation}
   \gamma^c_1 = \frac{\sqrt{\pi}(c_1 - \tan(\pi\kappa)c_2)}{2 \Gamma(3/2+\kappa)} \left(\frac{\omega}{2}\right)^\kappa, \qquad \gamma_c^2 = -\frac{\sqrt{\pi} c_2}{2 \cos(\pi\kappa) \Gamma(1/2-\kappa)}\left(\frac{\omega}{2}\right)^{-\kappa-1},
\end{equation}
which for $\kappa=0$ gives
\begin{equation}
     \gamma^c_1 = c_1, \qquad \gamma_2^c = -\frac{c_2}{\omega}.
\end{equation}
From here, we obtain 
\begin{equation}
    i c_1 = i a_1 \cos(\pi \omega\ell) + \frac{a_1}{\omega^2 r_e^2} \sin(\pi \omega\ell),
\end{equation}
\begin{equation}
    c_2 = -a_1 \omega^2 r_e^2 \sin(\pi \omega\ell) - i a_1 \cos(\pi \omega\ell),
\end{equation}
which gives us the reflection coefficient,
\begin{equation}
   \mathcal{R} = \frac{\alpha_{out}}{\alpha_{in}} =  \frac{c_2 + ic_1}{c_2 - ic_1} = \frac{\frac{\sin(\pi \omega\ell)}{\omega^2 r_e^2}\left(\omega^4 r_e^4 -1\right)}{2i\cos(\pi \omega\ell) + \frac{\sin(\pi \omega\ell)}{\omega^2 r_e^2}\left(\omega^4 r_e^4 +1\right)},
\end{equation}
which, of course, agrees with the calculation in the main text \eqref{refl}. 

\paragraph{Case with generic $\kappa \neq 0$.} The solution in the wormhole background is given by 
\begin{equation}
    R_{WH}(\rho) = \hat{b}_1 P_{\kappa}^{(\Omega)}(i \rho )+ \hat{b}_2 Q_{\kappa}^{(\Omega)}(i\rho ),
\end{equation}
with the same meaning of functions as in the black hole case. It turns out that to best match with our results for $\kappa=0$, we should use only the associated Legendre functions of the first kind, that is $P^\mu_\kappa(z)$. It is known that $Q^\mu_\kappa(z)$ can be expressed as a linear combination of $P^\mu_\kappa(z)$ and $P^{-\mu}_\kappa(z)$, so we will work with 
\begin{equation}
    R_{WH}(\rho) = b_1 P^\mu_\kappa(i\rho) + b_2 P^{-\mu}_\kappa(i\rho), \qquad \mu = \Omega.
\end{equation}
We will use the hypergeometric representation, as it is the one best suited for these limits. In Abramowitz-Stegun \cite{Abramowitz1964}, one can find the relevant representations (see 8.1.5), but as these hypergeometrics will always depend on $z^{-2}$, they will all tend to 1 in the large $\rho$ limit, and we will be left only with the prefactors. Here we write all 4 relevant functions,
\begin{equation}
    P_\kappa^\mu(i\rho) = \bar{p}_1 \frac{(i \rho)^{\mu-\kappa-1}}{(\rho^2 + 1)^{\mu/2}} + \bar{p}_2 \frac{(i\rho)^{\kappa+\mu}}{(\rho^2 + 1)^{\mu/2}}, \quad  P_\kappa^\mu(-i\rho) = \bar{p}_1 \frac{(-i \rho)^{\mu-\kappa-1}}{(\rho^2 + 1)^{\mu/2}} + \bar{p}_2 \frac{(-i\rho)^{\kappa+\mu}}{(\rho^2 + 1)^{\mu/2}},
\end{equation}
\begin{equation}
     P_\kappa^{-\mu}(i\rho) = \hat{p}_1 \frac{(i \rho)^{-\mu-\kappa-1}}{(\rho^2 + 1)^{-\mu/2}} + \hat{p}_2 \frac{(i\rho)^{\kappa-\mu}}{(\rho^2 + 1)^{-\mu/2}}, \, P_\kappa^{-\mu}(-i\rho) = \hat{p}_1 \frac{(-i \rho)^{-\mu-\kappa-1}}{(\rho^2 + 1)^{-\mu/2}} + \hat{p}_2 \frac{(-i\rho)^{\kappa-\mu}}{(\rho^2 + 1)^{-\mu/2}},
\end{equation}
where we introduced a shorthand notation,
\begin{equation}
    \bar{p}_1 = \frac{2^{-\kappa-1} \;\Gamma(-\kappa-1/2)}{(-1)^{\mu/2}\; \Gamma(-\kappa-\mu)}, \quad \bar{p}_2  = \frac{2^\kappa \Gamma(\kappa+1/2)}{(-1)^{\mu/2} \Gamma(1+\kappa-\mu)},
\end{equation}
\begin{equation}
    \hat{p}_1 = \frac{2^{-\kappa-1} \;\Gamma(-\kappa-1/2)}{(-1)^{-\mu/2}\; \Gamma(-\kappa+\mu)}, \quad \hat{p}_2  = \frac{2^\kappa \Gamma(\kappa+1/2)}{(-1)^{-\mu/2} \Gamma(1+\kappa+\mu)}.
\end{equation}
At the exit, we will have
\begin{equation}
    \tilde{\gamma}_1^a = \tilde{\gamma}_1^b, \qquad \tilde{\gamma}_2^a = \tilde{\gamma}_2^b,
\end{equation}
where, when setting $a_2 = i a_1$, we have
\begin{equation}
   \tilde{\gamma}_1^a = \frac{\sqrt{\pi}(1 - i\tan(\pi\kappa))a_1}{2 \Gamma(3/2+\kappa)} \left(\frac{\omega}{2}\right)^\kappa = n_1 a_1, \quad \tilde{\gamma}_2^a = -\frac{i\sqrt{\pi} a_1}{2 \cos(\pi\kappa) \Gamma(1/2-\kappa)}\left(\frac{\omega}{2}\right)^{-\kappa-1} = n_2 a_1.
\end{equation}
We can now take the $\rho \to \infty$ limit of all 4 functions, and see the behaviour at both ends of the wormhole, where $L$ denotes the exit and $R$ entrance,
\begin{equation}
    R_L(\rho\to - \infty) = \tilde{\gamma}_1^b r^\kappa + \tilde{\gamma}_2^b r^{-\kappa-1},
\end{equation}
with 
\begin{equation}
    \tilde{\gamma}_1^b = \left(\frac{\ell}{r_e^2}\right)^\kappa e^{-i\pi\kappa/2}\left(b_1 \bar{p}_2 e^{-i\pi\mu/2} + b_2 \hat{p}_2 e^{i\pi\mu/2}\right) = n_1 a_1,
\end{equation}
\begin{equation}
     \tilde{\gamma}_2^b = \left(\frac{\ell}{r_e^2}\right)^{-\kappa-1} e^{i\pi(\kappa+1)/2}\left(b_1 \bar{p}_1 e^{-i\pi\mu/2} + b_2 \hat{p}_1 e^{i\pi\mu/2}\right) = n_2 a_1.
\end{equation}
We can solve for the $b$ coefficients in terms of $a_1$ to obtain
\begin{equation}
    b_1 = \frac{e^{i\pi\mu/2} \hat{p}_2}{\bar{p}_1 \hat{p}_2 - \hat{p}_1 \bar{p}_2} \left(\left(\frac{\ell}{r_e^2}\right)^{\kappa+1} e^{-i\pi(\kappa+1)/2} n_2 - \left(\frac{\ell}{r_e^2}\right)^{-\kappa} e^{i\pi\kappa/2} \frac{\hat{p}_1}{\hat{p}_2} n_1\right)a_1,
\end{equation}
\begin{equation}
    b_2 = \frac{-e^{-i\pi\mu/2} \bar{p}_2}{\bar{p}_1 \hat{p}_2 - \hat{p}_1 \bar{p}_2} \left(\left(\frac{\ell}{r_e^2}\right)^{\kappa+1} e^{-i\pi(\kappa+1)/2} n_2 - \left(\frac{\ell}{r_e^2}\right)^{-\kappa} e^{i\pi\kappa/2} \frac{\bar{p}_1}{\bar{p}_2} n_1\right)a_1
\end{equation}
Now we can write the right side,
\begin{equation}
    R_R(\rho\to\infty)  = \gamma_1^b r^\kappa + \gamma_2^b r^{-\kappa-1},
\end{equation}
with 
\begin{equation}
    \gamma_1^b = \frac{\sin(\pi\mu)}{\bar{p}_1 \hat{p}_2 - \hat{p}_1 \bar{p}_2}\left(-i n_1 e^{i \pi \kappa} (\bar{p}_1 \hat{p}_2 + \hat{p}_1 \bar{p}_2) + \left(\frac{\ell}{r_e^2}\right)^{2\kappa+1} 2 n_2 \bar{p}_2 \hat{p}_2\right) a_1 + \cos(\pi\mu) n_1 e^{i\pi\kappa} a_1, 
\end{equation}
\begin{equation}
    \gamma_2^b = \frac{-\sin(\pi\mu)}{\bar{p}_1 \hat{p}_2 - \hat{p}_1 \bar{p}_2} \left(i n_2 e^{-i \pi \kappa} (\bar{p}_1 \hat{p}_2 + \hat{p}_1 \bar{p}_2) + \left(\frac{\ell}{r_e^2}\right)^{-2\kappa-1} 2 n_1 \bar{p}_1 \hat{p}_1\right) a_1 -\cos(\pi\mu) n_2 e^{-i\pi\kappa} a_1.
\end{equation}
At the entrance, as before, we will have
\begin{equation}
    \gamma_1^b = \gamma_1^c, \qquad \gamma_2^b = \gamma_2^c,
\end{equation}
with 
\begin{equation}
   \gamma^c_1 = \frac{\sqrt{\pi}(c_1 - \tan(\pi\kappa)c_2)}{2 \Gamma(3/2+\kappa)} \left(\frac{\omega}{2}\right)^\kappa, \qquad \gamma_c^2 = -\frac{\sqrt{\pi} c_2}{2 \cos(\pi\kappa) \Gamma(1/2-\kappa)}\left(\frac{\omega}{2}\right)^{-\kappa-1}.
\end{equation}
From here, we can find the expressions for $c_1$ and $c_2$, which go into the calculation for the reflection coefficient. For $\kappa=0$, the result agrees with our previous calculation. For generic $\kappa$, we have
\begin{equation}
\begin{aligned}
\label{eq:hefty}
c_1
= \frac{e^{-i\pi\kappa}\left(
\alpha_1 + \alpha_2 + \alpha_3
\right)}{2\left(1+e^{2 i \pi \kappa}\right)
\left(\csc\!\bigl(\pi(\ell \omega-\kappa)\bigr)+\csc\!\bigl(\pi(\kappa+\ell \omega)\bigr)\right)},
\end{aligned}
\end{equation}

\begin{equation}
\begin{aligned}
\alpha_1 &=
\frac{i \sin(\pi\ell \omega)}{2^{-4\kappa-3}\pi^3}\left(1+e^{2 i \pi \kappa}\right)^2
\Gamma\!\left(\kappa+\tfrac12\right)^2
\Gamma\!\left(\kappa+\tfrac32\right)^2
\,
\left(\frac{\ell}{\omega r_e^2}\right)^{2\kappa+1}
\Gamma(-\ell \omega-\kappa)\Gamma(\ell \omega-\kappa),
\\[1em]
\alpha_2 &=
\frac{i \sin{(\pi\ell \omega)}}{2^{4\kappa} \pi^3}\left(-1+e^{2 i \pi \kappa}\right)
\cos^2(\pi\kappa)\,
\Gamma\!\left(-\kappa-\tfrac12\right)^2
\Gamma\!\left(\tfrac12-\kappa\right)^2
\,
\left(\frac{\omega r_e^2}{\ell}\right)^{2\kappa+1}\times\\
&\times\Gamma(-\ell \omega+\kappa+1)\Gamma(\ell \omega+\kappa+1),
\\[1em]
\alpha_3 &= \frac{8 e^{2 i \pi \kappa}\cos(\pi\kappa)\sin(2\pi \ell \omega)}
{\cos(2\pi\kappa)-\cos(2\pi \ell \omega)}(1-2i \tan(\pi \kappa)).
\end{aligned}
\end{equation}

\begin{equation}
\begin{aligned}
c_2
&=
\frac{\beta_1 + \beta_2}{2\left(\csc\!\bigl(\pi(\ell \omega-\kappa)\bigr)+\csc\!\bigl(\pi(\kappa+\ell \omega)\bigr)\right)},
\end{aligned}
\end{equation}

\begin{equation}
\begin{aligned}
\beta_1 &=
\frac{e^{-i\pi\kappa}
\Gamma\!\left(\tfrac12-\kappa\right)
\Gamma\!\left(-\kappa-\tfrac12\right)
\left(\frac{\omega r_e^2}{\ell}\right)^{2\kappa+1}
\sin(\pi \ell \omega)\,
\Gamma(\ell \omega+\kappa+1)\Gamma(-\ell \omega+\kappa+1)}
{2^{4\kappa} \pi\,
\Gamma\!\left(\kappa+\tfrac12\right)
\Gamma\!\left(\kappa+\tfrac32\right)},
\\[1em]
\beta_2 &= \frac{4i\,\sin(2\pi \ell \omega)}
{\cos(2\pi \ell \omega)-\cos(2\pi\kappa)}.
\end{aligned}
\end{equation}
From here, it is easy to see that
\begin{equation}\label{eq:chargedscalarsRfull}
    \mathcal R = e^{-i\pi\kappa} \frac{c_2 + ic_1}{c_2 - i c_1} =e^{-i\pi\kappa} \frac{2\cos(\pi\kappa) \beta_i + i e^{-2i\pi \kappa} \alpha_i }{2\cos(\pi\kappa) \beta_i - i e^{-2i\pi \kappa} \alpha_i} \equiv e^{-i\kappa \pi} \frac{N}{D},
\end{equation}
where summation over the $i$ index is assumed. We can start analyzing this form of the reflection coefficient to look for resonant frequencies. If we work in the small $r_e \omega$ limit, we can see that, both in the numerator and the denominator, the leading term is given by $\alpha_1$, whereas the subleading term is given by the combination
\begin{align}
    \zeta_{N}=2\cos(\pi\kappa) \beta_2 + i e^{-2i\pi \kappa} \alpha_3=\frac{16 \sin (\pi  \kappa ) \sin (2 \pi  \omega  \ell )}{\cos (2 \pi  \kappa )-\cos (2 \pi  \omega  \ell )},
\end{align}
in the numerator, and in the denominator, we have
\begin{align}
    \zeta_{D}=2\cos(\pi\kappa) \beta_2 - i e^{-2i\pi \kappa} \alpha_3=-\frac{16 \cos (\pi  \kappa ) (\tan (\pi  \kappa )+i) \sin (2 \pi  \omega  \ell )}{\cos (2 \pi  \kappa )-\cos (2 \pi  \omega  \ell )}.
\end{align}
So, the numerator and denominator, with the leading and next-to-leading terms, are
\begin{align}
    N=&-\frac{2^{4 \kappa +5}\cos ^2(\pi  \kappa ) }{\pi ^3} \Gamma \left(\kappa +\frac{1}{2}\right)^2 \Gamma \left(\kappa +\frac{3}{2}\right)^2  \Gamma (-\kappa -\ell  \omega ) \Gamma (\ell  \omega -\kappa ) \left(\frac{\ell }{\omega  r_e^2}\right)^{2 \kappa +1}+\\
    &-\frac{16}{\pi^2}\sin (\pi  \kappa )\cos ( \pi  \omega  \ell )
\Gamma(1+\kappa+\ell\omega)
\Gamma(1+\kappa-\ell\omega)
\Gamma(-\kappa-\ell\omega)
\Gamma(\ell\omega-\kappa)
\end{align}
and
\begin{align}
    D=&-\frac{2^{4 \kappa +5}\cos ^2(\pi  \kappa ) }{\pi ^3} \Gamma \left(\kappa +\frac{1}{2}\right)^2 \Gamma \left(\kappa +\frac{3}{2}\right)^2 \Gamma (-\kappa -\ell  \omega ) \Gamma (\ell  \omega -\kappa ) \left(\frac{\ell }{\omega  r_e^2}\right)^{2 \kappa +1}+\\
    &+32 \cos (\pi  \kappa ) (\tan (\pi  \kappa )+i) \cos ( \pi  \omega  \ell )
\Gamma(1+\kappa+\ell\omega)
\Gamma(1+\kappa-\ell\omega)
\Gamma(-\kappa-\ell\omega)
\Gamma(\ell\omega-\kappa),
\end{align}
where in both cases we rewrote the expressions using gamma function identities and simplified factors of $\sin(\pi \omega \ell)$ that appear in both numerator and denominator.

We now want to look for resonances, i.e., frequencies where $\mathcal R \propto N/D$ vanishes. These can come from two sources: either from poles of the gamma functions in the denominator, which are not balanced by poles in the numerator, or frequencies at which the numerator genuinely vanishes. Notice that while in the numerator both terms are real (so if the signs work right, they can cancel), in the denominator this does not happen, since one of the terms is imaginary. So when the numerator vanishes, we do not have to worry about checking that the denominator is also doing so. Let us first look at singularities coming from the gamma function poles. If either a single or double poles come from $\Gamma(-\kappa-\omega \ell)\Gamma(\omega \ell-\kappa)$, then these cancel between numerator and denominator. The situation changes if the pole comes from $\Gamma(1+\kappa-\omega \ell)$, which happens for $\ell \omega =n+\kappa$ with $n>0$ an integer. In this case, the poles from the denominator and numerator would cancel, \textit{unless} $\kappa$ is an integer, in which case, in the numerator this term is multiplied by a $\sin(\pi \kappa)$ that removes one pole. Notice that when this is the case, the $\Gamma(-\kappa-\omega \ell)$ also has a pole, so the pole counting is: one in the numerator (two coming from the gamma functions, but one is removed by the sine), but two in the denominator. We therefore conclude that for an integer $\kappa$, there are resonant frequencies at
\begin{align}
    \omega \ell=n+\kappa\,,\quad n,\kappa\in \mathbb{N}\,,\;n>0.
\end{align}
The second origin of resonances can be traced in places where the numerator vanishes. Since we have two terms and one is suppressed by a small number, the only way to get them to cancel is to enhance the subleading term by going close to a pole of $\Gamma(1+\kappa-\omega \ell)$, i.e. at $\omega \ell=n+\kappa+\epsilon,$ with $n\in \mathbb{N},\;n>0.$ where $\epsilon$ is a small number. If one expands the numerator around these points, one finds 
\begin{equation}
  \begin{aligned}
N\simeq \frac{8\sin(\pi\kappa)}{\pi\epsilon}
&-\frac{2^{5+4\kappa}}{\pi^3}\,
(n+\kappa)^{1+2\kappa}
\cos^2(\pi\kappa)\,
\Gamma(n)\Gamma(-n-2\kappa)\times
\\
&\qquad\times
\Gamma\!\left(\kappa+\tfrac12\right)^2
\Gamma\!\left(\kappa+\tfrac32\right)^2
\sin\!\bigl[\pi(n+\kappa)\bigr]\,
(\omega r_e)^{-2-4\kappa},
\end{aligned}
\end{equation}
which can have a zero for
\begin{align}
    \epsilon=\frac{\pi ^2 4^{-2 \kappa -1} (-1)^n \sec ^2(\pi  \kappa ) (\kappa +n)^{-2 \kappa -1} \left(\omega  r_e\right){}^{4 \kappa +2}}{\Gamma \left(\kappa +\frac{1}{2}\right)^2 \Gamma \left(\kappa +\frac{3}{2}\right)^2 \Gamma (n) \Gamma (-n-2 \kappa )}.
\end{align}
Notice that $\epsilon$  scales with the small $\omega r_e$ parameter, so it has to be very small. Moreover, for integer $\kappa$, one sees that $\epsilon=0$, so for integer $\kappa$, we get back only the resonant frequencies we found above. We therefore conclude that for $\kappa$ non-integer, there are still resonant frequencies at
\begin{align}
    \omega \ell=n+\kappa+\epsilon\,,\quad n\in \mathbb{N}\,,\;n>0, \quad \epsilon \sim (\omega r_e)^{4\kappa+2}\,.
\end{align}
Away from these frequencies, one can expand the absolute value of the reflection coefficient for small $\omega r_e$ getting
\begin{align}
    |\mathcal R|^2 \simeq 1-\frac{\pi ^2 2^{-8 \kappa } \cos ^4(\pi  \kappa ) \Gamma \left(-\kappa -\frac{1}{2}\right)^2 \Gamma \left(\frac{1}{2}-\kappa \right)^2 (\omega  \ell )^{-4 \kappa -2} \left(\omega  r_e\right){}^{8 \kappa +4}}{\Gamma \left(\kappa +\frac{1}{2}\right)^2 \Gamma \left(\kappa +\frac{3}{2}\right)^2 (\cos (2 \pi  \kappa )-\cos (2 \pi  \omega  \ell ))^2 \Gamma (-\kappa -\ell  \omega )^2 \Gamma (\ell  \omega -\kappa )^2}\,.
\end{align}
This expression defines the function $g(\omega \ell,\kappa)$ in \eqref{eq:integerkawayfromzeroes}, where we see that we have almost total reflection, and the transmission coefficient scales as $\omega r_e$ to a power larger than the neutral case, yielding a much smaller cross section. All of these results can be confirmed numerically as well.
\paragraph{Comments on the regime of validity.} Notice that we said that there are resonances for $\omega \ell > \kappa$, but we must check if this is consistent with the other approximations we used. We have $\mu=\frac{q_e Q_m}{2}$, which implies $\lambda =j(j+1)-\mu^2$. The smallest value of $\lambda$ is $\mu$. Then the value of $\kappa$ is $\kappa=\frac{1}{2}(\sqrt{4 \lambda+1}-1)$, which for large magnetic charge is just $\kappa \approx \sqrt{\lambda}=\sqrt{\frac{q_e Q_m}{2}}$. Now the condition says
\begin{align}
    \omega \ell> \kappa \implies \frac{16 r_e^3}{G_N Q_m}\omega > \sqrt{\frac{q_e Q_m}{2}}.
\end{align}
The largest possible value of $\omega$ we can look at is $r_e^{-1}$, so we have
\begin{align}
    \omega_{\max} \ell> \kappa \implies \frac{16 r_e^2}{G_N Q_m} > \sqrt{\frac{q_e Q_m}{2}}
\end{align}
which we can rewrite as (since $r_e^2\equiv \frac{\pi^2 Q_m^2 G_N}{g^2}$)
\begin{align}
    \frac{\pi^2 16}{g^2}\,Q_m> \sqrt{\frac{q_e Q_m}{2}}
\end{align}
which is satisfied if $Q_m/g \gg q_e$, so in this regime we can have these resonant frequencies.


\section{Details for the charged fermion scattering}
\label{app:charged-fermion}


As explained in the review on monopoles in Sec.~\ref{sec:monopole}, the relevant sector of interest will be the $s$-wave sector, as higher angular momentum modes obtain an effective barrier, as we will show\footnote{One can also directly calculate the probability for the higher angular modes to be near the monopole core and obtain basically zero; see \cite{Bolognesi:2024kkb} for more details.}. Therefore, we will first reduce our action to 2d, and then see immediately what the consequences are for fermion scattering in this setup. We start from a 4D action, which is given by
\begin{equation}
    I = \int d^4x \sqrt{-G}\left(\frac{R^{(4)}}{16\pi G_N} - \frac{1}{4g^2} F_{\mu\nu} F^{\mu\nu} - i\bar{\chi}\gamma^\mu\left(D_\mu -iA_\mu\right)\chi\right),
\end{equation}
with $ F_{\mu\nu} = \partial_\mu A_\nu - \partial_\nu A_\mu$ and $G_{\mu\nu}$ is the 4D metric which defines $R^{(4)}$. We will perform a spherically symmetric reduction, with the metric ansatz
\begin{equation}
    G_{PQ} dy^P dy^Q = g_{\mu\nu}dx^\mu dx^\nu + e^{2\phi(x)}d\Sigma_N,
\end{equation}
where $P, Q$ run over $D + N$ dimensions, $\mu, \nu$ over $D$ and $\Sigma_N$ is the space on which one is reducing, with dimensionality $N$. In our case, $N = 2$, since we are reducing on an $S^2$. 

\paragraph{Einstein-Hilbert term.} Then, one can show that the Ricci scalar is given by \cite{Kolanowski:2023hvh}
\begin{equation}
    R^{(4)} = R^{(2)} - 4 \Box \phi - 6 (\partial \phi)^2 + 2 e^{-2\phi},
\end{equation}
and the volume measure is then
\begin{equation}
    \int_{\mathcal{\tilde{M}}} d^4x \sqrt{-G}  = \Omega_2 \int_{\mathcal{M}} d^2x \sqrt{-\bar{g}} e^{2\phi},
\end{equation}
where $\Omega_2$ is the unit area of a sphere. In order to have a more canonical form, we will perform a redefinition of the field,
\begin{equation}
    e^{2\phi} = \Phi.
\end{equation}
In the expression, we will obtain a $\Box \Phi$ term, but through Stokes' theorem, this will cancel with another boundary term. The final result is then 
\begin{equation}
    \frac{\Omega_2}{16 \pi G_N} \int d^2x \sqrt{-\bar{g}} \left(\Phi \bar{R} + \frac{(\nabla \Phi)^2}{2\Phi} + 2\right) + \frac{\Omega_2}{8\pi G_N} \int dx \sqrt{-\bar{h}} \Phi \bar{K},
\end{equation}
where we dropped the index $(2)$ on the Ricci scalar, and included the boundary term as well. The prefactor is just the effective Newton's coupling in 2d, which we define as $G_2 = \frac{G_N}{\Omega_2}$. Also, we would like to have the form with a dilaton potential and no kinetic terms. Therefore, we perform a Weyl rescaling, which can be found, for instance, in appendix A of \cite{Svesko:2022txo},
\begin{equation}
    \bar{g}_{\mu\nu} \to \omega^2 g_{\mu\nu},
\end{equation}
for which 
\begin{equation}
    \bar{R} = \omega^{-2} R - 2\omega^{-3} \Box \omega + 2\omega^{-4} (\nabla \omega)^2,
\end{equation}
\begin{equation}
    \bar{K} = \omega^{-1} K + \omega^{-2} n_\mu \nabla^\mu \omega,
\end{equation}
where $n_\mu$ is the normal vector. The term with $\Box \omega$, after partial integration, will cancel with the boundary term, and to get rid of the kinetic term, we set $\omega = \alpha \Phi^\beta$, for which we see that $\beta  = -1/4$ in order to cancel the kinetic term ($\alpha$ can remain unconstrained in this context for now. The final action is then
\begin{equation}
    I = \frac{1}{16 \pi G_2} \int d^2x \sqrt{-g} \left(\Phi R + U(\Phi)\right) + \frac{1}{8\pi G_2} \int dx \sqrt{-h} \Phi K,
\end{equation}
with 
\begin{equation}
    U(\Phi) = \frac{2\alpha^2}{\sqrt{\Phi}}.
\end{equation}
To match the standard notation in the literature \cite{Emparan:2023ypa}, we will set $\alpha = 1/\sqrt{2}$.\footnote{The authors of \cite{Svesko:2022txo} set it to $\alpha = (d-1)^{-1/2}$, but they also rescaled the size of the sphere.}

\paragraph{Maxwell term.} We have a magnetically charged black hole, so we have $F_{\theta\phi}$ components. Therefore, the natural ansatz for the vector potential is
\begin{equation}
    A = A_\mu(x) dx^\mu + \tilde{A}_\phi \cos\theta d\phi, \hspace{15pt}  \tilde{A}_\phi = const.
\end{equation}
Note that this is not the same reduction as in equations 2.19 and 2.20 from \cite{Iliesiu:2020qvm} because they had an electrically charged black hole in that case, so no $F_{\theta\phi}$ terms. In our case, the spherical term will give some magnetic flux through the S$^2$. We conveniently parametrized the above ansatz, so that upon reduction, we have
\begin{equation}
   \int d^4x F^2 = \Omega_2 \int d^2x \sqrt{-g}\; \Phi^{3/2} F_{\mu\nu}F^{\mu\nu} + \Omega_2 \int d^2x  \sqrt{-g} \;\Phi^{-3/2}  \tilde{A}_\phi^2,
\end{equation}
where we implemented the Weyl rescaling, and noted $F^2 \to \omega^{-4} F^2$. 

\paragraph{Fermions.} Here, we have to be a bit more careful. We will follow the notation of \cite{Bintanja:2021xfs, Maldacena:2018milekhinpopov}. We will parametrize our background as 
\begin{equation}
    ds^2 = e^{2\sigma(t,x)}(-dt^2 + dx^2) + R^2(x)d\Omega^2, \hspace{15pt} A = \frac{q}{2} \cos\theta d\phi.
\end{equation}
To write down the Dirac equation on this background, we will work in the vielbein formalism. The vielbein defines a local rest frame, allowing the constant $\gamma$ matrices to act at each spacetime point. We choose
\begin{equation}
    e^1 = e^\sigma dt, \hspace{10pt} e^2 = e^{\sigma} dx, \hspace{10pt} e^3 = R d\theta, \hspace{10pt} e^4 = R \sin\theta d\phi.
\end{equation}
To read off the spin connection terms, we employ Cartan's first structure equation,
\begin{equation}
    de^A + \omega^A_{\;B}\; \wedge \; e^B = 0, \hspace{15pt} \omega^{AB} = -\omega^{BA},
\end{equation}
where $e^A = e^A_\mu dx^\mu$, or, in components, $e^\mu_A e^B_\nu = \delta^\mu_{\;\nu}$, and
\begin{equation}
    \omega^{AB}_\mu = e^A_\nu \nabla_\mu e^{B\nu} = e^A_\nu \left(\partial_\mu e^{B\nu} + \Gamma_{\mu\lambda}^\nu e^{B\lambda}\right).
\end{equation}
Using the above equations, we determine the non-zero spin connections,
\begin{equation}
    \omega^{12} = \sigma' dt + \dot{\sigma} dx, \hspace{10pt} \omega^{32} = R' e^{-\sigma} d\theta, \hspace{10pt} \omega^{42} = R'\sin\theta e^{-\sigma} d\phi, \hspace{10pt} \omega^{43} = \cos\theta d\phi,
\end{equation}
where we used, for instance with $A=3$,
\begin{equation}
    R' dx \; \wedge \; d\theta + \omega^3_{\; 2} \; \wedge \; e^2 = R' dx \; \wedge \; d\theta - e^\sigma dx \; \wedge \; \omega^3_{\;2},
\end{equation}
from which one obtains the above result for $\omega^{32}$. Here, prime denotes a derivative w.r.t. $x$, and a dot w.r.t. $t$.

Now, we want to write down the Dirac equation, but first, we need to choose the correct representation of our $\gamma$ matrices. Since we have a 4D Dirac spinor, we want to have a tensor product representation which splits into two parts. Namely, a 4D spinor naturally factorizes into a 2D spinor (living on the (t,x) part) times a 2D spinor (living on the S$^2$). This makes the decomposition of the 4D Dirac operator into external and internal parts clean. Therefore, we can choose a representation for our $\gamma$ matrices as
\begin{equation}
    \gamma^1 = i\sigma_x \otimes \mathbbm{1}, \hspace{10pt} \gamma^2 = \sigma_y \otimes \mathbbm{1}, \hspace{10pt} \gamma^3 = \sigma_z \otimes \sigma_x, \hspace{10pt} \gamma^4 = \sigma_z \otimes \sigma_y,
\end{equation}
where we used a Pauli matrix representation
\begin{equation}
    \sigma_x = \begin{pmatrix} 0 & 1 \\ 1  & 0
    \end{pmatrix}, \hspace{15pt}  \sigma_y = \begin{pmatrix} 0 & -i \\ i  & 0
    \end{pmatrix}, \hspace{15pt}  \sigma_z = \begin{pmatrix} 1 & 0 \\ 0  & -1
    \end{pmatrix},
\end{equation}
and with which one reproduces the $\gamma$-matrix Clifford algebra,
\begin{equation}
    \{\gamma_i, \gamma_j\} = 2\eta_{ij} \mathbbm{1}.
\end{equation}

Now that we have the proper $\gamma$ matrix representation, we can construct the Dirac operator, defined as 
\begin{equation}
    \slashed{D} = \gamma^\mu D_\mu = \gamma^a e_a^\mu D_\mu,
\end{equation}
where
\begin{equation}
    D_\mu = \partial_\mu + \Gamma_\mu, \hspace{15pt} \Gamma_\mu = \frac{1}{8} \eta_{ac} \eta_{bd} \omega^{ab}_\mu \left[\gamma^c, \gamma^d\right].
\end{equation}
We can now calculate the various components:
\begin{equation}
    \gamma^1 e_1^t D_t = e^{-\sigma} \left(i\sigma_x \otimes \mathbbm{1} \right) \,\partial_t + \frac{\sigma'}{2} e^{-\sigma} \left(\sigma_y \otimes \mathbbm{1}\right),
\end{equation}
\begin{equation}
    \gamma^2 e_2^x D_x = e^{-\sigma} \left(\sigma_y \otimes \mathbbm{1}\right)\, \partial_x,
\end{equation}
\begin{equation}
    \gamma^3 e_3^\theta D_\theta = R^{-1} \left(\sigma_z \otimes \sigma_x \right)\,\partial_\theta + \frac{R' e^{-\sigma}}{2 R} \left(\sigma_y \otimes \mathbbm{1}\right),
\end{equation}
\begin{equation}
    \gamma^4 e_4^\phi D_\phi = \frac{1}{R \sin\theta} \left(\sigma_z \otimes \sigma_y\right)\,\partial_\phi + \frac{1}{2 R} \left(R' e^{-\sigma} \left(\sigma_y \otimes \mathbbm{1}\right) + \cot\theta \left(\sigma_z \otimes \sigma_x\right)\right).
\end{equation}
Taking into account that $\slashed{\nabla} = \slashed{D} - i \slashed{A}$, we have
\begin{equation}
    \slashed{\nabla} = e^{-\sigma} \left(i \sigma_x \partial_t + \sigma_y \left(\partial_x + \frac{\sigma'}{2} + \frac{R'}{R}\right)\right) \otimes \mathbbm{1} + \frac{\sigma_z}{R} \otimes \left(\frac{\sigma_y - i A_\phi}{\sin\theta} \partial_\phi + \sigma_x \left(\partial_\theta + \frac{\cot\theta}{2}\right)\right).
    \label{nabla}
\end{equation}
Now we apply this to our tensor product solution, which we normalize as 
\begin{equation}
   \chi(t,x,\theta,\phi)=\frac{e^{-\sigma/2}}{R}
\sum_n \sum_{m=-j_n}^{j_n}
\psi_{n,m}(t,x)\otimes \eta_{n,m}(\theta,\phi).,
\end{equation}
where
\begin{equation}
\mathcal{D}_2\eta_{n,m}=\lambda_n\eta_{n,m},\qquad
\int d\Omega_2\,\eta_{n,m}^\dagger\eta_{n',m'}=\delta_{nn'}\delta_{mm'}.
\end{equation}
Here $n$ labels the Landau level, $m = -j_n, -j_n+1, \dots, j_n$ labels the degeneracy inside that level, and $d_n = 2 j_n + 1$ is the degeneracy of the $n$-th level. We then have (suppressing the sums)
\begin{align}
    \slashed{\nabla} \chi =& \frac{e^{-3\sigma/2}}{R} \left(i\sigma_x \partial_t \psi + \sigma_y \partial_x \psi\right) \otimes \eta +\nonumber\\
    &+\frac{e^{-\sigma/2}}{R^2} \sigma_z \psi \otimes \left(\frac{\sigma_y}{\sin\theta} \left(\partial_\phi \eta - i A_\phi \eta\right) + \sigma_x \left(\partial_\theta \eta + \frac{\cot\theta}{2} \eta\right)\right).
\end{align}
We see that the Dirac operator on the sphere is given by the last part of the equation,
\begin{equation}
    \mathcal{D}_2 \eta_{n,m} = \left(\frac{\sigma_y}{\sin\theta}\left(\partial_\phi  - i A_\phi\right) + \sigma_x \left(\partial_\theta + \frac{\cot\theta}{2}\right)\right)\eta_{n,m} = \lambda_n \eta_{n,m}.
\end{equation}
Using 
\begin{equation}
    \bar{\chi} = \chi^\dagger \gamma^1 = (\frac{e^{-\sigma/2}}{R}\psi \otimes \eta)^\dagger (i\sigma_x \otimes \mathbbm{1}) = \frac{e^{-\sigma/2}}{R} \bar{\psi} \otimes \eta^\dagger,
\end{equation}
we can immediately obtain the action 
\begin{equation}
    \int d^2x\, d\Omega^2_2 \sqrt{-g}\,R^2 \bar{\chi} (\slashed{D} - i \slashed{A}) \chi = \int d^2x\, d\Omega^2_2 \sqrt{-g}
 \left(\bar{\psi} \alpha_1 \psi \otimes \eta^\dagger \eta + \bar{\psi} \alpha_2 \psi \otimes \eta^\dagger \mathcal{D}_2 \eta\right),
\end{equation}
where
\begin{equation}
    \alpha_1 = e^{-2\sigma} \left(i\sigma_x (\partial_t - i A_t) + \sigma_y(\partial_x - i A_x)\right), \qquad \alpha_2 = \frac{e^{-\sigma}}{R} \sigma_z.
\end{equation}
Using the orthonormality of the $\eta$ functions, we obtain 
\begin{equation}
\sum_n\sum_{m=-j_n}^{j_n}
\int d^2x\,\sqrt{-g}\;
\bar\psi_{n,m}\big(\alpha_1+\lambda_n \alpha_2\big)\psi_{n,m}.
\end{equation}
Here we see that the $\alpha_1$ term provides the kinetic term for the 2d fermions, whereas the second term gives the mass to the fermions. Given that the metric in 2d is parametrized here as $\sqrt{-g} = e^{2\sigma}$, and $e^{2\phi} = R^2$, we see that the kinetic term sees flat space. We will be interested only in the lowest ($n=0$) Landau level, which gives the zero modes of the 2d Dirac operator on the sphere, $\lambda_0 = 0$; all the non-zero Landau levels will give a mass contribution, $m \sim \frac{\lambda_n}{R}$. As explained in the main text, the index theorem gives us the number of zero modes to be equal to $|Q_m| = N$, where $N$ is a large number. Therefore, the effective fermion action becomes
\begin{equation}
   \sum_{m=1}^N \int d^2x \sqrt{-g}\, \bar{\psi}_{m} \alpha_1 \psi_m.
\end{equation}
We can also obtain the solutions for the angular functions for the lowest Landau level, since then we have
\begin{equation}\label{eq:eometazerolandau}
    \frac{\sigma_y}{\sin\theta} \left(\partial_\phi \eta - i A_\phi \eta\right) + \sigma_x \left(\partial_\theta \eta + \frac{\cot\theta}{2} \eta\right) = 0.
\end{equation}
Denoting
\begin{equation}
    \eta = \begin{pmatrix} \eta_+ \\ \eta_-
    \end{pmatrix},
\end{equation}
we obtain
\begin{align}
   \left( \frac{\sigma_y}{\sin\theta} \left(\partial_\phi  - i A_\phi \right) + \sigma_x \left(\partial_\theta + \frac{\cot\theta}{2} \right)\right) \begin{pmatrix} \eta_+ \\ \eta_-
    \end{pmatrix} = 0.
\end{align}

We can parametrize our solution as $\eta_\pm = \sum_{m\in\mathbb{Z}} \eta_\pm^m$, and set all $\eta_+^m = 0$\footnote{If we had kept it, we would have seen that it would not be able to satisfy all the conditions for $m$ and $j$, for $q > 0$ (and we would set the other solution to zero had we had $q<0$).}. The negative component can be written to satisfy the above equation of motion, which leads to the solution
\begin{equation}\label{eq:etasolution}
   \eta^m_-(\theta, \phi) = c_m \left(\sin\left(\frac{\theta}{2}\right)\right)^{j-m}\left(\cos\left(\frac{\theta}{2}\right)\right)^{j+m} e^{im\phi}, \qquad \chi=\frac{e^{-\sigma/2}}{R}
\begin{pmatrix}
0\\
\psi_+\eta_-\\
0\\
\psi_-\eta_-
\end{pmatrix},
\end{equation}
where $j = \frac{q-1}{2}$. We will determine the constant $c_m$ by demanding
\begin{equation}
\label{normalization}
    \sum_{m\in \mathbb{Z}}\int \sin{\theta} d\theta d\phi \;\eta^{\dagger m}_- \eta^n_- = 1.
\end{equation}
This gives then
\begin{equation}
    |c_m|^2 = \frac{1}{4\pi} \frac{\Gamma(2j+2)}{\Gamma(1+ j - m) \Gamma(1 + j +m)},
\end{equation}
under the condition that both $\text{Re}(j\pm m) > -1$. Therefore, our full 2d action has the form
\begin{align}
\label{rel-action}
    I =\int d^2x \sqrt{-g}\left(\frac{1}{16\pi G_2}\Phi R + U(\Phi) - \frac{\Omega_2\Phi^{3/2}}{4 g^2} F_{\mu\nu}F^{\mu\nu} +- \frac{\Omega_2\Phi^{-3/2}}{4 g^2}  \tilde{A}_\phi^2 + \sum_{i=1}^N\bar{\psi_i} \alpha_1 \psi_i\right),
\end{align}
where 
\begin{equation}
    U(\Phi) = \frac{1}{16\pi G_2} \frac{1}{\sqrt{\Phi}},
\end{equation}
and additionally we have a boundary term
\begin{align}
    I_{bdy} = \frac{1}{8\pi G_2} \int dx \sqrt{-h} \Phi K.
\end{align}
\section{Recent developments on the monopole-fermion scattering}
\label{komargodski}

In this appendix, we will provide a brief review of the charge-flavor puzzle in the monopole-fermion scattering, and outline the possible resolution. 

\subsection*{Different fermions, different story}

In our calculations, we focused on a single massless four-dimensional Dirac fermion, since this was the explicit model studied in MMP. However, much of the same analysis can be repeated for Weyl fermions, or for multiple Dirac fermions. These cases differ significantly in the magnetic monopole literature, and here we briefly explain why; for further details, see \cite{vanBeest:2023dbu, vanBeest:2023mbs}. Throughout, we consider minimally charged monopoles and fermions.

A useful point is that fermion--monopole scattering can be formulated entirely in an effective two-dimensional language \cite{AFFLECK1994374}. This is due to the special $s$-wave channel, in which only charged massless\footnote{Note that the final state puzzle exists only for massless fermions \cite{Shnir:2005vvi}: for massive fermions, chiral symmetry is explicitly broken, and there is no charge-flavor puzzle, as explained in \cite{Kitano:2021pwt}. In non-Abelian monopole backgrounds, however, this distinction can become subtle, as the GUT Higgs profile can vanish near the monopole core.} 
fermions propagate radially. One may therefore rephrase the problem as a boundary CFT (BCFT), with the monopole replaced by a boundary or defect at $r=0$, endowed with a boundary condition inherited from its UV completion. The appropriate boundary condition depends on the UV completion because, when the fermion enters the monopole core, it excites internal degrees of freedom that react in a model-dependent way. 

Viewed this way, it becomes easier to organize which types of matter admit a sensible scattering problem and which do not: the answer depends on whether the relevant boundary condition can be imposed consistently. In particular, if the two-dimensional reduction has an 't~Hooft anomaly for a symmetry, then no boundary condition can preserve that symmetry \cite{Jensen:2017eof}. Hence, any such symmetry must either be broken by the monopole, or else the monopole must exchange the corresponding charge or energy with the incoming state. In that situation, the usual notion of elastic scattering is obstructed. For this reason, we will be primarily interested in setups for which the effective two-dimensional theory is free of such 't~Hooft anomalies.

\paragraph{4d Weyl fermions.}
To illustrate how severe the anomaly obstruction can be, let us first consider a single Weyl fermion of either chirality. As is well known, this theory is anomalous and therefore should not be taken as a consistent UV-complete example. In four dimensions, the relevant obstructions are the cubic gauge anomaly and the mixed gauge--gravitational anomaly, proportional respectively to $\sum_i q_i^3$ and $\sum_i q_i$. In this case, the effective two-dimensional Dirac equation contains only ingoing modes or only outgoing modes. By contrast, in our Dirac example, both sectors were present. If only one sector exists, then there is no ordinary scattering problem, since there are no asymptotic states in the other sector---this is the ``missing state'' puzzle. Physically, the monopole must absorb the charge and energy carried by the incoming fermion, reflecting the underlying anomalies.

Even though the case of a single Weyl fermion is inconsistent, a version of the missing state puzzle can also arise for collections of Weyl fermions that are free of perturbative anomalies. For example, consider five left-handed Weyl fermions with charges $q=\{1,5,-7,-8,9\}$ for which $\sum_i q_i=\sum_i q_i^3=0$.\footnote{In fact, an infinite family of anomaly-free sets of five charges is
\[
q=\left\{
2p^2,\,-k^2,\;2p^2-k^2,\;-(2p^2-k^2+pk),\;-(2p^2-k^2-pk)
\right\},
\qquad p,k\in\mathbb Z .
\]
} Since dimensional reduction does not generate new local anomalies, the resulting 2d theory is also anomaly-free. Nevertheless, there is still a mismatch between the number of ingoing and outgoing modes. Without anomalies, BCFT reasoning suggests that some consistent monopole boundary condition should exist, so that a meaningful scattering problem can, in principle, be defined. The question is, then, whether such boundary conditions are easy to find.

In the language of \cite{vanBeest:2023dbu, vanBeest:2023mbs}, this case is subtle because the total number of ingoing and outgoing fermionic modes (here 15) exceeds the rank of the gauge group, $SU(2)\times U(1)$. In that regime, the BCFT analysis becomes harder to control because one does not yet know how to construct the appropriate monopole states systematically. To our knowledge, this case is not yet fully understood and remains an open question; see \cite{vanBeest:2023dbu, vanBeest:2023mbs, Bolognesi:2024kkb} for further discussion. Recent work indeed treats chiral gauge versions of the monopole--fermion problem as still open in important respects \cite{Bolognesi:2024kkb}.

\paragraph{4d Dirac fermions.}

For Dirac fermions, the situation is more controlled. Since a Dirac fermion contains both chiralities, one always has both ingoing and outgoing modes, as we saw in the effective 2d equation \eqref{eq:etasolution}. However, one must still be careful, because not all quantum numbers need to be preserved. Let us first review the case of a single massless Dirac fermion in 4d. Such a fermion can be decomposed into a left-handed Weyl fermion $\psi_L$ and a right-handed Weyl fermion $\psi_R$ of the same gauge charge $q$,
\begin{equation}
    \Psi_D=\begin{pmatrix}
        \psi_L\\[2pt]
        \psi_R
    \end{pmatrix}
    =
    \begin{pmatrix}
        \psi\\[2pt]
        \tilde\psi^\dagger
    \end{pmatrix}.
\end{equation}
Here we have also rewritten everything in terms of left-handed Weyl fields by defining $\tilde\psi\equiv (\psi_R)^c$, so that $\tilde\psi^\dagger=\psi_R$ and $\tilde\psi$ has gauge charge $-q$. (The gauge charge $q$ should not be confused with the corresponding Noether charge $Q$.) Upon reduction to 2d, these give rise to left- and right-moving two-dimensional Weyl fermions. In the lowest partial wave, chirality is correlated with ingoing versus outgoing propagation, so a consistent monopole boundary condition must relate opposite chiralities. This is allowed because the axial symmetry is anomalous. One can show \cite{vanBeest:2023dbu} that there exists a consistent boundary condition for a single Dirac fermion,
\begin{equation}
\psi_L|_{r=0}=e^{i\theta}\psi_R|_{r=0},
\qquad\text{equivalently}\qquad
\psi|_{r=0}=e^{i\theta}\tilde\psi^\dagger|_{r=0},
\end{equation}
so there is no puzzle in this setup. Thus, a single incoming massless charged fermion can scatter into an outgoing massless charged fermion with the same gauge charge but opposite chirality.

Things become more interesting when we consider several Dirac fermions. Let us take $N_f$ massless Dirac fermions\footnote{Strictly speaking, we require that $N_f$ is even, since it is tied to an $SU(2)$ UV completion, where one must avoid the Witten anomaly \cite{Witten:1982fp}.}. The continuous non-anomalous global symmetry is
\begin{equation}
U(1)_V\times SU(N_f)_L\times SU(N_f)_R,
\end{equation}
where we omit the anomalous axial symmetry $U(1)_A$ for the moment. Writing the theory in terms of left-handed Weyl fermions, the relevant fields are $\Psi_L$ and $\Psi_R^c$, which transform independently under $SU(N_f)_L$ and $SU(N_f)_R$. In the monopole background, the boundary condition can preserve a diagonal $SU(N_f)_V$ subgroup in two inequivalent ways: either $\Psi_L$ and $\Psi_R^c$ both transform in the fundamental representation, or one transforms in the fundamental while the other transforms in the anti-fundamental. Physically, only a diagonal subgroup can be preserved because the boundary condition must relate ingoing and outgoing modes, so any unbroken symmetry must act consistently on both sectors at once, rather than separately on the two chiralities. Following \cite{vanBeest:2023dbu}, these two possibilities are denoted by QED$(\Box,\Box)$ and QED$(\Box,\bar{\Box})$, respectively.

Let us now see how these different representations lead to different outcomes. Let $T^a$ be the Hermitian generators of $SU(N_f)$ in the fundamental representation $\Box$. Then $\psi\sim\Box$ means $\psi\to (1+i\alpha_aT^a)\psi$. Since the anti-fundamental representation $\bar{\Box}$ has generators $\bar T^a=-(T^a)^*$ \cite{Bastianelli2024GaugeTheories}, if $\tilde\psi\sim\bar{\Box}$ then $\tilde\psi^\dagger\sim\Box$. Thus in QED$(\Box,\bar{\Box})$ one has
\begin{equation}
\psi\sim\Box, \qquad \tilde\psi^\dagger\sim\Box,
\end{equation}
so the boundary condition
\begin{equation}
\psi|_B=e^{i\theta}\tilde\psi^\dagger|_B
\end{equation}
preserves $SU(N_f)_V$. By contrast, in QED$(\Box,\Box)$ one has
\begin{equation}
\tilde\psi\sim\Box,
\qquad
\tilde\psi^\dagger\sim\bar{\Box},
\end{equation}
so the same untwisted boundary condition is not, in general, $SU(N_f)_V$-invariant\footnote{When $N_f=2$, the fundamental is pseudoreal and $\Box\cong\bar{\Box}$, so this distinction disappears. Thus the puzzle requires $N_f>2$.}. If instead we tried to impose
\begin{equation}
    \psi \to \tilde{\psi},
\end{equation}
so as to preserve flavor, then gauge charge would fail to be preserved, since $\psi$ carries charge $q$ whereas $\tilde{\psi}$ carries charge $-q$. The fact that one seems forced to give up either flavor conservation or charge conservation is the modern form of the \textit{Callan--Rubakov puzzle}. More generally, one cannot find a monopole boundary condition that preserves all quantum numbers simultaneously.

The same tension is visible after bosonization, in a way closely analogous to the discussion in the previous section. Writing $\psi_i\to e^{iX_i}$ and $\tilde\psi_i\to e^{i\tilde X_i}$, one finds
\begin{align*}
\text{QED}(\Box,\bar{\Box}): \qquad & e^{iX_i}\to e^{-i\tilde X_i},\\
\text{QED}(\Box,\Box): \qquad & e^{iX_i}\to e^{i\tilde X_i-\frac{2i}{N_f}\sum_j\tilde X_j}.
\end{align*}
In the latter case, fractionally charged states appear in the bosonized description; see \cite{vanBeest:2023dbu} for the derivation. The recent monopole-scattering literature emphasizes exactly this topological dressing of the outgoing states. 

\subsection*{The proposed resolution of the Callan--Rubakov puzzle}

To understand the proposed resolution of this puzzle, we must return to the symmetries of the monopole, and in particular to the subtle role of the broken axial symmetry\footnote{See \cite{Harlow:2018tng} for an early discussion on this point from the algebraic QFT point of view. See also \cite{Benedetti:2023owa} for caveats regarding the non-invertible nature of the relevant defects.}. The basic idea is to replace the anomalous axial symmetry by a non-invertible symmetry, obtained by supplementing the topological operator with additional degrees of freedom; see \cite{Choi:2022jqy, Cordova:2022ieu, Karasik:2022kkq, GarciaEtxebarria:2022jky}. Historically, the axial symmetry is broken by quantum effects through the ABJ anomaly,
\begin{equation}
    d\star j_A=\frac{1}{8\pi^2}F\wedge F.
\end{equation}
One may nevertheless define an improved current,
\begin{equation}
    \star\hat j_A\equiv \star j_A-\frac{1}{4\pi^2}A\wedge dA,
    \qquad
    d\star\hat j_A=0,
\end{equation}
which is formally conserved but not globally gauge-invariant on topologically nontrivial backgrounds. From it one writes the would-be symmetry operator
\begin{equation}
    \hat U_\alpha(M)=\exp\left(i\alpha\oint_M \star j_A-\frac{1}{4\pi^2}A\wedge dA\right),
    \qquad
    \alpha\sim\alpha+2\pi,
\end{equation}
for a closed three-manifold $M$. When $M$ is topologically trivial, this operator is conserved and gauge-invariant; on a closed manifold with nontrivial topology, however, gauge invariance fails. The reason is that the Chern--Simons factor $\exp\!\left(\frac{iN}{4\pi}\int_M A\wedge dA\right)$ is gauge-invariant only when the level $N$ is properly quantized, whereas here the effective level is $\alpha/2\pi$. In the monopole problem, it is precisely the nontrivial topology sourced by the monopole that makes this obstruction physically relevant.

The obstruction lies in the non-integer nature of $\frac{\alpha}{2\pi}$, so why don't we just set $\alpha = 2\pi N$, with $N \in \mathbb{Z}$? This simple restriction cannot work because $\alpha$ is restricted to lie in $[0, 2\pi)$, as it goes in the exponential.  However, one may instead consider rational angles,
\[
\alpha=\frac{2\pi p}{N},
\qquad \gcd(p,N)=1.
\]
By itself, this still leads to a fractional Chern--Simons level, but one can restore gauge invariance by coupling to an additional topological sector, in close analogy with constructions familiar from fractional quantum Hall systems; see \cite{Shao:2023gho} for a review\footnote{The new field $a$ is an emergent three-dimensional gauge field living on the defect. It is partly Lagrange-multiplier-like in that it imposes the discrete magnetic-flux gauging condition, but it is more properly viewed as a genuine topological gauge field of the Chern--Simons wall theory, rather than as a purely auxiliary field.}. In this way, one obtains a well-defined topological operator, usually denoted $\mathcal D_{p/N}$. For $p=1$, this reproduces the fractional quantum Hall-type construction, while for general coprime $(p,N)$ one obtains the minimal $\mathbb Z_N^{(1)}$ TQFT $\mathcal A^{N,p}$. The surviving symmetry is then a discrete remnant of the axial symmetry.\footnote{Although see \cite{GarciaEtxebarria:2022jky, Karasik:2022kkq, Arbalestrier:2024oqg} for a different take on this construction. }

In the presence of magnetic monopoles, this symmetry becomes non-invertible\footnote{One might be worried about the fact that this operator, being non-invertible, will also be non-unitary. However, it was recently proven that the action
of a non-invertible symmetry operation can be represented by
a unitary operator, as long as we include twisted sectors, thereby preserving quantum probabilities \cite{Tachikawa:2026cxd, Bartsch:2026wqq}.}. The non-invertible nature of $\mathcal{D}_{p/N}$ is revealed by its action on extended operators. In particular, when $\mathcal{D}_{p/N}$ crosses an 't Hooft line, the latter is mapped not to itself times a phase, but to a dyon-like line operator attached to a topological surface. This is naturally interpreted using the Witten effect: a monopole in a background with $\theta$-angle carries electric charge, and the jump in the effective $\theta$-angle across $\mathcal{D}_{p/N}$ induces a fractional electric charge $p/N$ \cite{Shao:2023gho, Kaidi:2026urc}. In the monopole-scattering problem, this same topological structure reappears in the asymptotic out-state: the outgoing fermion is not an ordinary local excitation, but is dressed by a codimension-one topological surface ending on the monopole. In the effective two-dimensional description, this dressing is captured by a twist operator, and the outgoing radiation therefore lives in a twisted sector of the Hilbert space. 

We note that the above resolution relies on several crucial ingredients: first, one must have a magnetic monopole compatible with the symmetries that allow such topological surfaces; this is essentially a constraint on the monopole boundary conditions. Second, the type of matter mattered: clearly, the case with Weyl fermions or with non-Abelian symmetries will be more complicated. We also note that the interpretation of the outgoing mode remains unclear; see \cite{vanBeest:2023mbs, Tachikawa:2026cxd} for further discussion. Finally, there have been other proposed resolutions to the Callan-Rubakov puzzle that have some similar ingredients to the above discussion, but that ultimately rely on a different mechanism; see the list of relevant references in Sec.~\ref{sec:monopole}.

\bibliography{bibliography.bib}
\bibliographystyle{JHEP.bst}

\end{document}